\documentclass[12pt]{iopart}

\usepackage{amssymb}
\usepackage{amsbsy}
\usepackage{graphicx}
\usepackage{bbm}
\usepackage{bbold}
\usepackage{hyperref}

\usepackage{fancyhdr}
\newcommand{\vareps}{\varepsilon}

\begin{document}

\title[$\mathbf{k}\cdotp\mathbf{p}$ theory for 2D TMDCs]{$\mathbf{k}\cdotp\mathbf{p}$ theory for two-dimensional
transition metal dichalcogenide semiconductors}

\author{Andor Korm\'anyos$^1$, Guido Burkard$^1$}
\address{$^1$ Department of Physics, University of Konstanz, D-78464 Konstanz, Germany} 

\author{Martin Gmitra$^2$, Jaroslav Fabian$^2$}
\address{$^2$ Institute for Theoretical Physics, University of Regensburg, 
         D-93040 Regensburg, Germany}

\author{Viktor Z\'olyomi$^3$, Neil D. Drummond$^3$, Vladimir Fal'ko$^3$}

\address{$^3$ Department of Physics, Lancaster University, Lancaster LA1 4YB, United Kingdom}

\ead{andor.kormanyos@uni-konstanz.de}
\ead{guido.burkard@uni-konstanz.de}

\begin{abstract}

 We present  $\mathbf{k}\cdotp\mathbf{p}$ Hamiltonians parametrised by {\it ab initio} density 
functional theory calculations to describe the dispersion of the valence and conduction bands 
at their extrema (the $K$, $Q$, $\Gamma$,  and $M$ points of the hexagonal Brillouin zone) 
in atomic crystals of semiconducting monolayer transition metal dichalcogenides. 
We discuss the parametrisation of the essential parts of the  $\mathbf{k}\cdotp\mathbf{p}$ 
Hamiltonians for MoS$_2$, MoSe$_2$, MoTe$_2$, WS$_2$, WSe$_2$, and WTe$_2$, 
including the spin-splitting and spin-polarisation of the bands, and we
{briefly review} the vibrational properties of these materials. 
We then use $\mathbf{k}\cdotp\mathbf{p}$ theory to analyse optical transitions
in two-dimensional transition metal dichalcogenides over a broad spectral 
range that covers the Van Hove singularities in the band structure (the $M$
points). We also discuss the visualisation of scanning tunnelling microscopy maps.

\end{abstract}

\pacs{}
 
 

\tableofcontents

\section{Introduction}

Monolayers of  transition metal dichalcogenides (TMDCs) \cite{yoffe1972,mattheis}  
are truly two-dimensional (2D) semiconductors  
\cite{heinz2010,splendiani,korn,jones,xueato,sanfeng,menon,gedik,urbaszek2015a}, 
which hold great appeal for electronics and opto-electronics applications due to 
their direct band 
gap properties (which contrast  the indirect band gaps of three-dimensional layered 
crystals of TMDCs). Monolayer TMDCs have already been implemented in field-effect transistors 
\cite{kis-transistor1,kis-transistor2,hui-fang,hanwang,banerjee2013,banerjee2014,roelofs},
logical devices \cite{hanwang,kis-circuits}, and lateral and 
tunnelling optoelectronic structures 
\cite{britnell,pospischil,ross2014,jarillo-herrero2014,morpurgo2014a}.

Like graphene, the group-IVB monolayer TMDCs of chemical composition MX$_2$ 
(where M=Mo or W and X=S, Se and Te) considered in this work 
have hexagonal lattice structures, and 
the extrema  (valleys) in the dispersion relations of both the valence 
and conduction bands (VB and CB) can be found at the $K$ and $-K$ points of 
the hexagonal  Brillouin zone (BZ)\@. Unlike graphene, however, these 2D crystals 
do not have inversion symmetry. 
The minimalistic approach to the theoretical modelling of monolayer TMDCs is 
therefore based on mimicking 
them as graphene with a staggered 
sublattice potential that breaks inversion symmetry \cite{wyao,dxiao}. 
This approach captures certain
optical and transport effects related to the valley degree of freedom of the electrons 
\cite{dxiao,heinz2012,cui2012,cao,sallen,kioseoglou}. 
The staggered graphene analogue \cite{dxiao} has also been generalised 
to the tight-binding (TB) description of TMDCs
\cite{dxiao,rostami,zahid,liugb,capellutti,kosmider2,roldan-SOC,ghorbani-asl}, 
but this approach suffers from the large number of 
 atomic orbitals that have to be included on each site and the need for 
 beyond-nearest-neighbour hopping 
to account for the variation of the weight of individual atomic orbitals in the band 
wave functions across the BZ, as  revealed by detailed density 
functional theory (DFT) modelling 
(see, e.g., Figure \ref{fig:atomic-orbs}). 
{The accumulation of experimental data 
and the drive towards the implementation 
of monolayer TMDCs in practical devices call for theoretical
models of their electronic properties that are 
both detailed and compact, containing a limited number of 
parameters while still offering an accurate description.}

In this Review, we describe two complementary theoretical approaches that have
recently been used to achieve a 
detailed description 
of the electronic properties of these materials. 
One consists of {\it ab initio} DFT modelling of the band structure, which 
has the potential to be accurate. DFT can be combined with transport codes 
\cite{banerjee2014,ghorbani-asl,kaasbjerg2012,wook_kim,kaasbjerg2013,yao,changj,hongguo}
or used to calculate optical spectra \cite{ashwin,komsa,qiu,wirtz}, 
but \textit{ab initio} calculations 
are prohibitively expensive for many practical problems focused on modelling
devices and studies of, e.g.,  quantum dots \cite{sajat2,qd-yao}. 
Moreover, magnetic-field effects \cite{rostami,sajat2,liux,goerbig,yen-hung,rui-lin} 
and certain questions regarding neutral  and charged excitons \cite{AFG} 
cannot easily be addressed 
by  DFT-based techniques.  The second approach uses  
the $\mathbf{k}\cdotp\mathbf{p}$ methodology 
\cite{luttinger-kohn,BirPikus,dresselhaus-book,willatzen}, which exploits the 
symmetries of the system. 
This approach provides an accurate characterisation of the dispersion of the valence 
and conduction bands in the vicinity of, {e.g.,} the $K$ and $-K$ points 
{and other points of interest in} the BZ in terms of a relatively small 
number of parameters \cite{sajat1}. 
Magnetic-field and spin-orbit coupling effects can also be taken into account 
in a straightforward way \cite{sajat2}. 
In contrast to DFT modelling, this method is only valid in the vicinity of 
certain high-symmetry 
$\mathbf{k}$-space points; however, for those intervals, it enables one to quantify 
all the essential features of the electronic properties.
One can also relate a $\mathbf{k}\cdotp\mathbf{p}$ Hamiltonian
to a particular TB model \cite{dxiao,rostami,liugb}, although it is not necessary to 
set up a TB model in order to 
derive a  $\mathbf{k}\cdotp\mathbf{p}$  Hamiltonian. Here we present phenomenological
$\mathbf{k}\cdotp\mathbf{p}$ Hamiltonians derived for all extrema of the bands
(at the $K$, $Q$, $\Gamma$,  
and $M$ points of the BZ) using the symmetry properties of TMDC atomic crystals, with 
specific material parameters obtained by fitting them to the DFT
band structures of MoS$_2$, MoSe$_2$, MoTe$_2$, WS$_2$, WSe$_2$ and WTe$_2$. 
\footnote{
{Most of the recent theoretical and experimental work
has focused on the properties of MoS$_2$, MoSe$_2$, WS$_2$ and WSe$_2$, while 
MoTe$_2$ and WTe$_2$ have received much less attention. Bulk MoTe$_2$ 
with a trigonal prismatic coordination of the chalcogen atoms (see Figure \ref{fig:lattice}(a)) 
exists below $815^{\rm o}$C (known as $\alpha$-MoTe$_2$),  whereas 
above  $900^{\rm o}$C the crystal structure is 
monoclinic and the material becomes metallic ($\beta$-MoTe$_2$) \cite{bullett,pollmann}. 
Monolayer samples using liquid
exfoliation technique have been obtained from  $\alpha$-MoTe$_2$ \cite{nicolosi}, and 
the optical properties of monolayer \cite{heinz2014c} 
the transport properties of few-layer  $\alpha$-MoTe$_2$ \cite{morpurgo2014b,balicas} 
have been investigated recently, giving a clear motivation to include this material
in our review. Bulk WTe$_2$ has an orthorhombic crystal structure, where 
eight tellurium  atoms surround the tungsten atom in a distorted 
octahedral coordination \cite{brown,manzke}. 
Nevertheless, one would expect that it may be possible to grow monolayer WTe$_2$ with 
hexagonal prismatic coordination on a suitable substrate. For completeness, therefore,
we include this material as well, assuming that its hexagonal structure is stable.  
}}
The DFT calculations discussed 
in this Review were performed using the \textsc{vasp} \cite{vasp}  and 
\textsc{fleur} \cite{fleur} codes. The 
robustness of our results is well illustrated by the close agreement 
between the results obtained from these two  different first-principles 
codes and through  comparison to all available experimental results. 

{Finally, we note that the field of TMDCs, akin to that of 
graphene \cite{graphene-rev-1,graphene-rev-2,graphene-rev-3,graphene-rev-4}, 
has witnessed a large expansion over the last four years, encompassing both 
fundamental and more applications-oriented research directions. 
Here we focus on a particular topic that we think will be  important 
for the further development of this field. To limit the length of 
this review, some  fascinating topics related to, e.g., the valley-dependent 
optical selection rule 
or the exciton physics  are not discussed in detail here. 
We refer the interested reader to  complementary 
reviews 
\cite{naturenano-review,nature-review,kis-yazyev,chemsoc-review,deviceappl-review,fewlayermos2-review,annalsofphys-1,annalsofphys-2} instead. 
}

This Review is organised as follows. Section \ref{sec:lattice-geom} is 
devoted to the crystalline 
lattice parameters and vibrational properties of TMDCs. 
Sections \ref{sec:definitions} and \ref{sec:Dvb} discuss spin-splitting due to
spin--orbit coupling (SOC) and band width [relevant for angle-resolved 
photoemission spectroscopy (ARPES) studies of TMDCs].
Sections \ref{sec:K-point-main}, \ref{sec:Q-point-main},
\ref{sec:G-point-main}, and \ref{sec:M-point-main} describe the structure and 
parametrisation of $\mathbf{k}\cdotp\mathbf{p}$ Hamiltonians for 
$K$, $Q$, $\Gamma$, and $M$ points of the BZ, respectively.
Finally, we draw our conclusions in Section \ref{sec:conclusions}.


\section{Lattice parameters, band-structure calculations and vibrational properties} 
\label{sec:lattice-geom}

\begin{figure}[ht]
\begin{center}
 \includegraphics[scale=0.45]{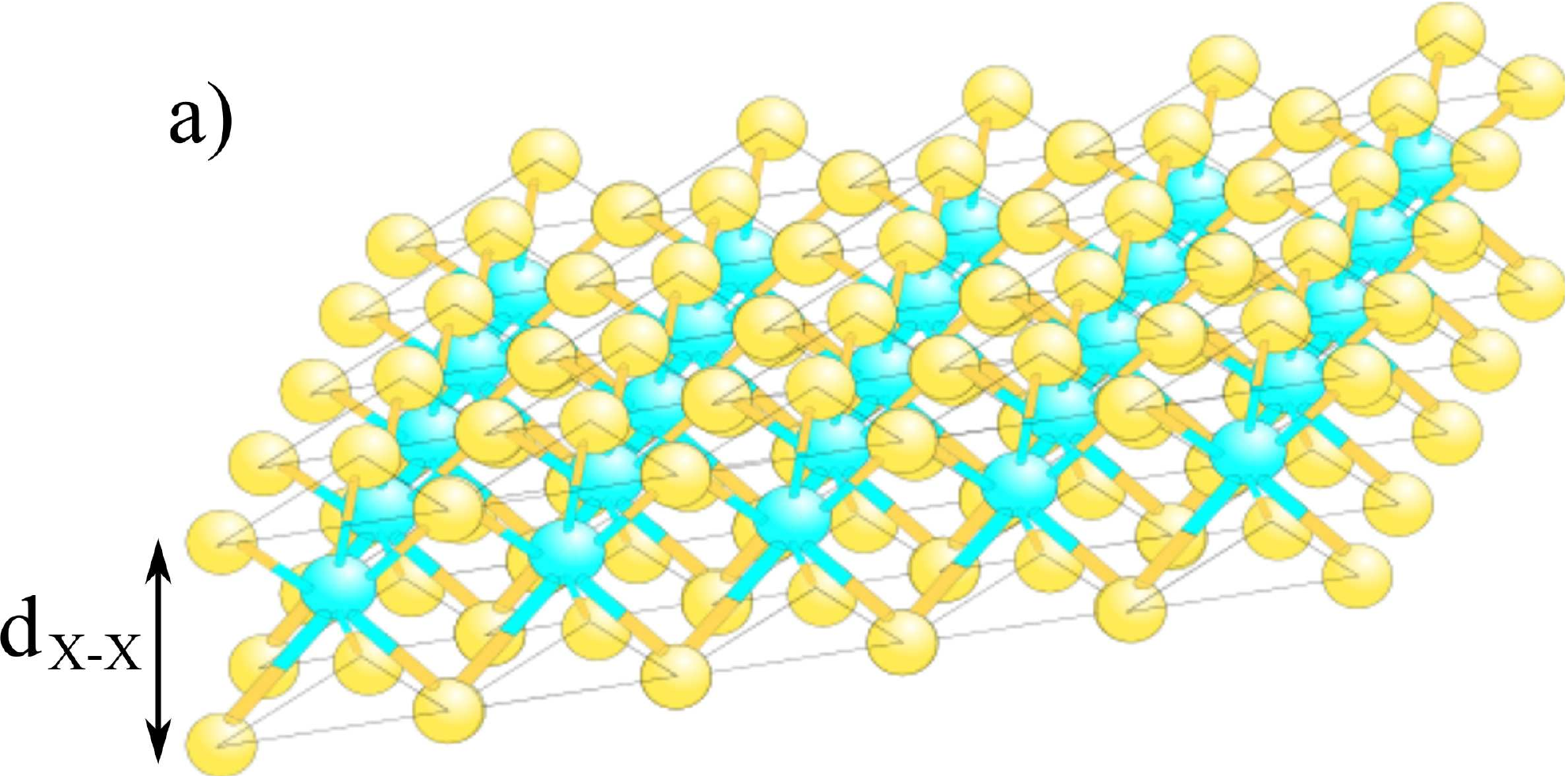}\,\,\,\,\,\,
 \includegraphics[scale=0.27]{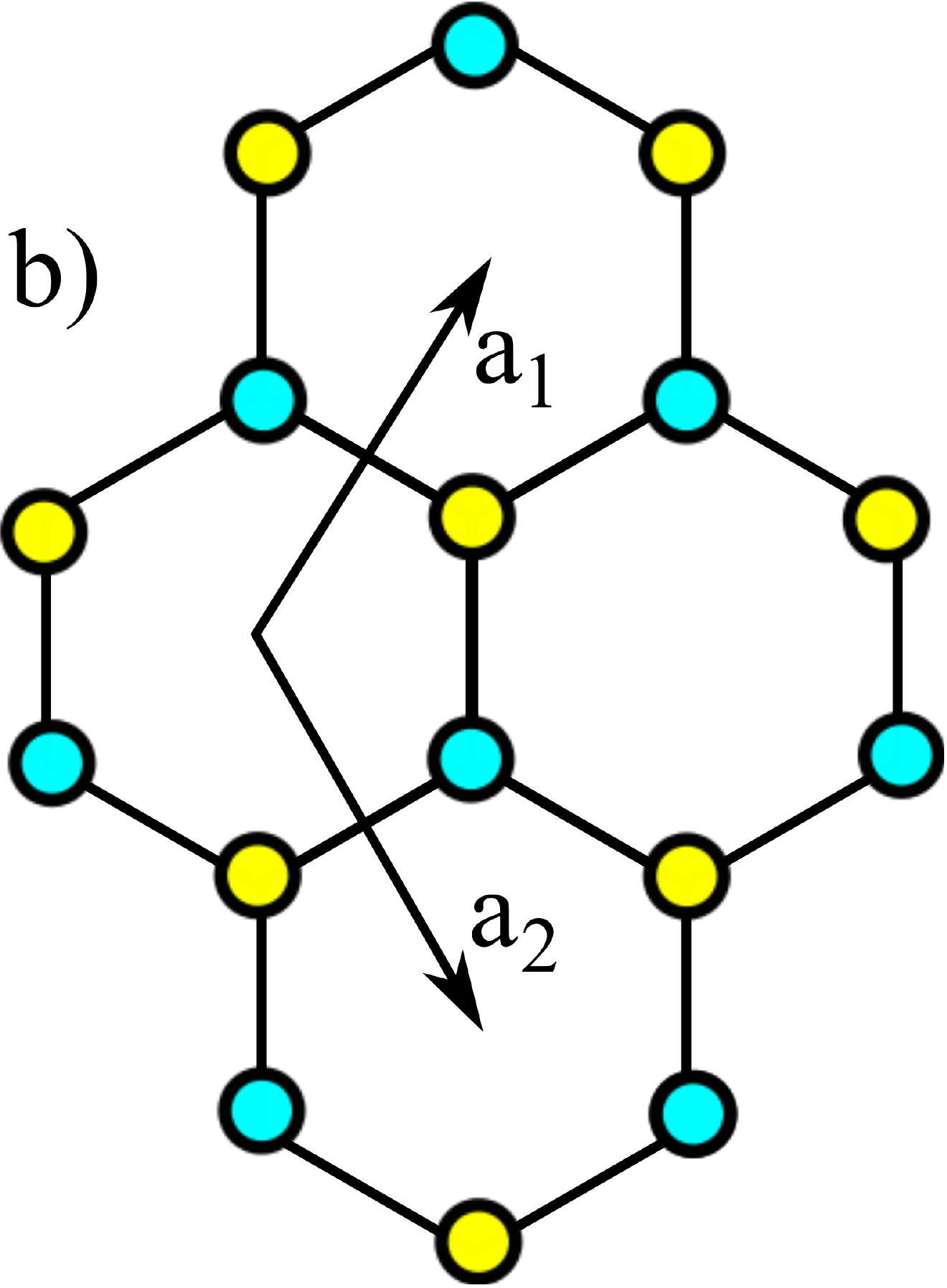}
 \end{center}
 \caption{Crystal structure of monolayer MX$_2$. a) Side view and b) top view. Metal atoms are 
 cyan and chalcogens are yellow. The lattice vectors $\mathbf{a}_1$ and $\mathbf{a}_2$ are 
             also shown.} 
 \label{fig:lattice}
\end{figure}

The crystal structure of each MX$_2$ monolayer  considered in this work
consists of three atomic layers, X--M--X. Within each layer the M or X atoms 
form a 2D hexagonal lattice: see Figure \ref{fig:lattice}.
The M atoms in the middle plane are surrounded by three nearest-neighbour X atoms
in both the bottom and the top layer so that the crystal has $D_{3h}$ symmetry.
The crystal structure is characterised by the in-plane lattice constant $a_0$ 
and the distance $d_{X-X}$ between the two chalcogen planes. 
It has already been noted \cite{yakobson} that certain details of the  band structure 
obtained from DFT calculations depend rather sensitively on $a_0$ and $d_{X-X}$. 
Indeed, we have also found that agreement with the available experimental results 
regarding, e.g., the  effective mass  $m^{\rm vb}_{\Gamma}$ at the $\Gamma$ point 
of the BZ or the energy difference  $E_{K \Gamma}$ between 
the top of the VB at the 
$K$ and $\Gamma$ points can only be achieved if the  
values of   $a_0$ and  $d_{X-X}$ fall in a rather narrow range.

As a first step, we have used two approaches to calculate the basic lattice parameters 
$a_0$ and $d_{X-X}$.  
The first approach  used \textsc{vasp} \cite{vasp}.
The \textsc{vasp} geometries were calculated using the Heyd--Scuseria--Ernzerhof 2006 
(HSE06) exact-exchange density functional \cite{hse06}. The plane-wave cutoff energy was 
set to 600 eV and the BZ was sampled by a $12 \times 12 \times 1$ 
Monkhorst--Pack grid. The vertical separation between the layers was set to $20$ \AA\@  
to make the interaction between the repeated images of the layer 
in the three-dimensional cell negligible. Optimisation was carried out until atomic 
forces fell below $0.005$ eV/\AA\@.
The second approach used the full-potential linearised augmented plane-wave (FLAPW) 
method as implemented in the \textsc{fleur} code \cite{fleur}. The FLAPW method is 
an all-electron method within DFT\@. The \textsc{fleur} code allows 2D systems to be studied
without constructing slabs in three-dimensionally periodic cells and the resulting electronic
spectra are free of plane-wave continua. All our \textsc{fleur} calculations were carried 
out with  a  cut-off ${\rm k_{max}}$ of $10.6$ eV$^{-1}$ for the plane-wave basis 
set and 144 ${\bf k}$ points corresponding to a $12\times 12\times 1$ Monkhorst--Pack grid 
in the irreducible wedge of the BZ\@. Muffin-tin radii of $1.0$, $1.21$, $1.27$, $1.27$, 
and $1.27$ \AA\,  were used for S, Se, Te, Mo, and W, respectively. 
We note that considering local orbitals for Mo ($s$, $p$), Se ($s$, $p$, $d$), 
and W ($s$, $p$, $f$)
to improve the linearised augmented plane-wave basis proved to be crucial for a 
correct description of the excited states.
We used the Perdew--Burke--Ernzerhof (PBE) generalised gradient approximation 
\cite{Perdew1996:PRL} to the exchange-correlation
potential. The structures were relaxed (with the effects of SOC included) until the forces 
were less than $0.0005$ eV/\AA\@.

The calculated values of $a_0$ and $d_{S-S}$  for monolayer TMDCs
are shown in Table \ref{tbl:geomparam} and compared
to measured values for the corresponding \emph{bulk} materials. 
The lattice  parameters obtained from the first of the DFT approaches described above
are shown in the rows  labelled by ``(HSE)'', the ones from the second approach 
are in the rows  labelled by ``(PBE)''. 
``(Exp)'' indicates experimental results found in the literature. 
\begin{table}[htb]
\caption{\label{tbl:geomparam}  Lattice vector $a_0$ and chalcogen--chalcogen distance 
$d_{X-X}$ as obtained from DFT calculations. Experimental values for the 
corresponding bulk material are shown in rows labelled by ``Exp''. For WTe$_2$ experimental results 
are only available for the orthorhombic structure and are therefore not shown.}
\begin{indented}
\lineup
\item[]
\begin{tabular}{@{}lllllll}
\br
       & MoS$_2$  & MoSe$_2$ & WS$_2$  & WSe$_2$ & MoTe$_2$ & WTe$_2$\\
\hline
$a_0$ [\AA] (HSE) & $3.1565$ & $3.289$ & $3.16$ & $3.291$ & $3.516$&  $3.521$\\
$a_0$ [\AA] (PBE) & $3.1854$ & $3.319$ & $3.18$ & $3.316$ & $3.557$&  $3.553$\\ 

\hline
$a_0$ [\AA] (Exp) & $3.1604^{a}$ & $3.288^{a}$ &   $3.154^{a,b}$ &  $3.286^{a}$  & $3.519^{j}$ & --\\ 
                  & $3.14^{c}$   & $3.299^{c}$ &   $3.1532^{d}$  &   $3.282^{c}$ & $3.522^{c,k}$& --\\
                  & $3.1602^{e}$ & $3.289^{e}$ &                 &   $3.282^{d}$  & $3.517^{a}$& --\\
                  & $3.1475^{f}$ & $3.290^{g}$ &                 &                 &  & --\\
                  &              &             &                 &                 & & --\\
\br
$d_{X-X}$ [\AA] (HSE) & $3.0996$ & $3.307$ & $3.1176$ & $3.327$ &  $3.5834$ & $3.5999$\\ 
$d_{X-X}$ [\AA] (PBE) & $3.1246$ & $3.4371$ & $3.1529$ & $3.471$ & $3.6195$ & $3.6394$ \\ 
\hline
$d_{X-X}$ [\AA] (Exp) & $3.17^{e}$ & $3.335^{e}$ & $3.14^{d}$ & $3.34^{d}$ & $3.604^{k}$ & --\\ 
\br
\end{tabular}
\item[] $^{a}${\cite{yoffe1969}, $^{b}$\cite{yang},  $^{c}$\cite{evans1972}, $^{d}$\cite{jellinek}, 
$^{e}$\cite{jellinek2}, $^{g}$\cite{evans1971}, $^{j}$\cite{poutinen},
$^{k}$\cite{pollmann}.}
\item[] $^{f}${\cite{young}, measurement at 293 K\@.}
\end{indented}
\end{table}
Although there is some scatter in the experimental data, 
Table \ref{tbl:geomparam} suggests that  using the 
HSE06 functional to  relax the monolayer crystal structure leads to  a good agreement 
with the \emph{room-temperature} empirical \emph{bulk}  $a_0$ values.  On the other hand, 
the PBE functional seems to slightly overestimates  $a_0$. However, the situation is less 
clear in the case of $d_{X-X}$.   We note that both the HSE06 and the PBE 
results are in good agreement with Reference \cite{kang}.

Recent experiments show  that the energy of the  photoluminescence peak is 
quite sensitive to the temperature \cite{korn,lagarde,rana}, which  
can be understood in terms of the dependence of the band structure on $a_0$ and $d_{X-X}$. 
Indeed, a recent computational study \cite{soklaski} was able to qualitatively reproduce the 
redshift of the photoluminescence peak of MoS$_2$ as a function of temperature 
by assuming  a thermal  expansion of the lattice. 
The good agreement between the calculated lattice parameters and 
the corresponding experimental ones suggests that, interestingly, the predictions
based on our DFT results are expected to be most accurate at room temperature 
(except for the band gap, which is known to be underestimated by DFT)\@.
To our knowledge systematic measurements of the temperature-dependence of the 
lattice parameters of 
bulk MX$_2$ have not been performed, except for MoS$_2$ \cite{young}.

As in the case of the lattice parameters, we have used both the 
\textsc{vasp} and the \textsc{fleur}  
codes to calculate  the band structures of monolayer TMDCs. 
For the  \textsc{vasp} calculations we used the HSE lattice parameters as input. 
The band structures were calculated in the local 
density approximation (LDA)\@. SOC was taken into account 
in the non-collinear magnetic structure approach  with the symmetry turned off. 
The charge density was obtained 
self-consistently using a $12 \times 12 \times 1$ ${\bf k}$-point grid and a 600 eV cutoff 
energy.  The results obtained by this method are shown in rows denoted by ``(HSE,LDA)'' in
Tables \ref{tbl:D_VB}--\ref{tbl:dft-at-G} below.  
For the \textsc{fleur}  calculations the charge densities obtained from the geometry 
relaxation calculations (see Section \ref{sec:lattice-geom})
were used for further calculation of the band structure and spin expectation values. 
SOC in \textsc{fleur} is included within the second variational method for the
valence electrons, whereas the core electrons are treated fully relativistically.
These results are in rows denoted by ``(PBE,PBE)'' 
in Tables \ref{tbl:D_VB}--\ref{tbl:dft-at-G} below.

One possibility, which we did not explore,  
is to use the HSE lattice parameters and the HSE06 functional for 
band-structure calculations, as in Reference \cite{kang}. 
We note that the results of Reference \cite{kang} seem to 
indicate that the HSE06 functional gives larger VB spin-splittings 
than found experimentally.

In addition to the band structure  of the TMDCs, which is our main focus in this work,  
electron--phonon coupling is also essential in order to understand transport 
\cite{kaasbjerg2012,wook_kim,kaasbjerg2013} and relaxation \cite{dery} processes.   
For completeness, we give a brief review  of  the  vibrational characteristics of 
monolayer TMDCs. 
{\it Ab initio} lattice-dynamics calculations indicate that single layers of
the TMDCs MoS$_2$, MoSe$_2$, WS$_2$, and WSe$_2$
are dynamically stable \cite{Molina_2011,Horzum_2013,Sahin_2013}, in 
agreement with experiments. 

A comprehensive group-theory analysis of the different polytypes 
and stacking arrangements of few-layer TMDCs is presented 
in Reference \cite{jorio}. The symmetry of few-layer structures determines 
which phonon modes are Raman-active, 
and therefore provides an important means of characterising samples.
As mentioned earlier, monolayer MX$_2$ has $D_{3h}$ point-group symmetry 
(see Table \ref{tbl:charactd3h} for the character table and irreducible representations). 
The six zone-centre optical phonon modes may be classified according to the irreducible
representations under which their eigenvectors transform: in the
twofold-degenerate $E^{\prime \prime}$ modes the metal atom remains stationary
while the chalcogen atoms vibrate in opposite in-plane directions; in the
twofold-degenerate $E^\prime$ modes the chalcogen atoms vibrate together
in-plane in the opposite direction to the metal atom; in the non-degenerate
$A_1^\prime$ mode the metal atom remains stationary while the chalcogen atoms
vibrate in opposite out-of-plane directions; finally, in the non-degenerate
$A_2^{\prime \prime}$ mode the chalcogen atoms vibrate together out-of-plane
in the opposite direction to the metal atom.  Of these vibrations, all but the
$A_2^{\prime \prime}$ mode are Raman-active.  Only the $E^\prime$ and
$A_2^{\prime \prime}$ modes are infrared-active.

DFT-LDA and DFT-PBE results for the phonon frequencies are summarised in Table
1 of Reference \cite{Ding_2011}.  There is a reasonable degree of agreement
between the LDA and PBE results, suggesting that the DFT phonon frequencies
are accurate. Subsequent theoretical studies
\cite{Molina_2011,Horzum_2013,Sahin_2013} have reproduced the results of
Reference \cite{Ding_2011} for the monolayer. 
Regarding WTe$_2$, we note that our calculations give real phonon 
frequencies in the whole BZ, indicating that the assumed hexagonal  
structure may  indeed be stable. 
In experimental studies of thin
films of WS$_2$, WSe$_2$, and MoS$_2$ it is found that modes that were Raman
inactive in the bulk become active in thin films and that there are small
shifts in the phonon frequencies on going from the bulk to a thin film
\cite{Luo_2013,Zhao_2013,Li_2012}.  Where comparison is possible, the
experimental Raman frequencies of thin films are in agreement with the DFT
results.


\section{Band-edge energy differences and spin-splittings}
\label{sec:definitions}

\begin{figure}[ht]
\begin{center}
 \includegraphics[scale=0.8]{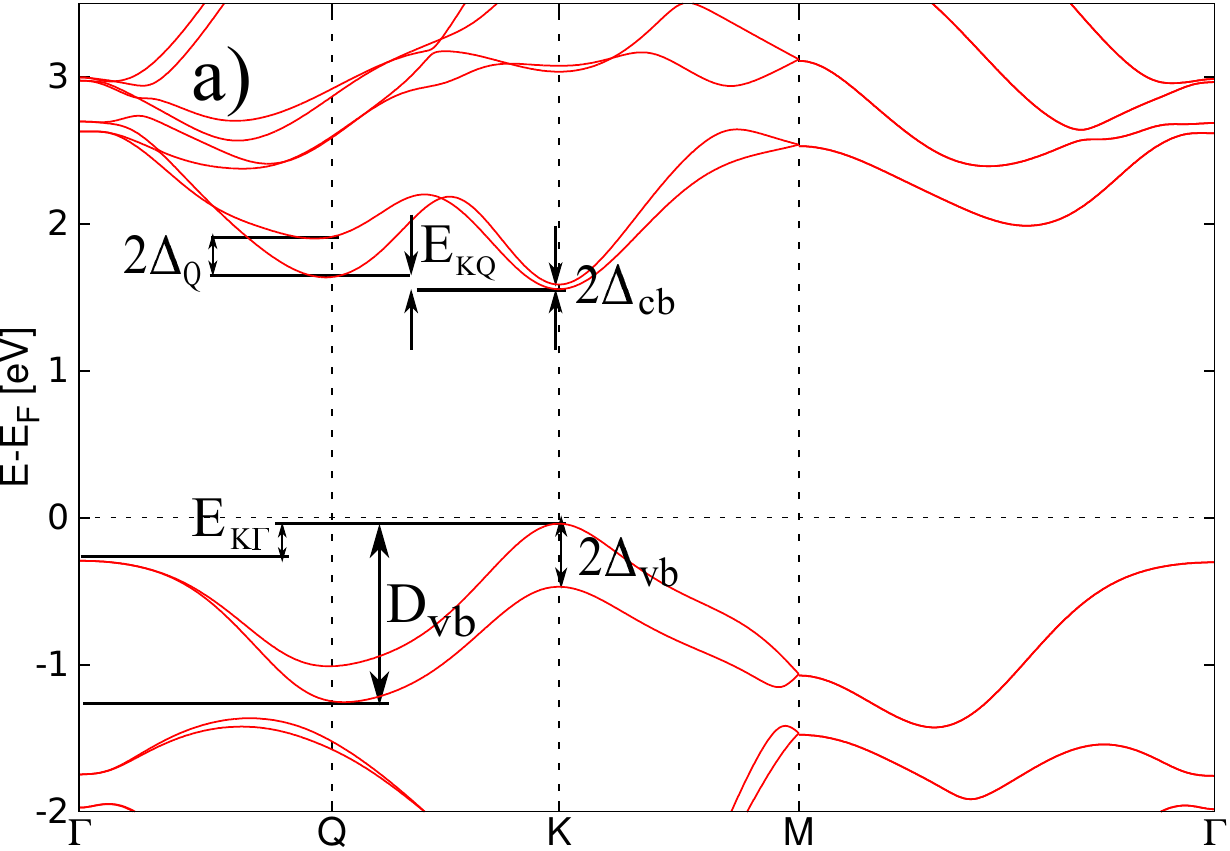}\\
 \includegraphics[scale=1.1]{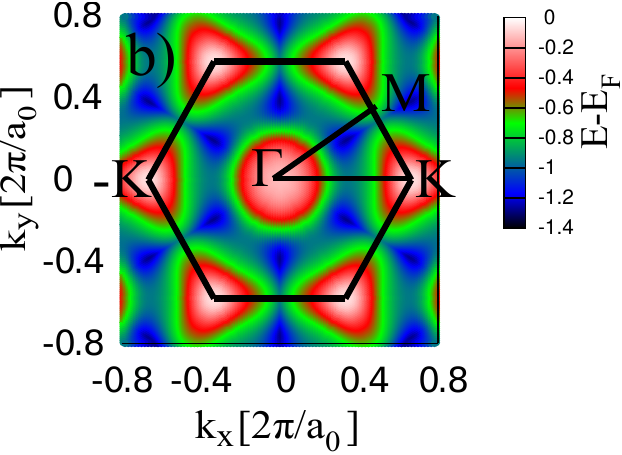}\,
 \includegraphics[scale=1.1]{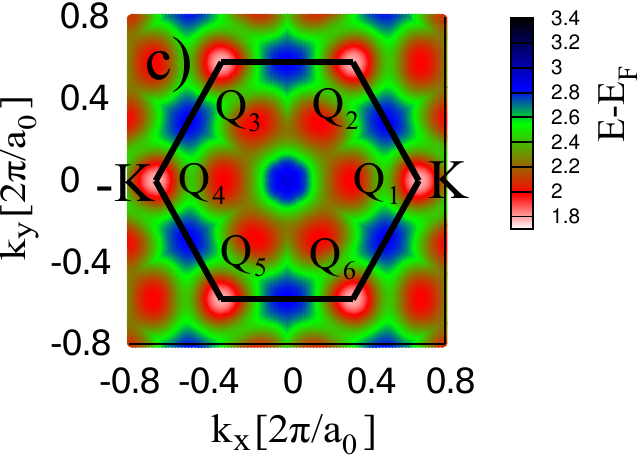}
 \end{center}
 \caption{
 Overview of the band structure of  monolayer TMDCs as obtained from DFT calculations. 
 a) Dispersion along the $\Gamma$--$K$--$M$--$\Gamma$ line in the BZ\@. 
    SOC is taken into account. 
    Various band-edge energy differences and spin-splittings are also indicated; for 
    definitions see the main text. 
     b) Dispersion of the VB  as a function of the wavevector $\mathbf{k}$ in the whole BZ\@. 
        The hexagonal BZ is denoted by thick black lines. 
     c) The same as b)  for the CB\@.  In b) and c) SOC is neglected.} 
 \label{fig:defs}
\end{figure}

Detailed discussion of the conduction and valence band dispersions  in the vicinity of the 
$\mathbf{k}$-space points of interest ($K$, $Q$, $\Gamma$, and $M$) will be given in 
Sections \ref{sec:K-point-main}, \ref{sec:Q-point-main}, \ref{sec:G-point-main}, 
and \ref{sec:M-point-main}. 
In this section we briefly introduce the various band-splittings and  band-edge 
energy differences that 
we use to characterise the band structure. 
An overview of the band structure obtained from DFT calculations is shown 
in Figure \ref{fig:defs}.
The direct band gap $E_{\rm bg}$ of monolayer TMDCs can be found at the $K$ and $-K$ points 
of the BZ\@. Due to the lack of inversion symmetry, all bands are split by the intrinsic 
SOC except at the time-reversal 
invariant points $M$ and $\Gamma$. We denote by $2\Delta_{\rm vb}$ and $2\Delta_{\rm cb}$  
the spin-splitting of the  VB and CB, respectively. 
There are another six minima in the CB that might be important, e.g., for 
transport or relaxation processes in certain compounds. We denote these  
points by $Q_{i}$, $i=1\dots 6$. 
They can be found roughly half way between the $K$ ($-K$)  and the $\Gamma$ points. 
The spin-splitting of the CB at  $Q_{i}$  given by $2\Delta_{Q}$. 
The importance of the $Q_{i}$ points depends, amongst other things, on the energy difference
between the bottom of the CB at the $K$ and $Q_{i}$ points. This energy difference is denoted 
by $E_{\rm KQ}$. Looking at the VB now, 
the energy difference between the top of the VB at  $K$ and $\Gamma$  is
denoted by $E_{\rm K\Gamma}$. Finally, since it is directly available in recent 
ARPES measurements 
\cite{osgood,shen,hasan}, we also record the width 
of the VB, which we define as the energy difference between the maximum of 
the VB at $K$ and the minimum that can be found on the $\Gamma$--$K$ line.

\begin{figure}[ht]
\begin{center}
 \includegraphics[scale=0.65]{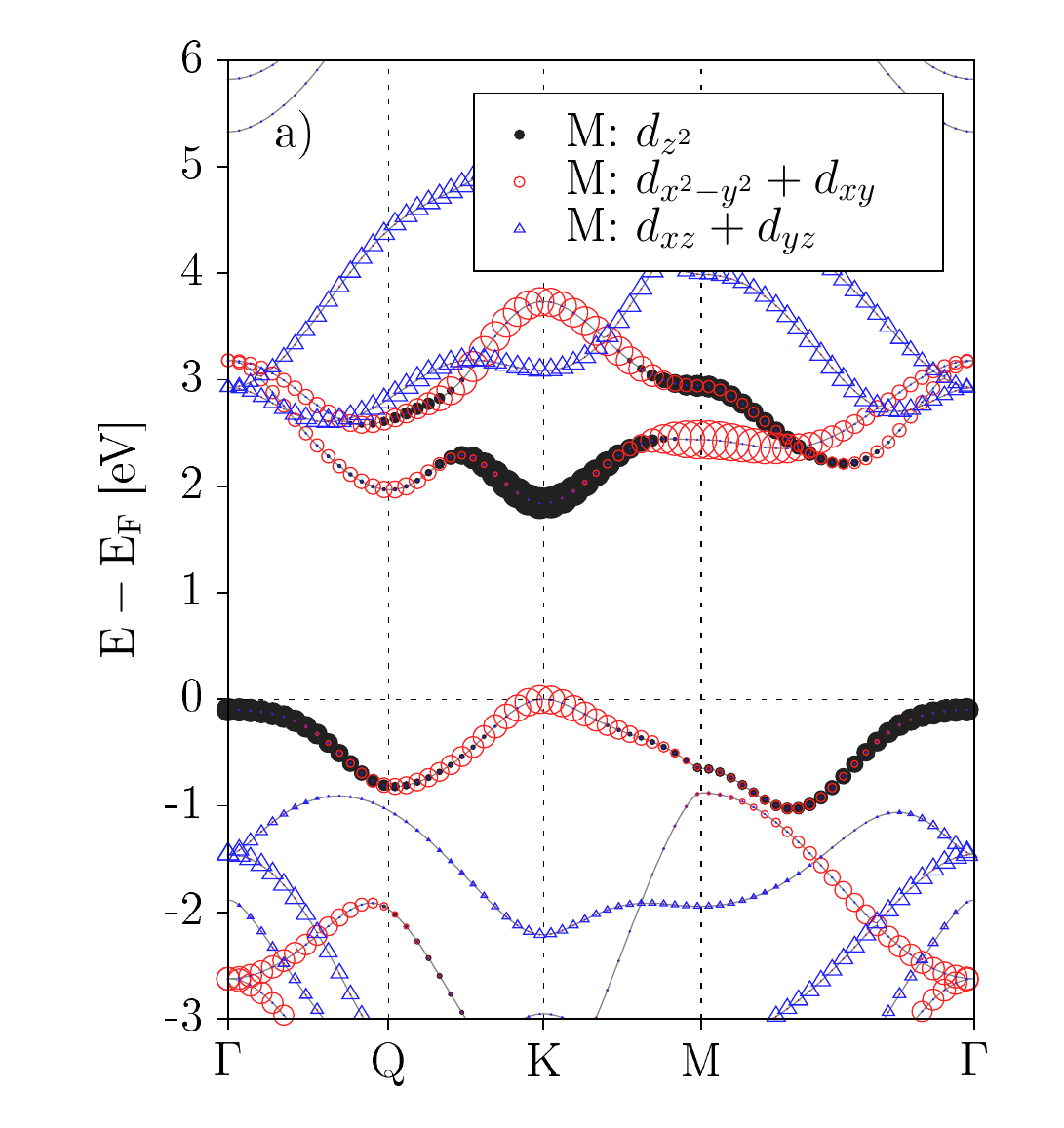}
 \includegraphics[scale=0.65]{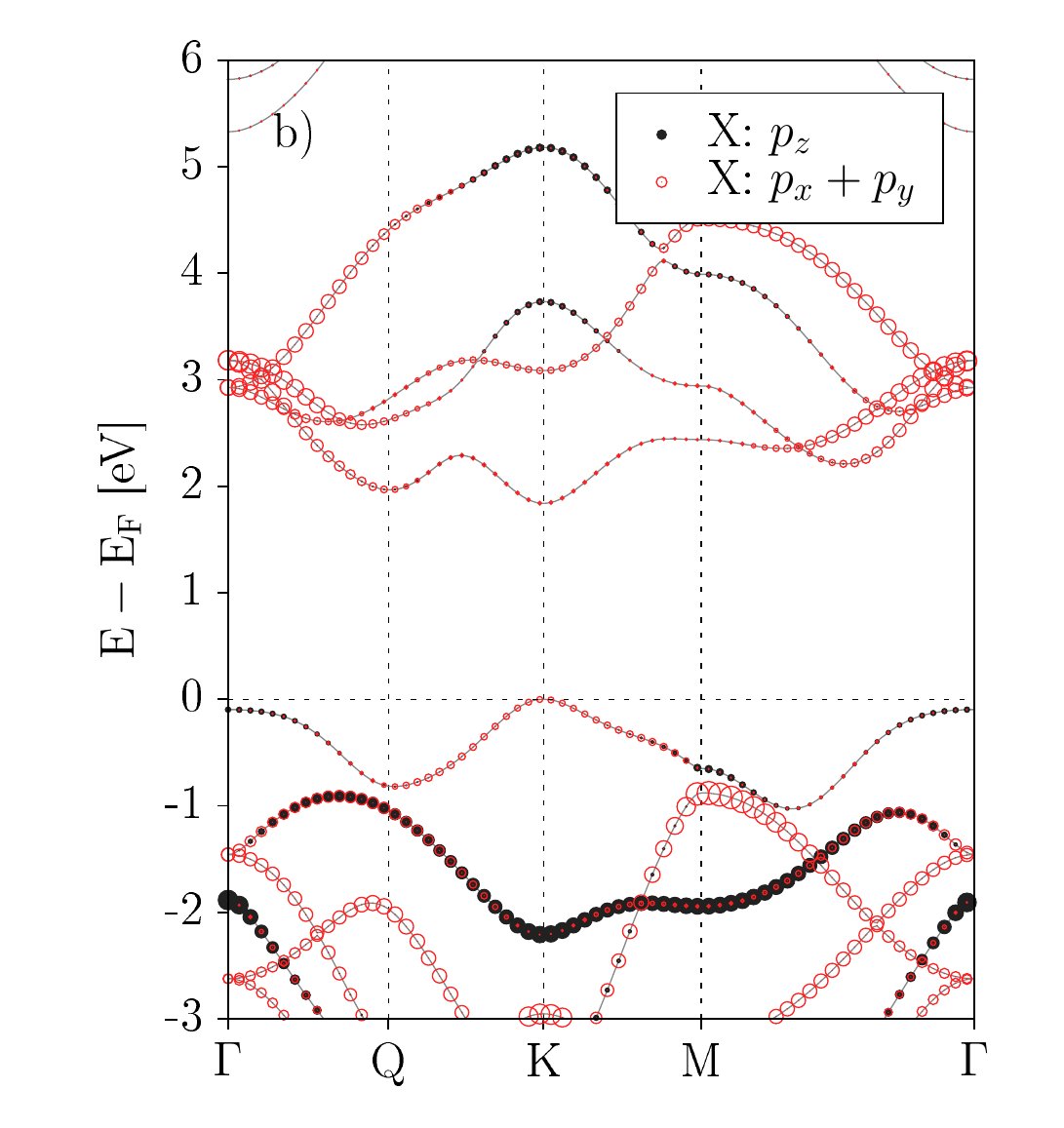}
 \end{center}
 \caption{Atomic orbital weights in the energy bands of MX$_2$. a) $d$ orbitals of 
         the metal atom, and
   b) $p$ orbitals of the chalcogen atoms. The size of each symbol is proportional 
      to the weight of the
      atomic orbital. SOC was neglected in these calculations. } 
 \label{fig:atomic-orbs}
\end{figure}
Certain  properties of TMDCs are easier to understand if one considers  which 
atomic orbitals contribute to a given band at a given $\mathbf{k}$-space point.
For example, as pointed out in, e.g., References 
\cite{dxiao,kosmider2,roldan-SOC,schwingenschlogl},
the different atomic orbital composition can explain the difference in the 
spin-splitting magnitude of the CB and VB at the $K$ point. 
{Furthermore, the atomic orbital composition of the energy bands underlies the 
 tight-binding modelling of TMDCs \cite{dxiao,rostami,zahid,capellutti} and 
  was also important in developing the $\mathbf{k}\cdot\mathbf{p}$ model 
  \cite{sajat2,sajat1}.}
The contribution of individual atomic orbitals to a given band 
 is shown in Figure \ref{fig:atomic-orbs} for the 
$d$ orbitals of the metal atoms and the $p$ orbitals of the chalcogens 
(the weights of other  atomic orbitals are much smaller). 
Comparing  Figures \ref{fig:atomic-orbs}(a) and (b) we find
that in general more than one type of atomic 
orbital contributes to both the CB and the VB and the weight of the 
atomic orbitals changes throughout the BZ\@. 
Setting up   a  consistent tight-binding model for TMDCs is therefore
more difficult than is the case for, e.g., graphene.


\section{Valence band width $D_{\rm vb}$}
\label{sec:Dvb}

An observable that can be directly compared to experimental ARPES measurements 
\cite{osgood,shen,hasan} 
is the width of the VB $D_{\rm vb}$. In order to be able to 
compare the experimental and theoretical results,
we define $D_{\rm vb}$ to be the difference between the top of the VB  at the  
$K$ point and the minimum, which lies between the $\Gamma$ and $K$ points: 
see Figure \ref{fig:defs}. 
(Note that the absolute minimum of the VB is not at this $\mathbf{k}$-space  point. 
However, Reference \cite{shen} shows 
the dispersion only between $\Gamma$ and $K$; therefore we use the definition 
of $D_{\rm vb}$ given above.) 
Comparison between the calculated and experimental values is given in Table \ref{tbl:D_VB}.

\begin{table}[htb]
\caption{\label{tbl:D_VB} The  width of the VB as obtained from DFT calculations.
Experimental values are shown in the row denoted by ``Exp''.}
\begin{indented}
\lineup
\item[]
\begin{tabular}{@{}lllllll}
\br
   D$_{\rm vb}$ [eV]     & MoS$_2$  & MoSe$_2$ & WS$_2$  & WSe$_2$ & MoTe$_2$ & WTe$_2$\\
\hline
  (HSE,LDA)  & $0.911$ & $0.84$  & $1.215$ & $1.132$  & $0.657$ & $0.933$ \\ 
  (PBE,PBE)  & $0.896$ & $0.84$ &  $1.207$ & $1.136$  & $0.688$ & $0.965$  \\ 
\hline
  Exp          & $\approx 0.8^{\rm a}$,    & $\approx 1.0^{\rm c}$  &  &  & &\\
               & $\approx 0.9$--$1.0^{\rm b}$ &  &  & & & \\
\br
\end{tabular}
\item[] $^{\rm a}${\cite{osgood}, exfoliated samples  on a SiO substrate. }
\item[] $^{\rm b}${\cite{hasan}, samples grown by chemical 
                  vapour deposition on a highly oriented pyrolytic graphite (HOPG) substrate. }
\item[] $^{\rm c}${\cite{shen},  samples grown by molecular beam epitaxy (MBE) on bilayer 
                   graphene on top of SiC (0001).}
\end{indented}
\end{table}

{In the case of MoS$_2$, Reference \cite{osgood} reported that the VB is narrower than 
the  calculated one by $\approx 10 \%$, whereas for MoSe$_2$ \cite{shen} the opposite 
seems to be  true. Reference \cite{osgood} also provides a comparison between calculations and 
the ARPES band structures of bilayer, trilayer and bulk MoS$_2$, showing a better agreement
than is found for monolayer MoS$_2$. Furthermore, a good  agreement between DFT calculations and 
ARPES measurements  for the VB was observed for bulk MoS$_2$ and MoSe$_2$ \cite{pollmann,mahatha} 
and for MoTe$_2$ \cite{pollmann}. 
The orbital composition of the VB away from the $K$ point 
is not purely of Mo $d$ orbital type:
$p$ orbitals of X atoms are also admixed (see Figure \ref{fig:atomic-orbs}); hence
$D_{\rm vb}$ in monolayers  can  be  sensitive 
to interactions with substrates, which are not 
considered in our calculations and which might explain some 
of the differences with  respect to  measurements.}


\section{Effective  model at the $K$  and $-K$ points}
\label{sec:K-point-main}

\subsection{$K$ and $-K$ points}

The physics around the $K$ and $-K$  points has attracted the most 
attention both experimentally  and theoretically so far. This is mainly due to the 
exciting optical properties of these materials at the direct band gap, 
which can be found at the $K$ and $-K$ points. Moreover, it turns out that the effect of 
SOC is strong at this BZ point, leading to spin-split and spin-polarized bands. 
Since the $K$ and $-K$ points are connected by time-reversal symmetry, 
the polarization of the bands has to be opposite 
at  $K$ and $-K$, i.e., the spin  and the valley degrees of freedom are coupled \cite{dxiao}. 
We start our discussion  in Section \ref{subsec:K-point-params} with a basic  
characterization of the band structure in terms of effective masses and spin-splittings. 
Then, in Section \ref{subsec:kp-at-K}, a detailed  $\mathbf{k}\cdot\mathbf{p}$ 
theory is presented which captures the salient features of the 
DFT band structure and allows us to interpret the results of 
recent experiments \cite{iwasa,macneill,atac,aivazian,heinz2014b}.

\subsection{Basic characterization and material parameters}
\label{subsec:K-point-params}

The aim of this section  is twofold. 
First, we want to point out that there is a difference between 
the MoX$_2$ and WX$_2$ materials regarding the sign 
of the SOC constant in the CB 
(for a microscopic explanation see References 
\cite{liugb}, \cite{kosmider2} and \cite{sajat2}).
This difference is important for the interpretation of  
experiments in which properties of A and B excitons \cite{heinz2012,sallen} are compared 
(for introduction to exciton physics see e.g., \cite{ivchenko-pikus}).
Second, we  report  effective masses and spin-splittings   extracted
from  our DFT calculations and compare them to experimental results, where available; see 
Tables \ref{tbl:K-dftparams-cb} and  \ref{tbl:K-dftparams-vb}.

One of the phenomena that first sparked strong interest in monolayer TMDCs was 
the pronounced effect  of  SOC on the  VB  around the $K$ and $-K$  points. 
SOC leads to the spin-splitting and spin-polarization of the VB and the 
energy scale associated with SOC is several hundreds of meVs: see Table \ref{tbl:K-dftparams-vb}.
SOC in the VB was first studied using 
DFT calculations \cite{schwingenschlogl,lambrecht,hawrylak,kosmider1},
but  it can be readily understood using, e.g, a tight-binding model and 
first-order perturbation theory \cite{dxiao,kosmider2,roldan-SOC}. 
An experimental signature of the spin-splitting of the VB is 
the energy difference of the A and B excitons \cite{heinz2012,sallen}.

SOC also affects the CB\@. This was initially neglected, mainly because 
in MoS$_2$, which is the most widely studied of the TMDCs, it is indeed a small effect 
and it was assumed that the situation would be similar in other monolayer TMDCs. 
In general the magnitude of the spin-splitting of the CB is 
only 7--10\% of that of the VB, with the exception of MoS$_2$, 
where it is only $\approx 2\%$:  see Table \ref{tbl:K-dftparams-cb}.
However, in absolute terms it is an energy scale that can be important 
at low temperatures  and in ballistic samples. Note that the SOC 
in the CB  at the $K$ point is a more subtle effect than in the VB\@. 
In the simplest theoretical approximation,
which assumes that 
it is sufficient to consider only the $d_{z^2}$ atomic orbitals of the metal atoms, 
the SOC vanishes. DFT calculations, on the other hand, indicate that there 
is a finite spin-splitting in the CB at the $K$ point
\cite{sajat1,schwingenschlogl,lambrecht,hawrylak,kosmider1}.  

\begin{figure}[ht]
\begin{center}
 \includegraphics[scale=0.5]{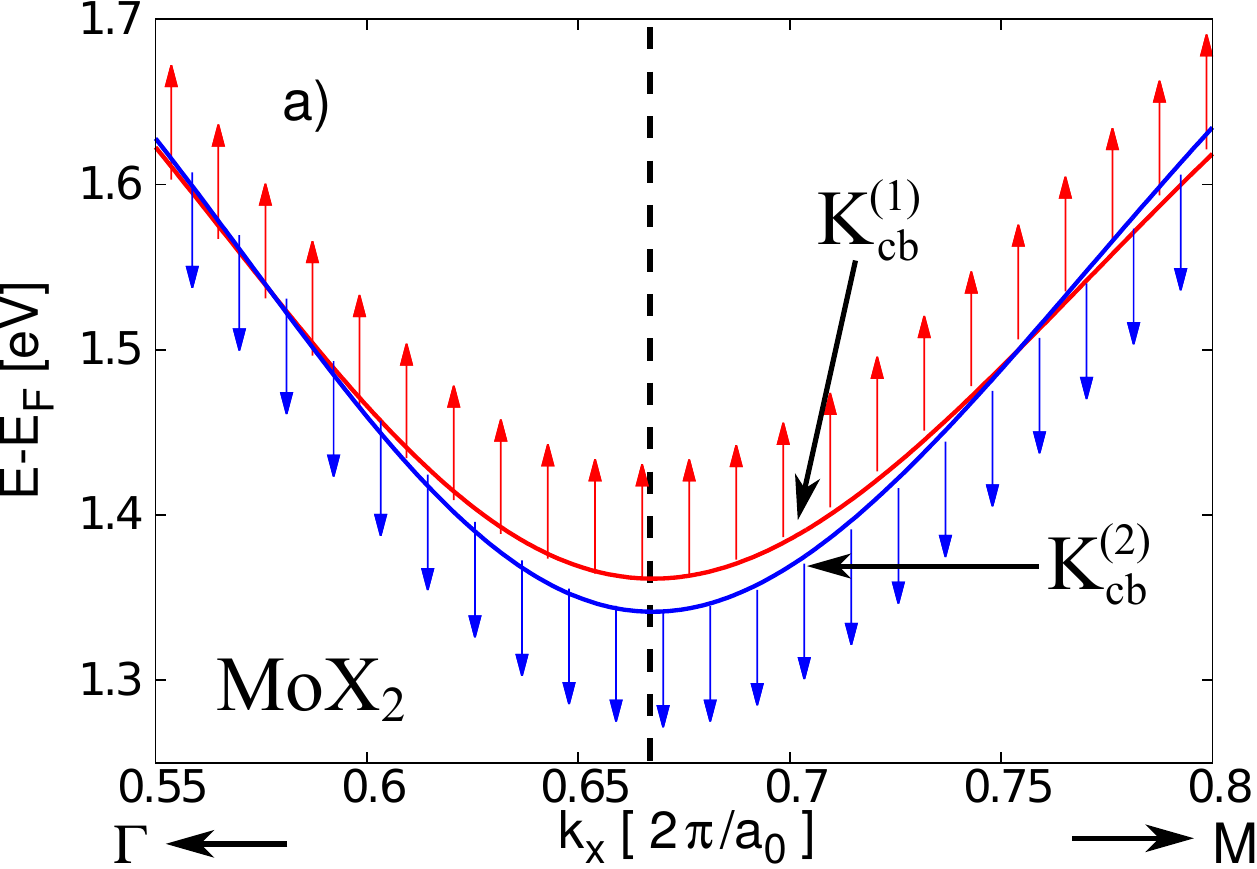}
 \includegraphics[scale=0.5]{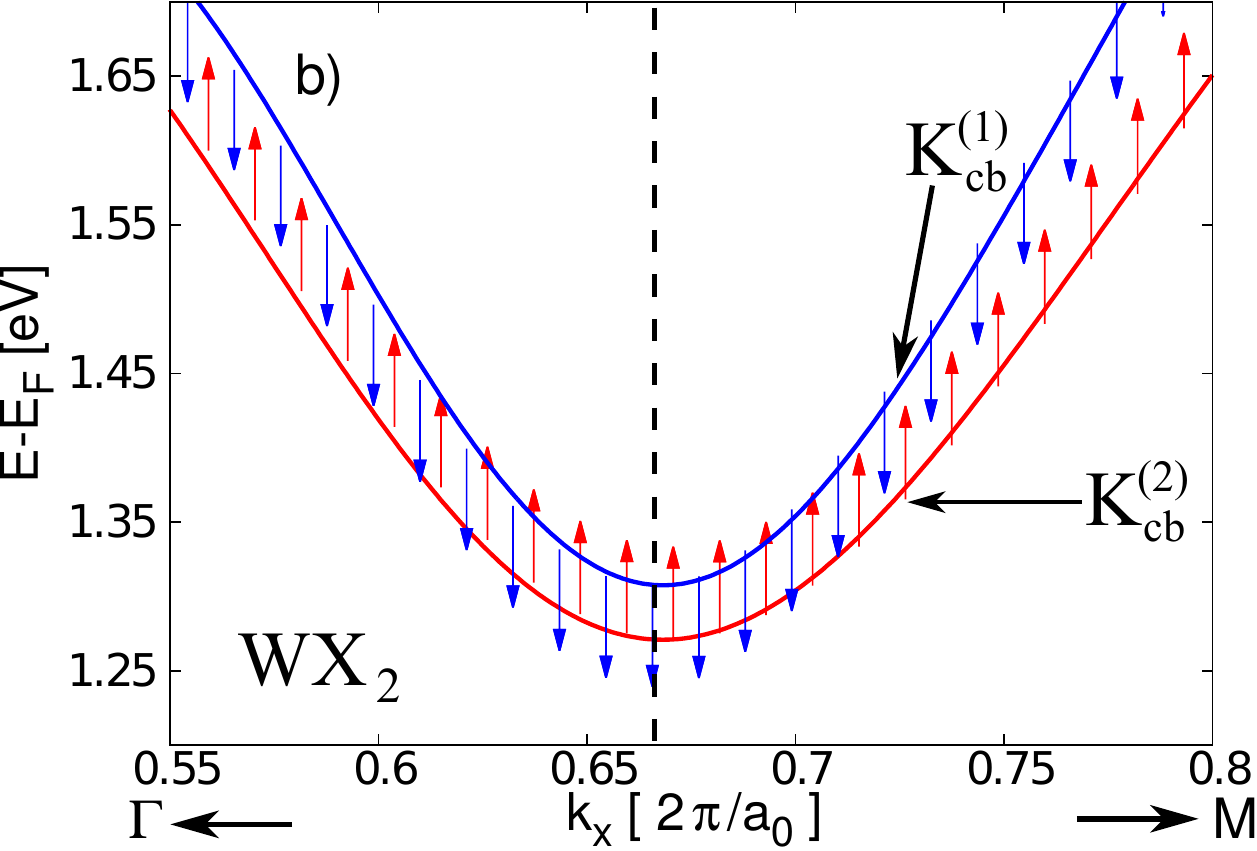}
 \includegraphics[scale=0.5]{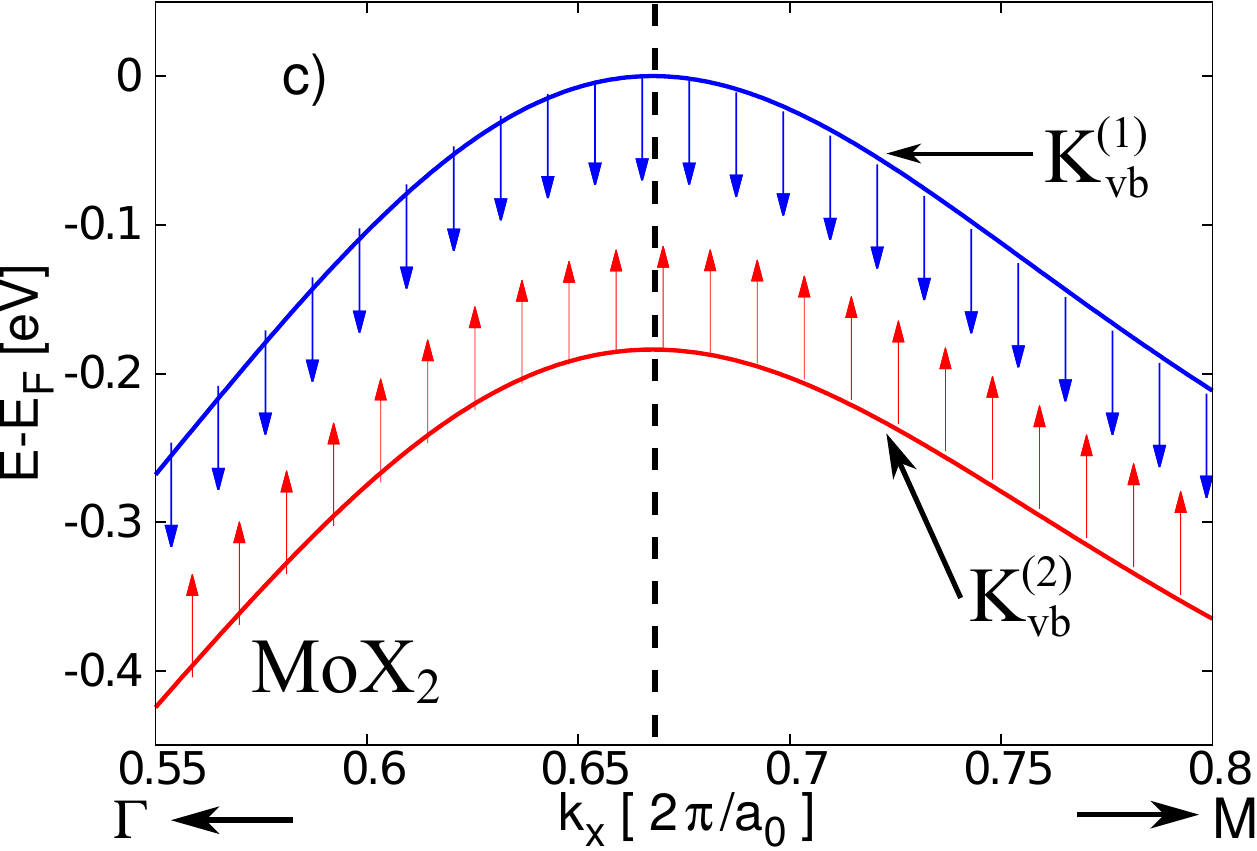}
 \includegraphics[scale=0.5]{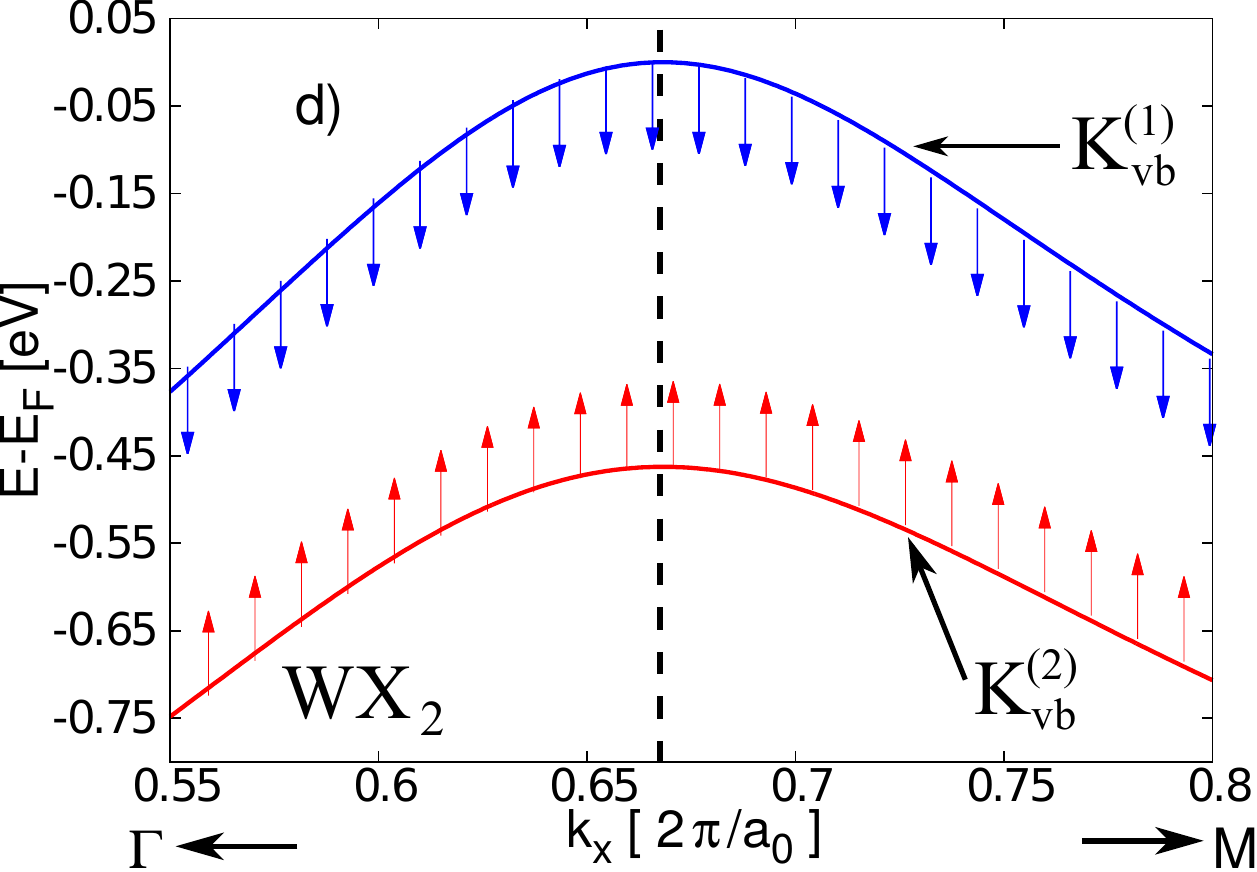}
 \end{center}
 \caption{Spin polarisation and dispersion of the spin-split CB and VB in the vicinity of the 
          $K$ point  from DFT calculations. Arrows show the direction of the 
           spin expectation values 
          (red: spin-up, blue: spin-down). a) and c) results for MoX$_2$; b) and d) 
           results for  WX$_2$. 
          Note that the order of spin-up and spin-down bands in the CB is different 
          for MoX$_2$  and WX$_2$.  
          The vertical dashed line shows the position of the $K$ point.
          The actual calculations were performed for MoSe$_2$ and WSe$_2$ 
          using the ``(PBE,PBE)'' approach. 
 \label{fig:spins-at-K}
 }
\end{figure}

As it turns out, the SOC in the CB can be understood in terms of a competition 
between two contributions \cite{liugb,kosmider2,sajat2,sajat1,ochoa-roldan}:
i) a first-order contribution  from the chalcogen atoms, which have a small, but 
finite weight \cite{liugb,roldan-SOC} and 
ii) a second-order contribution due to  the coupling to other bands 
\cite{liugb,kosmider2,sajat2,sajat1}, where the
$d_{xz}$ and $d_{yz}$ atomic orbitals have large  weights; see Figure \ref{fig:atomic-orbs}. 
Due to this competition  the spin-polarisation of the spin-split CBs is 
different in MoX$_2$ and WX$_2$. 
Our latest results were obtained using the \textsc{fleur} code, which allows the explicit 
calculation of the spin expectation 
value $\langle s_z \rangle $  in a given band. We find that the spin-split CB with
$\langle s_z \rangle >0$  ($\langle s_z \rangle <0$) 
is higher (lower) in energy in  MoX$_2$, while the opposite is true for WX$_2$:
see Figures~\ref{fig:spins-at-K}(a) and \ref{fig:spins-at-K}(b), in which
the CBs of MoSe$_2$ and WSe$_2$ are shown, respectively.
By contrast,  as shown in  Figures~\ref{fig:spins-at-K}(c) and \ref{fig:spins-at-K}(d), 
in the VB the sign of $\langle s_z \rangle $ is the same for both  
MoX$_2$ and  WX$_2$. {Furthermore, as can be seen in 
Figures~\ref{fig:spins-at-K}(a) and ~\ref{fig:spins-at-K}(b),  
the band with the lighter effective mass is lower in energy for 
MoX$_2$, leading to band crossing of the two spin-split bands in the vicinity of the
$K$ and $-K$ points \cite{liugb,kosmider2,sajat2}, whereas for WX$_2$ the 
lighter spin-split band is higher in energy and therefore there is no band crossing. 
(MoTe$_2$ is somewhat special in that the crossing of the spin-split bands 
on the $\Gamma$--$K$ line is absent. 
The other band crossing, on the $K$--$M$ line, is present).}
These differences notwithstanding, there is a spin--valley coupling in 
the CB  similar to the VB\@. 
In Figure  \ref{fig:spins-at-K} we  also introduce the notation 
$K_{\rm vb}^{(1)}$ ($K_{\rm vb}^{(2)}$) for the higher-in-energy (lower-in-energy)
spin-split VB, and similarly for the CB\@.  
As a consequence of   the spin polarisation of the bands in optical experiments
the lowest-energy spin-allowed transition is $K_{\rm vb}^{(1)} \rightarrow K_{\rm cb}^{(2)}$ 
for  MoX$_2$ and $K_{\rm vb}^{(1)} \rightarrow K_{\rm cb}^{(1)}$ for  WX$_2$. 
We note that very recently the first spin-resolved ARPES measurement on bulk 
WSe$_2$ has appeared \cite{king2014b} and seems to indicate an out-of-plane 
spin polarisation of 
the spin-split VB around $K$ and $-K$ points.  Assuming that the measurements  
predominantly probe the top layer \cite{king2014b}, i.e., effectively a monolayer sample, 
they are in agreement with the DFT calculations presented here.

The dispersion around the $K$ and $-K$ points is not simply parabolic \cite{sajat1},
which has to be borne in mind when  fitting the band structure to obtain 
the effective masses and other band parameters. This can already be appreciated 
in Figures \ref{fig:defs}(b) and (c), where a trigonal warping (TW) of the dispersion 
around the  $K$ and $-K$ points can clearly be seen.  The TW is more pronounced in the 
VB than in the CB\@. 
In the simplest approximation this can be taken into account by a 
cubic term in the dispersion. 
Therefore the dispersion of each spin-split band in the VB and the CB can be described by 
\begin{equation}
 E_{K}(\mathbf{q})=\frac{\hbar^2 \mathbf{q}^2}{2 m_{\rm eff}} + C_{3w} |\mathbf{q}|^3 \cos (3\varphi_{\mathbf{q}}), 
 \label{eq:K-dispersion}
\end{equation}
where the wavevector $\mathbf{q}=(q_x,q_y)$ is measured from the $K$ point, 
$\varphi_{\mathbf{q}}=\arctan({q_y}/{q_x})$, 
$m_{\rm eff}$ is the effective mass of the given band, and  $C_{3w}$ is a 
parameter describing the TW\@. 
The derivation of  $E_{K}(\mathbf{q})$ based on a multi-band 
$\mathbf{k}\cdot\mathbf{p}$ model is presented 
in Section \ref{subsec:kp-at-K} and \ref{sec:seven-band-at-K}. We note that a similar 
model was recently used in Reference \cite{hongyiyu}.

The values of the   $m_{\rm eff}$ and $C_{3w}$ that we have extracted from our DFT calculations 
for each band and material are given in Tables \ref{tbl:K-dftparams-cb} 
and \ref{tbl:K-dftparams-vb}. 
 We note that several works have already presented  tables of, e.g., 
 effective masses \cite{changj,ashwin,yakobson,peelaers,lake,kuc,wenxu} 
 for different monolayer TMDCs.
 However, the effects of SOC have often been neglected leading to, e.g.,  the conclusion that 
 the effective masses  of the spin-split VBs are the same. Recent experimental 
 evidence shows that this is not  the case \cite{shen,dowben}.  
  Moreover, due to the presence of the TW,
  some care has to be taken when defining the effective mass and, especially, when choosing 
  the fitting range that is  used to  obtain it from a DFT  band structure. 
All our DFT band-structure calculations were performed along the $\Gamma$--$K$--$M$ 
line in the BZ\@.  
We first fitted  $m_{\rm eff}$, i.e., we set $C_{3w}=0$  in Equation \eref{eq:K-dispersion}. 
The fitting  range  corresponded to $5\%$ of the $\Gamma$--$K$ distance.  
The dispersion over such range was considered   to be  isotropic and the difference 
in the effective masses along $K$--$\Gamma$ and  $K$--$M$ was neglected. 
Therefore the effective masses shown in Tables \ref{tbl:K-dftparams-cb} 
and \ref{tbl:K-dftparams-vb} 
characterise, strictly speaking, a rather narrow  vicinity of the band edge.  
The non-parabolicity of the band structure and the trigonal distortion of the 
constant energy contours, 
described by the second term in Equation \eref{eq:K-dispersion}, was taken into 
account in a second step,
whereby Equation \eref{eq:K-dispersion} was fitted over a wider range 
(typically $\approx 10\%$ of the  $\Gamma$--$K$ distance),
but $m_{\rm eff}$, obtained in the previous step, was kept fixed. 
This two-step fitting was needed
to obtain coherent parameter sets between the simple approach outlined here and a 
more accurate model presented in Section \ref{subsec:kp-at-K}. 
Further details of the fitting procedure are 
discussed in \ref{sec:fitting-at-K}.  Looking at 
Tables \ref{tbl:K-dftparams-cb} and \ref{tbl:K-dftparams-vb} one can  
see that the effective masses and spin-splittings obtained from the two  different 
DFT calculations are in almost perfect agreement, while there are some differences 
in the extracted values of $C_{3w}$.

Considering first the CB, the extracted  band parameters 
and SOC splittings $2\Delta_{\rm cb}$ 
for different monolayer TMDCs are shown in  Table \ref{tbl:K-dftparams-cb}. 
{To our knowledge there are no direct measurements of  $\Delta_{\rm cb }$ or 
$m_{\rm cb}$ for any of these materials yet; therefore it is difficult to tell how 
reliable these DFT-based predictions are.}
In addition we show 
the charge density $n_{cb}$ at which the upper spin-split CB $K_{\rm cb}^{(1)}$ starts 
to be populated. 
This charge density is calculated using 
the effective mass of the $K_{\rm cb}^{(2)}$ band  given in  Table \ref{tbl:K-dftparams-cb} 
and assuming a simple parabolic dispersion (i.e., neglecting $C_{3w}$), 
which is a good approximation in the CB\@. 
\begin{table}[htb]
\caption{Band dispersion parameters  and spin-splittings at the $K$ and $-K$ points in the CB 
from DFT calculations. $m_{\rm cb}^{(1)}$ ($m_{\rm cb}^{(2)}$) is the effective 
mass of the $K_{\rm cb}^{(1)}$  ($K_{\rm cb}^{(2)}$) band, and similarly for 
$C_{3w}^{(1)}$ ($C_{3w}^{(2)}$).
$m_e$ is the free electron mass.
$n_{cb}$ is the electron density above which the upper spin-split 
CB starts to fill.}
\label{tbl:K-dftparams-cb}
\begin{indented}
\lineup
\item[]
\begin{tabular}{@{}lrrrrrr}
\br
       & MoS$_2$  & MoSe$_2$ & WS$_2$  & WSe$_2$ & MoTe$_2$& WTe$_2$\\
\mr
$m_{\rm cb}^{(1)}/m_e$ (HSE,LDA) & $0.46$ & $0.56$ & $0.26$ & $0.28$ & $0.62$ & $0.26$\\  
$m_{\rm cb}^{(1)}/m_e$ (PBE,PBE)  & $0.47$  & $0.58$ &  $0.27$ &  $0.29$ & $0.61$ & $0.25$ \\ 
\br
$m_{\rm cb}^{(2)}/m_e$  (HSE,LDA) & $0.43$ & $0.49$ & $0.35$ & $0.39$ & $0.53$ & $0.39$ \\
$m_{\rm cb}^{(2)}/m_e$ (PBE,PBE)  & $0.44$ & $0.50$ & $0.36$  & $0.40$ & $0.51$ & $0.38$\\
\br
 $C_{3w}^{(1)}$ [eV\AA$^3$]   (HSE,LDA) & $-3.36$ & $-3.11$ &  $-2.8$  & $-3.02$ & $-3.85$& $-5.86$\\
 $C_{3w}^{(1)}$ [eV\AA$^3$]  (PBE,PBE) & $-3.57$ & $-2.94$ & $-1.8$  & $-2.44$ & $-3.95$& $-17.54$\\
\mr
 $C_{3w}^{(2)}$ [eV\AA$^3$]  (HSE,LDA) & $-3.34$ & $-3.12$  & $-3.14$   & $-3.23$ & $-3.86$& $-4.90$\\
 $C_{3w}^{(2)}$ [eV\AA$^3$]  (PBE,PBE) & $-3.49$ & $-2.86$  & $-2.54$   &  $-2.97$ & $-4.04$& $-9.67$\\
\br 
$2\Delta_{\rm cb}$ [meV] (HSE,LDA)  & $3$ & $22$ & $-32$ & $-37$ &  $ 36$ &  $-52$ \\ 
$2\Delta_{\rm cb}$ [meV] (PBE,PBE)  & $3$ & $20$  & $-31$ & $-37$ & $ 32$ &  $-54$ \\ 
\br
$n_{\rm cb}$ [$10^{12}$ cm$^{-2}$] (HSE,LDA) & $0.54$ & $4.5$  & $4.68$  &  $6.03$ & $7.97$ & $8.48$\\
\br
\end{tabular}
\end{indented}
\end{table}
Note that typical charge densities achieved by gating in   MoS$_2$  
are reported to be $\sim 4\cdot 10^{12}$ cm$^{-2}$--$3.6\cdot 10^{13}$ cm$^{-2}$ 
\cite{radisavljevic}, a few times $10^{12}$ cm$^{-2}$ for monolayer samples 
\cite{jarillo-herrero2013} and  few-layer samples \cite{tutuc}),  
and up to $10^{14}$ cm$^{-2}$ in few-layer WS$_2$ using ionic 
liquid gating \cite{morpurgo2012}. 

Turning now to the VB, the  band parameters  and SOC splitting $2\Delta_{\rm vb}$ 
obtained from our DFT 
calculations are shown in Table \ref{tbl:K-dftparams-vb}.   
In the case of MoSe$_2$,  very recent  high-resolution ARPES measurements \cite{shen} 
allow for a direct comparison with the calculations, because 
the difference between the effective masses of $K_{\rm vb}^{(1)}$ and  $K_{\rm vb}^{(2)}$ 
could be directly observed. We show two theoretical values for the effective masses in 
the VB of MoSe$_2$. The first one is obtained using the fitting procedure described 
above, i.e., by averaging the values along the
$K$--$\Gamma$ and $K$--$M$ directions. The second value, shown in parenthesis, 
is obtained by following
the fitting procedure that was used for the experimental data \cite{priv-com-effmass}. 
This latter procedure involves fitting only along the $K$--$\Gamma$ direction, and a fitting 
range of $\approx 13\%$ of the $K$--$\Gamma$ distance. One can see that the 
theoretical and experimental effective masses that were obtained using the same fitting range 
are in good agreement. Moreover, the calculated value of $2\Delta_{\rm vb}$  
also corresponds rather  well to the measured one. 
\begin{table}[hbt]
\caption{Effective masses and spin-splittings at the $K$ point 
in the VB from DFT calculations. 
$m_{\rm vb}^{(1)}$ ($m_{\rm vb}^{(2)}$) is the effective 
mass of the $K_{\rm vb}^{(1)}$  ($K_{\rm vb}^{(2)}$) band, and similarly for 
$C_{3w}^{(1)}$ ($C_{3w}^{(2)}$).  $m_e$ is the free electron mass. 
The values in brackets were obtained using a slightly different fitting range, 
as explained in the text. 
Experimental values are shown in rows denoted by ``Exp''.}
\label{tbl:K-dftparams-vb}
\begin{indented}
\lineup
\item[]
\begin{tabular}{@{}lrrrrrr}
\br
       & MoS$_2$  & MoSe$_2$ & WS$_2$  & WSe$_2$ & MoTe$_2$ & WTe$_2$ \\
\mr
$m_{\rm vb}^{(1)}/m_e$ (HSE,LDA)  & $-0.54$ & $-0.59$ ($-0.64$) & $-0.35$ &  $-0.36$ & $-0.66$& $-0.34$\\  
$m_{\rm vb}^{(1)}/m_e$ (PBE,PBE)  & $-0.54$ & $-0.60$ ($-0.60$) & $-0.36$ &  $-0.36$ & $-0.62$& $-0.32$\\
\mr
Exp                        & $-0.6\pm 0.08^{a}$ & $-0.67\pm0.4^{b}$ &  & $-0.35\pm0.01^{z}$ & &\\
\br
$m_{\rm vb}^{(2)}/m_e$ (HSE,LDA)  & $-0.61$ & $-0.7$ ($-0.72$) & $-0.49$ &  $-0.54$ & $-0.82$ & $-0.58$\\  
$m_{\rm vb}^{(2)}/m_e$ (PBE,PBE) & $-0.61$ & $-0.7$ ($-0.69$) & $-0.50$ &  $-0.54$ & $-0.77$ & $-0.54$\\
\mr
Exp                         & $-0.6 \pm 0.08^{a}$ & $-0.75\pm0.3^{b}$ &  &  $-0.49\pm0.05^{z}$& &\\  
\br
 $C_{3w}^{(1)}$ [eV\AA$^3$]   (HSE,LDA) & $6.16$  & $5.67$  & $4.59$ & $6.47$  & $5.44$ & $6.77$\\
 $C_{3w}^{(1)}$ [eV\AA$^3$]   (PBE,PBE) & $6.08$  & $5.21$ & $6.07$ &  $5.79$  & $5.46$ & $17.61$ \\
\mr
 $C_{3w}^{(2)}$ [eV\AA$^3$]   (HSE,LDA) & $5.78$ & $5.42$  & $5.50$  & $5.18$ & $5.14$ & $4.83$ \\
 $C_{3w}^{(2)}$ [eV\AA$^3$]   (PBE,PBE) & $5.71$ & $5.064$ & $5.04$ & $4.78$  & $5.09$ & $9.08$\\
\br 
$2\Delta_{\rm vb}$ [meV] (HSE,LDA)  &   $148$ & $186$ & $429$ & $466$ & $219$ & $484$ \\ 
$2\Delta_{\rm vb}$ [meV] (PBE,PBE)  & $148$ & $184$ & $425$ &  $462$ & $213$ &  $480$\\ 
\mr
 Exp [meV]   &  $\approx 140^{c}$ & $\approx 180^{e}$ & $\approx 400^{g}$, & $\approx 400^{g}$ & $250^{y}$&\\
             & $\approx 150^{d}$  & $\approx 180^{f}$ & $380^{h}$  & $\gtrsim 460^{k}$ & &\\
             & $160^{m}$          & $\approx 200^{m}$ & $410^{i}$  &  $400^{l}$ & &\\
             & $140^{n}$           &                   & $400^{j}$  &  $510^{m}$ & &\\
             &  $140^{p}$            &                 & $400^{l}$  & $\sim 500^{q}$ & &\\ 
             &  $138^{r}$          & $202^{r}$          &   $379^{r}$ &  $404^{r}$ & &\\
             &                     &                    &   $391^{t}$ &  $412^{t}$ & &\\
             &                     &                    &             &  $430^{x}$ & &\\
             & 			   &			&	      &  $513^{z}$ &  & \\		
\br
\end{tabular}
\item[] $^{a}${\cite{king2014a}, sodium intercalated sample and ARPES measurement.}
\item[] $^{b}${Private communication by Yi Zhang based on ARPES measurements; see \cite{shen}.}
\item[] $^{c}${\cite{heinz2012}, $^{f}$\cite{ross2013}, $^{i}$\cite{zhang}, $^{j}$\cite{heinz2014a},
              $^{l}$\cite{eda2013}, $^{r}$\cite{eda2014}, from differential reflectance.}
\item[] $^{e}${\cite{shen}, from ARPES measurement.}
\item[] $^{d}${\cite{steiner}, $^{g}$\cite{cui2013}, $^{p}$\cite{nguyen}, from photoluminescence.}
\item[] $^{h}${\cite{cui2014}, from differential transmission.}
\item[] $^{k}${\cite{jarillo-herrero2014}, from electroluminescence.}
\item[] $^{m}${\cite{bolotin}, from photocurrent spectroscopy of suspended samples.}
\item[] $^{n}${\cite{kis-peaks}, from absorbance measurement.}
\item[] $^{q}${\cite{king2014b}, from spin-resolved ARPES measurement.}
\item[] $^{t}${\cite{jonker}, from reflectivity measurement.}
\item[] $^{x}${\cite{shan}, from linear absorption.}
\item[] $^{y}${\cite{heinz2014c}, from reflectance measurements.}
\item[] $^{z}${\cite{dowben}, ARPES meauserement.}
\end{indented}
\end{table}
MoS$_2$ is the only other monolayer TMDC for which ARPES measurements are available to
extract the effective mass. 
However, the ARPES data of Reference \cite{king2014a} do not resolve $
K_{\rm vb}^{(1)}$ and  $K_{\rm vb}^{(2)}$ separately; therefore 
the reported effective mass is the  average of $m_{\rm vb}^{(1)}$ and $m_{\rm vb}^{(2)}$. 
Taking into account  the experimental uncertainty, our results are in reasonable
agreement with the measurements of Reference \cite{king2014a}. 
The available data  for MoS$_2$,  MoSe$_2$ and WSe$_2$  suggest that DFT can capture the VB effective 
masses quite well  even without  $GW$ corrections, such as those found in Reference \cite{qiu}. 
For the other three monolayers, to our knowledge, no ARPES data are yet available.

In optical experiments the difference  of the A and B exciton
energies are usually identified  with $2\Delta_{\rm vb}$ providing the 
results shown in Table \ref{tbl:K-dftparams-vb}. 
We note that there are two assumptions behind the identification of the A and B exciton energy
difference with $2\Delta_{\rm vb}$: 
i) that the spin-splitting in the CB is negligible and 
ii) that the binding energies of the A and B excitons are the same. 
Regarding i),  one can see in Table \ref{tbl:K-dftparams-cb} that $\Delta_{\rm cb}$ 
is small, but finite, and for quantitative comparisons between theory and experiment  
it should not be neglected. 
As for ii), we note that the binding energy of the A and B excitons depends on their reduced 
mass, which, according to Table \ref{tbl:K-dftparams-vb}, should be different for the 
different exciton species. 
With these caveats the agreement between the calculations and the experiments is 
qualitatively good, especially for MoS$_2$ and  MoSe$_2$. 

Comparing the DFT-calculated effective masses in  Tables \ref{tbl:K-dftparams-cb} 
and \ref{tbl:K-dftparams-vb} 
for the VBs and CBs that have the same spin-polarisation, one can observe  that
there is no electron--hole symmetry in the band structure. 
The first experimental evidence to support this observation, coming from 
magnetoluminescence experiments, has appeared very recently 
\cite{macneill,atac,aivazian,heinz2014b,urbaszek2015b}. 
Regarding the experimental relevance of TW, it has been argued \cite{iwasa} that 
it leads to measurable effects in the polarisation of electroluminescence in 
p--n junctions. We note that due to the heavier effective mass in the VB 
and the larger values of  $C_{3w}$, the TW is more pronounced 
{in our DFT calculations} in the VB than in 
the CB\@. In the latter a simple parabolic approximation is often adequate.

We finish Section \ref{subsec:K-point-params} with a brief  discussion of the 
quasiparticle band gap $E_{\rm bg, K}$,  which we define as the difference 
between the maximum of the $K_{\rm vb}^{(1)}$ and $K_{\rm cb}^{(2)}$ 
bands at the $K$ and $-K$ points.
DFT calculations for monolayer TMDCs  underestimate the band gap 
(see Table \ref{tbl:K-bandgap}) and its evaluation requires the use of 
$GW$ methodology \cite{ashwin,komsa,qiu,yakobson,lambrecht}.
\begin{table}[htb]
\caption{Band gap $E_{\rm bg, K}$ at the $K$  point from DFT calculations, 
from $GW$ calculations, and from measurements. $E_{\rm bg,K}$ is defined as the 
energy difference between the bands $K_{\rm vb}^{(1)}$ and $K_{\rm cb}^{(2)}$ at $K$.
The $GW$ ``flavour'' used in the calculations is also shown. 
Experimental values are shown in rows denoted by ``Exp''. All values are in eV\@.}
\label{tbl:K-bandgap}
\begin{indented}
\lineup
\item[]
\begin{tabular}{@{}lllllll}
\br
       & MoS$_2$  & MoSe$_2$ & WS$_2$  & WSe$_2$ & MoTe$_2$ & WTe$_2$\\
\mr
 (HSE,LDA)  & $1.67$ &  $1.40$ & $1.60$ &  $1.30$ & $0.997$ & $0.792$\\  
(PBE,PBE)  &  $1.59$ &  $1.34$ &  $1.58$ &  $1.27$ & $0.947$ & $ 0.765$\\
\mr
$GW$                      &  $2.84^{h}$ & $2.41^{l,m}$        & $2.88^{l,q}$ &  $2.42^{l}$ & $1.77^{l}$ & $1.77^{q}$\\
                          &  $2.76^{j,q}$ & $2.26$ $(2.13)^{n}$  & $2.70^{p}$ & $2.38^{q}$ &  $1.79^{m}$ & $1.79^{x}$\\     
                          &  $2.80^{k}$ & $2.33^{q}$           & $3.11^{k}$ &   $2.51^{x}$ &   $1.82^{q}$ & \\
                          &  $2.82^{l}$ &  $2.31^{x}$          &  $2.91^{x}$ &            &   $1.77^{x}$  & \\
                          &  $2.97^{m}$ &                      &           &   &  &  \\
\mr
Exp                      & $2.5^{a}$ & $2.18^{b}$  & $2.14^{c}$ &  $2.51\pm0.04^{r}$ & &\\
                         & $2.14\pm0.08^{g}$       &   $2.02^{s}$,$2.22^{s}$  & $2.41^{d}$ & $2.0^{s}$,$2.18^{s}$ & &\\
Exp (ARPES)             & $1.86^{e}$ & $1.58^{f}$            &            &  & & \\
\br
\end{tabular}
\item[] $^{a}${\cite{bolotin},  photocurrent spectroscopy on suspended samples, lower bound.}
\item[] $^{b}${\cite{crommie}, scanning-tunnelling experiments, on bilayer graphene substrate.}
\item[] $^{c}${\cite{morpurgo2014a}, transport measurements using ionic liquid gating.}
\item[] $^{d}${\cite{heinz2014a},  differential reflectance, on SiO$_2$ substrate.}
\item[] $^{e}${\cite{king2014a}, $^{f}$\cite{shen}, from ARPES, heavily doped sample.}
\item[] $^{g}$\cite{shih2014a}, $^{s}$\cite{shih2014b}, { scanning-tunnelling experiments, on graphite substrate.}
\item[] $^{r}${\cite{lainjong}, scanning-tunnelling experiments.}
\item[] $^{h}${\cite{qiu}, $G_1W$.}
\item[] $^{j}${\cite{lambrecht}, quasiparticle self-consistent $GW$.}
\item[] $^{k}${\cite{yakobson}, self-consistent $GW_0$.}
\item[] $^{l}${\cite{ashwin}, $^{m}$\cite{komsa}, $^{q}$\cite{liang}, $^{x}$\cite{Ding_2011}, $G_0W_0$ method.}
\item[] $^{n}${\cite{crommie}, $G_1W$,  without (with) substrate screening taken into account. }
\item[] $^{p}${\cite{zhang}, $G_1W$.}

\end{indented}
\end{table}
Experimental evidence that supports the conclusions of the $GW$ calculations is now 
also emerging. 
Apart from its fundamental importance, the main reason for discussing $E_{\rm bg, K}$ 
and showing
our DFT results is that $E_{\rm bg, K}$ enters into the fitting procedure that we use 
to obtain the parameters of the $\mathbf{k}\cdot\mathbf{p}$ Hamiltonian that describes
the dispersion in the vicinity of the band edge. The details of the $\mathbf{k}\cdot\mathbf{p}$ 
model and the fitting procedure are given in Section \ref{subsec:kp-at-K} and 
\ref{sec:fitting-at-K}.
As one can see, our DFT calculations significantly underestimate the experimental 
quasiparticle band gaps. 
We also note that  in heavily doped samples, which were used in the ARPES measurements 
\cite{shen,king2014a}, the observed band gap is reduced with respect to results obtained by 
other methods \cite{morpurgo2014a,heinz2014a,bolotin,crommie}, hinting at the crucial 
importance of screening in monolayer TMDCs.


\subsection{$\mathbf{k}\cdot\mathbf{p}$ Hamiltonian}
\label{subsec:kp-at-K}

We  now present a low-energy effective  $\mathbf{k}\cdot\mathbf{p}$ 
Hamiltonian that describes the coupled  dynamics of the VB and CB\@. 
Part of the theory was previously discussed  in References \cite{sajat1} and \cite{sajat2}; 
in the present work we both overview and extend  
these earlier results. 

To obtain a model that captures the most important features of the  dispersion of the 
VB and CB one can start from a seven-band model, which was introduced in Reference 
\cite{sajat1,sajat2}; {motivation and  details of the model are given 
 in \ref{sec:seven-band-at-K}}. 
An effective low-energy Hamiltonian can be derived {from the seven-band model} 
by systematically eliminating all degrees of freedom other than
the ones corresponding to  the VB and CB using L\"owdin partitioning \cite{winkler-book}. 
We keep terms up to third order in the off-diagonal coupling elements of the 
original seven-band model 
and  use the spinful basis  $\{ |\Psi^{\rm vb},s\rangle, |\Psi^{\rm cb},s\rangle\}$, where 
$| \Psi^{\rm vb}\rangle$  ($|\Psi^{\rm cb}\rangle$) are spinless Bloch 
wave functions in the VB (CB) 
and $| \Psi_{}^{\rm b}, s\rangle =  | \Psi_{}^{\rm b}\rangle \otimes | s \rangle$, 
with $\rm{b}=\{\rm{cb}, \rm{vb}\}$ and
$s=\{ \uparrow,\downarrow \}$ denoting  the band spin degree of freedom, respectively.  
One finds that the low-energy effective Hamiltonian 
\begin{equation}
 H_{\rm eff}^{\tau,s}=H_0+H_{\mathbf{k}\cdot\mathbf{p}}^{\tau,s}+H_{\rm so}^{\tau,s} 
 \label{eq:full-eff-Ham-at-K}
\end{equation}
is the sum of the following terms: 
\begin{itemize}
\item[i)] The free-electron term $H_0=\frac{\hbar^2 \mathbf{q}^2}{2 m_e}(\mathbb{1}_2 \otimes s_z)$, where 
$\mathbb{1}_2$ is a unit matrix in the electron--hole space, $s_z$ is a spin Pauli matrix,
and $m_e$ is the free electron mass. Here and in 
Equations \eref{eq:kp-massDirac}--\eref{eq:kp-cub2} 
the wavevector $\mathbf{q}=(q_x,q_y)$  is measured from the $K$ or $-K$ points. 
We note that $H_0$ is usually neglected in the GaAs literature on account of 
the light effective mass in this material,  but here we want to keep it. 
\item[ii)] The SOC  Hamiltonian $H_{\rm so}^{\tau,s}$,  which contains the 
diagonal and $\mathbf{q}$-independent contributions of the SOC\@.  It reads 
\begin{eqnarray}
 H_{\rm so}^{\tau,s}=
 \left(
 \begin{array}{cc}
     \tau \Delta_{\rm vb} s_z & 0\\
  0 &  \tau \Delta_{\rm cb} s_z 
 \end{array}
 \right),
 \label{eq:H_so-K}
\end{eqnarray}
i.e.,  it is diagonal in  spin space and is proportional to the Pauli matrix $s_z$ 
(for further details see \ref{sec:seven-band-at-K}).  
$H_{\rm so}^{\tau,s}$ describes the spin-splittings of the CB and the VB, which are due to 
the absence of inversion symmetry in monolayer TMDCs. 
Since $H_{\rm so}^{\tau,s}$ is diagonal, one 
can also write it in terms of the eigenvalues $s=\pm 1$ of $s_z$; we will use the 
two notations interchangeably.  Moreover, the index $\tau=1$ ($\tau=-1$) denotes 
the valley $K$ ($-K$). 
Where it is more convenient, we will also use the matrix  $\tau_z$ which acts 
in the valley space. 
In the VB, the parameter $\Delta_{\rm vb}$ that describes the strength of the 
SOC can always be taken to be  positive. 
As explained in Section \ref{subsec:K-point-params}, the situation is more complicated in the 
CB \cite{liugb,sajat2,kosmider2}, because DFT calculations  show that, in the case of MoX$_2$, 
the spin-split bands cross close to the $K$ and $-K$  points, while there is no 
such band crossing for WX$_2$. This can be understood in terms of 
$\Delta_{\rm cb}$ having opposite
signs in MoX$_2$ and WX$_2$.
\item[iii)] Finally, the $\mathbf{k}\cdot\mathbf{p}$ 
Hamiltonian $H_{\mathbf{k}\cdot\mathbf{p}}^{\tau,s}$ 
in Equation \eref{eq:full-eff-Ham-at-K}  is given by
\numparts
\begin{eqnarray}
 H_{\mathbf{k}\cdot\mathbf{p}}^{\tau,s}=H_{\rm D}^{\tau,s}+
 H_{\rm as}^{\tau,s}+H_{3 w}^{\tau,s}+H_{\rm cub}^{\tau,s},
 \label{eq:Hkp-K}\\
 {\rm where}\nonumber\\
 H_{\rm D}^{\tau,s}=
 \left(
 \begin{array}{cc}
    {\vareps}_{\rm vb} & \tau \cdot\gamma_{\tau,s}^{} {q}_{-}^{\tau}\\
 \tau \cdot \gamma_{\tau,s}^{*} {q}_{+}^{\tau}& {\vareps}_{\rm cb}
 \end{array}
 \right), \label{eq:kp-massDirac}\\
H_{\rm as}^{\tau,s}=
 \left(
 \begin{array}{cc}
   \alpha_{\tau,s} \mathbf{q}^2 & 0\\
   0 &   \beta_{\tau,s} \mathbf{q}^2
 \end{array}
 \right), \label{eq:kp-asym}\\
 H_{3 w}^{\tau,s}=
 \left(
 \begin{array}{cc}
   0 & \kappa_{\tau,s} ({q}_{+}^{\tau})^{2} \\
   \kappa_{\tau,s}^{*} ({q}_{-}^{\tau})^{2} & 0
 \end{array}
 \right),\label{eq:kp-3w}\\
 H_{\rm cub,1}^{\tau,s}=-\tau\,\frac{1}{2} \mathbf{q}^2 
\left(
\begin{array}{cc} 
 0 &  \eta_{\tau,s}  q_{-}^{\tau} \\
  \eta_{\tau,s}^{*} q_{+}^{\tau} & 0\\
\end{array}
\right).\label{eq:kp-cub1}\\
H_{\rm cub,2}^{\tau,s}=-\tau\,\frac{\omega_{s}}{2} |\mathbf{q}|^3 \cos(3\varphi_{\mathbf{q}}) 
\left(
\begin{array}{cc} 
 1 & 0\\
 0 & 1\\
\end{array}
\right).
\label{eq:kp-cub2}
\end{eqnarray}
\endnumparts
Here ${q}_{\pm}^{\tau}$ is defined as ${q}_{\pm}^{\tau}={q}_x\pm i \tau {q}_y$,  
$\varphi_{\mathbf{q}}={\rm arctan}(q_y/q_x)$,  
${\vareps}_{\rm vb}$ and ${\vareps}_{\rm cb}$ are band-edge energies, 
$\gamma_{\tau,s}$, $\alpha_{\tau,s}$, $\beta_{\tau,s}$, 
$\kappa_{\tau,s}$, $\eta_{\tau,s}$, and 
$\omega_{s}$ are material  parameters discussed below. 
$H_{\mathbf{k}\cdot\mathbf{p}}^{\tau,s}$ is a generalisation of the results 
given in Reference \cite{sajat1} for the case in which the material parameters 
depend on the SOC\@. 
\end{itemize}

In general all off-diagonal material parameters 
appearing in $H_{\mathbf{k}\cdot\mathbf{p}}^{\tau,s}$  are complex numbers such that 
for $\tau=-1$ ($-K$ valley) they are the complex conjugate of the $\tau=1$ ($K$ valley) values. 
Concrete values of the material parameters for each MX$_2$ material 
can be obtained by, e.g., fitting a DFT band structure, 
see Tables \ref{tbl:K-kp-params-1} and \ref{tbl:K-kp-params-2}. 
Note however, that the fitting procedure (see \ref{sec:fitting-at-K}) yields real 
numbers for each parameter.
We now briefly discuss each of the terms [Equations \eref{eq:kp-massDirac}--\eref{eq:kp-cub2}]. 
\begin{itemize}
\item[i)]  \emph{Terms up to linear order in   ${q}_+$ and ${q}_{-}$} can be found in Equation
    \eref{eq:kp-massDirac}.   $H_{\rm D}^{\tau,s}$  is basically the massive Dirac fermion model
   introduced in Reference \cite{dxiao}.  It  describes an isotropic dispersion 
   around the band edge and it does not break the electron--hole symmetry.  
    The value of  $\gamma_{s,\tau}$ also depends on the SOC, but the L\"owdin-partitioning 
    calculations suggest that this dependence should be weak. This is indeed what we have found from fits
    to the DFT band structure.   Therefore in the following we suppress  both the spin index $s$ and,  
    since $\gamma$ is taken to be a real number, the valley index $\tau$. 
    
\item[ii)] \emph{Diagonal terms quadratic in ${q}_+$ and ${q}_{-}$} are given 
     in Equation \eref{eq:kp-asym}. 
     $H_{\rm as}^{\tau,s}$ breaks the electron--hole symmetry because in 
     general $\alpha_{\tau,s}\neq \beta_{\tau,s}$.
     The recent observation of photoluminescence peak splitting in magnetic 
     fields \cite{macneill,aivazian,atac,heinz2014b} suggests that electron--hole symmetry 
     is indeed broken. 
     Both $\alpha_{\tau,s}$ and $\beta_{\tau,s}$ can be written as, e.g., 
     $\alpha_{\tau,s}=\alpha_0+\tau\cdot s \cdot\tilde{\alpha}$ and hence 
     $\alpha_{\tau,s}=\alpha_{-\tau,-s}$, $\beta_{\tau,s}=\beta_{-\tau,-s}$.  
     
\item[iii)] \emph{Off-diagonal terms quadratic in ${q}_+$ and ${q}_{-}$} are given 
     in Equation \eref{eq:kp-3w}. $H_{3 w}^{\tau,s}$, in combination 
     with $H_{\rm D}^{\tau,s}$, leads to 
     the TW of the 
     energy contours that can be observed in Figures \ref{fig:defs}(b) and (c).  
      [For further details see Equation \eref{eq:eigenvl-cub-appndx} in \ref{sec:fitting-at-K}].
     The TW is expected to play an important role in the explanation of recent 
     electroluminescence  experiments \cite{iwasa}.  {It may facilitate the generation
     of valley and spin currents that are second order in the applied
      bias \cite{hongyiyu}.}
     Moreover, it was observed in ARPES measurements \cite{hasan,king2014b}. 
     
\item[iv)] \emph{Off-diagonal terms cubic in ${q}_{+}$ and ${q}_{-}$} appear in Equation \eref{eq:kp-cub1}. 
    $H_{\rm cub,1}^{\tau,s}$ is important for obtaining a good fit to the DFT band structure away from the 
    $K$  point in a two-band model that describes the coupled dynamics of the VB and CB\@. 
    {They also play role when one uses the eigenvalues of Hamiltonian 
     \eref{eq:full-eff-Ham-at-K}  to fit the DFT band structure in order to 
      extract material parameters (see Tables \ref{tbl:K-kp-params-1} and 
      \ref{tbl:K-kp-params-2} below). In particular, 
       combined with the off-diagonal first-order terms, they contribute in 
       second order in the wavenumber to the eigenvalues 
      (for details see  \ref{sec:fitting-at-K}).}
\item[v)] \emph{Diagonal terms cubic in ${q}_{+}$ and ${q}_{-}$}. 
     In some cases  it is more  convenient to work with  a model that gives the dispersions 
     of the VB and CB separately.  Cubic terms in $\mathbf{q}$ are needed to 
      capture the non-parabolicity of the bands, and such a model is given
     by Equation \eref{eq:K-dispersion}.  It can easily be obtained by applying L\"owdin partitioning 
     to Equation \eref{eq:full-eff-Ham-at-K} and eliminating either the electron or the hole degrees of freedom. 
     In this case, for consistency,  the term  $H_{\rm cub,2}^{\tau,s}$ in Equation \eref{eq:kp-cub2} 
     also has to be taken into account.   
\end{itemize}
        {We  note that, starting from a TB Hamiltonian, 
        a model  containing the terms  \eref{eq:kp-massDirac}--\eref{eq:kp-3w} and the 
         VB spin-splitting was also obtained in Reference~\cite{rostami}.} 
         
       {In comparison to Equations \eref{eq:H_so-K} and \eref{eq:kp-massDirac}--\eref{eq:kp-cub2}, 
        the widely used gapped Dirac Hamiltonian model introduced in Reference \cite{dxiao}
         contains only the terms linear in ${q}$ and the spin-splitting
        in the VB\@. It can be  written as}
\begin{equation}
   \tilde{H}_{D}=\gamma(\tau q_x \sigma_{x}+q_y\sigma_{y})+\frac{E_{\rm bg}}{2}\sigma_{z}
   +\Delta_{\rm vb}\tau s_z \frac{\sigma_z-1}{2}.
\label{eq:gapped-Dirac}        
\end{equation}
        {Here the Pauli matrices $\sigma_{x,y,z}$ act in the electron--hole space. 
        This simple model correctly captures the large spin-splitting of the VB, 
         that the dispersion in the close vicinity of 
        the $K$ valley is quadratic, and predicts the valley-dependent optical 
        selection rule \cite{dxiao} 
        in accordance with experiments \cite{heinz2012,cui2012,cao}. 
        However, the preceding discussion of the various  terms 
        in Equation \eref{eq:full-eff-Ham-at-K}
        clearly indicates the limitations of Equation \eref{eq:gapped-Dirac} 
        in the interpretation of certain experimental results and DFT calculations: 
        it cannot describe, for example, the spin-splitting of the CB, the 
        electron--hole asymmetry and the trigonal warping of the spectrum. }

The eigenstates and eigenvalues of the $\mathbf{k}\cdot\mathbf{p}$ Hamiltonian 
\eref{eq:full-eff-Ham-at-K} 
can also be used as a starting point for analytical calculation of the 
\emph{Berry curvature} \cite{berry-curvature-review}. 
The Berry curvature is relevant for the quantum transport characteristics of TMDCs, such as 
the  valley Hall effect \cite{dxiao} and weak 
localisation \cite{OchoaFalko}, while a related quantity, the spin Berry curvature \cite{wanxiang},
gives rise to a finite spin Hall conductivity for moderate hole doping.

Finally, we show the $\mathbf{k}\cdot\mathbf{p}$ parameters obtained from fitting of the 
DFT band structure (see Tables \ref{tbl:K-kp-params-1} and \ref{tbl:K-kp-params-2}) 
using the model that explicitly contains
the coupling between the VB and the CB\@.  In this case the diagonal cubic
term [Equation \eref{eq:kp-cub2}] is not 
important for obtaining a good fit to the band structure and therefore the  $\omega_{s}$ parameter is not shown. 
Close to the band edge the $\mathbf{k}\cdot\mathbf{p}$ parameters given in 
Tables \ref{tbl:K-kp-params-1} and \ref{tbl:K-kp-params-2}
reproduce  the effective masses shown in Tables 
\ref{tbl:K-dftparams-cb} and \ref{tbl:K-dftparams-vb}.
The details of the fitting procedure are given  in \ref{sec:fitting-at-K}.  
Since the effective masses and $C_{3w}$  parameters 
extracted from the (HSE,LDA) and (PBE,PBE) approaches are rather  similar, 
we only show results that are based on  (HSE,LDA) DFT band-structure calculations.  
Due to the  SOC  all  parameters, with the exception of $\gamma$, are different for different spin indices $s$. 
Since the  Hamiltonian of Equation \eref{eq:Hkp-K} is diagonal in the spin space, i.e.,  it 
describes the coupled dynamics of the VB and CB having the same spin, it is convenient to  
introduce the notation $s=\uparrow$ ($s=\downarrow$) for $s=1$ ($s=-1$). 
Regarding the correspondence between the notation 
used in Section \ref{subsec:K-point-params} and  here, note that 
the order of the bands with  $\uparrow$ and  $\downarrow$ 
polarisation in the CB is different for MoX$_2$ and WX$_2$. 
Therefore in the VB the upper index ${(1)}$ (${(2)}$) is equivalent to 
$\downarrow$ ($\uparrow$), but in the CB the  relation  depends on which 
material is considered. 
We note that the parameter $\gamma$ can, in principle,  also be obtained  
directly as a momentum matrix element between the Kohn--Sham wave functions of the VB and CB\@. 
For these calculations we used the \textsc{castep} code \cite{castep}, where  the necessary
plane-wave coefficients of the wave functions at the band edges
are readily accessible. These values are denoted by $|\gamma_{\rm KS}|$ in 
Tables \ref{tbl:K-kp-params-1} and \ref{tbl:K-kp-params-2}. 
On the one hand, the good agreement between  $|\gamma|$  and  $|\gamma_{\rm KS}|$ indicates the consistency
of our  fitting procedure. This is not trivial, because the fitting involves a non-linear function of 
the $\mathbf{k}\cdot\mathbf{p}$ parameters. 
On the other hand, one has to bear in mind that  $|\gamma|$ is  obtained such that it would 
give the best fit to the DFT band structure over a certain range in the BZ\@. Therefore it may differ 
from the value of $|\gamma_{\rm KS}|$ that is calculated at a single point of the BZ\@. 
The valley index $\tau$ is suppressed in Tables \ref{tbl:K-kp-params-1} and  \ref{tbl:K-kp-params-2} 
because, as mentioned above,
from the fitting procedure we obtain real numbers for the off-diagonal terms.

\begin{table}[htb]
\caption{$\mathbf{k}\cdot\mathbf{p}$ parameters at the $K$ point. In columns labelled by ``DFT'' the parameters 
         obtained with the help of DFT band gap are shown, for the  columns labelled by ``$GW$'' the band gap 
         is taken from $GW$ calculations. 
}
\label{tbl:K-kp-params-1}
\lineup
\begin{tabular}{@{}lrrrrrrrr}
\br
       &\centre{2} {MoS$_2$}  & \centre{2} {MoSe$_2$} & \centre{2} {WS$_2$}  & \centre{2} {WSe$_2$} \\
       & DFT & $GW$  &  DFT  & $GW$  & DFT & $GW$ & DFT & $GW$ \\
\ns
       &  \crule{2} & \crule{2}  & \crule{2} & \crule{2} \\

$E_{\rm bg,K}$ [eV]  & $1.67$& $2.80$ & $1.40$ & $2.26$ &  $1.60$ & $2.88$ & $1.30$ & $2.42$ \\
\mr 
$|\gamma_{\rm KS}|$ [eV$\cdot$\AA] (HSE,LDA) & $3.00$  &   --   & $2.52$ &  --    & $3.85$ &  --    & $3.31$ & -- \\
$|\gamma|$ [eV$\cdot$\AA] (HSE,LDA)          & $2.76$ & $2.22$  & $2.53$ & $2.20$  & $3.34$ & $2.59$ & $3.17$ & $2.60$\\  
\mr
$\alpha_{\uparrow}$  [eV$\cdot$\AA$^2$] (HSE,LDA) & $-5.97$  & $-6.21$ & $-5.34$ & $-5.76$ & $-6.14$ & $-6.56$ & $-5.25$ & $-5.97$ \\  
$\alpha_{\downarrow}$ [eV$\cdot$\AA$^2$] (HSE,LDA) & $-6.43$ & $-6.65$ & $-5.71$ & $-6.20$  & $-7.95$ & $-7.96$ & $-6.93$ & $-7.58$ \\
\hline
$\beta_{\uparrow}$ [eV$\cdot$\AA$^2$] (HSE,LDA)   & $0.28$ & $0.52$  & $-0.95$ & $-0.54$ & $1.62$ & $2.03$ & $0.33$  & $1.08$\\ 
$\beta_{\downarrow}$ [eV$\cdot$\AA$^2$] (HSE,LDA) & $0.54$ & $0.76$  & $-0.52$ & $-0.03$ & $4.00$ & $4.00$  & $2.35$  & $3.0$  \\ 
\hline
$\kappa_{\uparrow}$ [eV$\cdot$\AA$^2$] (HSE,LDA)   & $-1.48$ & $-1.84$ & $-1.31$ & $-1.49$ & $-1.24$ & $-1.60$  & $-1.11$ & $-1.36$ \\  
$\kappa_{\downarrow}$ [eV$\cdot$\AA$^2$] (HSE,LDA) & $-1.45$ & $-1.80$  & $-1.23$ & $-1.40$  & $-1.09$ & $-1.41$ & $-0.93$ & $-1.14$ \\ 
\hline
$\eta_{\uparrow}$ [eV$\cdot$\AA$^3$] (HSE,LDA)  & $13.7$  & $17.74$ & $15.11$ & $18.28$ & $21.85$ & $29.49$ & $18.04$ & $23.78$\\  
$\eta_{\downarrow}$ [eV$\cdot$\AA$^3$] (HSE,LDA) & $21.1$ &  $26.95$ & $17.10$ &  $20.93$ & $31.73$ & $40.94$ & $26.17$ & $34.49$ \\ 
\br
\end{tabular}
\end{table}
\begin{table}[htb]
\caption{$\mathbf{k}\cdot\mathbf{p}$ parameters at the $K$ point. In columns labelled by ``DFT'' the parameters 
         obtained with the help of DFT band gap are shown, for the  columns labelled by ``$GW$'' the band gap 
         is taken from $GW$ calculations. 
}
\label{tbl:K-kp-params-2}
\lineup
\begin{tabular}{@{}lrrrr}
\br
       &\centre{2} {MoTe$_2$}  & \centre{2} {WTe$_2$}   \\
       & DFT & $GW$  &  DFT  & $GW$   \\
\ns
       &  \crule{2} & \crule{2}   \\

$E_{\rm bg,K}$ [eV]  & $0.997$& $1.82$ & $0.792$ & $1.77$   \\
\mr 
$|\gamma_{\rm KS}|$ [eV$\cdot$\AA] (HSE,LDA) & $2.12$  &   --   & $2.84$ &  --  \\
$|\gamma|$ [eV$\cdot$\AA] (HSE,LDA)          & $2.33$ & $2.16$  & $3.04$ & $2.79$  \\  
\mr
$\alpha_{\uparrow}$  [eV$\cdot$\AA$^2$] (HSE,LDA)  & $-4.78$  & $-5.31$ & $-3.94$ & $-5.02$  \\  
$\alpha_{\downarrow}$ [eV$\cdot$\AA$^2$] (HSE,LDA) & $-4.85$  &  $-5.78$ & $-5.20$ & $-7.31$   \\
\hline
$\beta_{\uparrow}$ [eV$\cdot$\AA$^2$] (HSE,LDA)   & $-2.19$ & $-1.66$  & $-0.9$ & $0.17$ \\ 
$\beta_{\downarrow}$ [eV$\cdot$\AA$^2$] (HSE,LDA) & $-1.78$ & $-0.84$  & $0.60$ & $2.72$  \\ 
\hline
$\kappa_{\uparrow}$ [eV$\cdot$\AA$^2$] (HSE,LDA)   & $-1.19$ & $-1.28$ & $-1.01$ & $-1.10$  \\  
$\kappa_{\downarrow}$ [eV$\cdot$\AA$^2$] (HSE,LDA) & $-1.01$ & $-1.09$  & $-0.96$ & $-1.04$   \\ 
\hline
$\eta_{\uparrow}$ [eV$\cdot$\AA$^3$] (HSE,LDA)  & $13.26$  & $15.18$ & $14.72$ & $17.61$ \\  
$\eta_{\downarrow}$ [eV$\cdot$\AA$^3$] (HSE,LDA) & $13.54$ &  $16.37$ & $19.41$ &  $27.12$  \\ 
\br
\end{tabular}
\end{table}
As explained in \ref{sec:fitting-at-K},  our fitting procedure 
involves the  quasiparticle band gap $E_{\rm bg,K}$.  
For this reason two sets of $\mathbf{k}\cdot\mathbf{p}$ 
parameters are reported in Tables \ref{tbl:K-kp-params-1} and \ref{tbl:K-kp-params-2}: 
one in which we used $E_{\rm bg,K}$ values obtained from our DFT calculations and one 
in which we used $E_{\rm bg,K}$ values found in $GW$ calculations; see Table \ref{tbl:K-bandgap}. 
In the latter case we  make the assumption that the bands above the Fermi energy 
are  rigidly shifted upwards  in energy such that  the effective masses and the TW 
in the VB and CB remain the same.
We believe that this is a reasonable assumption because the available experimental evidence 
(see Tables \ref{tbl:K-dftparams-vb} and \ref{tbl:dft-at-G}) suggests that, at least in the VB, the effective masses  
are captured quite well by the DFT calculations.


\section{Effective models  at the $Q$ (a.k.a.\ $\Lambda$) point}
\label{sec:Q-point-main}

\subsection{$Q_i$ points}

In addition to the $K$ and $-K$ points, there are six other minima in the CB which may be 
 important for, e.g., relaxation processes. 
We denote the BZ points  where these minima  are located by $Q_{i}$, ($i=1\dots 6$);
they are also known as $\Lambda$ points [see Figures \ref{fig:defs} and \ref{fig:Qcontour}(b)]. 
We note  that phonon scattering between the $K$ and $-K$ points  and  $Q_{i}$  points 
is symmetry-allowed \cite{dery} and that, depending on the energy difference 
$E_{\rm KQ}$ (see Figure \ref{fig:defs}), the electron mobility may be 
significantly affected by these scattering processes \cite{kaasbjerg2012,wook_kim,yao}. 
However, as we will show, understanding the  SOC at the $Q_{i}$ points is 
also  important when  considering 
the possible scattering processes, 
a fact which seems to have been overlooked in some recent publications. 
We start  in Section \ref{subseq:Q-point-params}   with a basic characterisation of the 
band structure in terms of the effective masses and point out 
an important effect of SOC on the spin 
polarisation of the bands. A detailed $\mathbf{k}\cdot\mathbf{p}$ theory is given 
in Section \ref{subsec:kp-at-Q}.


\subsection{Basic characterisation and material parameters}
\label{subseq:Q-point-params}

\begin{figure}[htb]
\begin{center}
\includegraphics[scale=0.6]{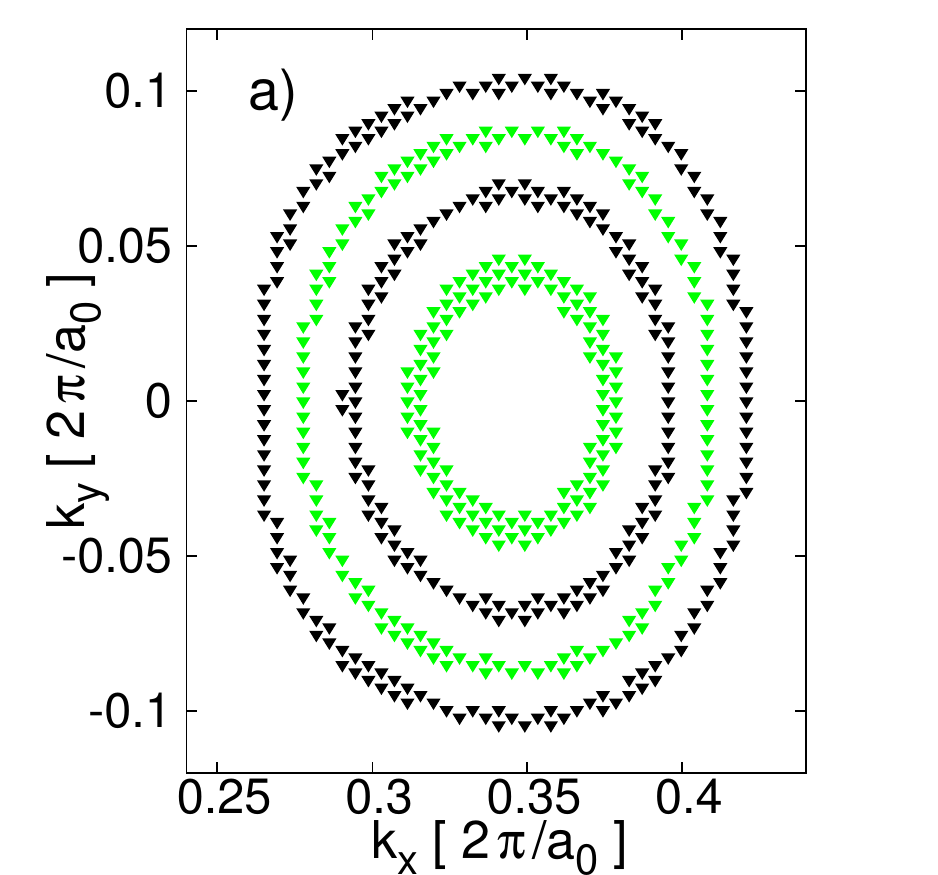}
\includegraphics[scale=0.55]{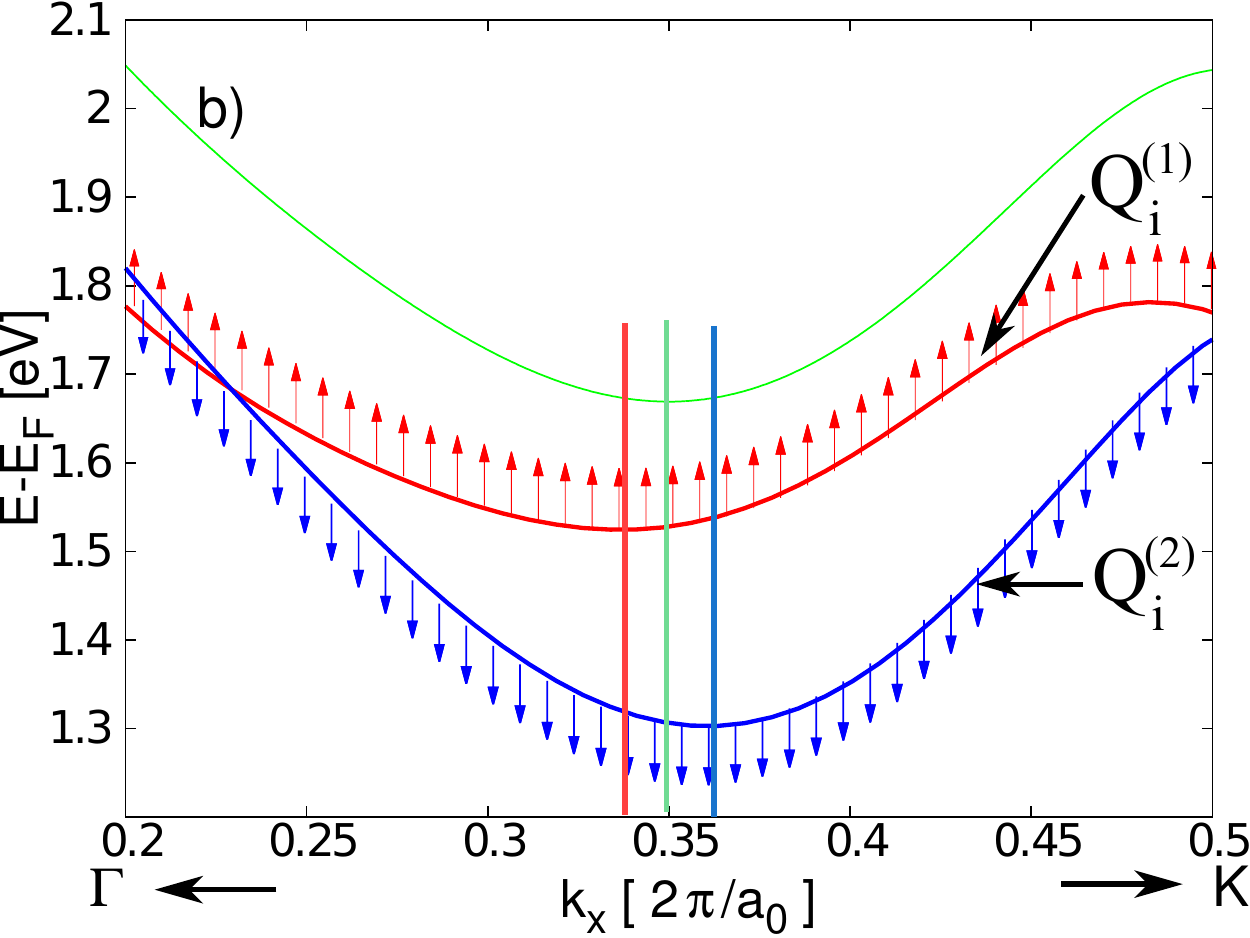}
\caption{a) Energy contours at the $Q$ point obtained from (HSE,LDA)
DFT calculations for MoS$_2$. SOC is not taken into account. The energy difference between 
the energy contours is $0.04$ eV\@.  b)  Band structure of WSe$_2$ 
along the $\Gamma$--$K$ direction 
around the $Q$ point with SOC (red and blue lines) and without SOC (green line). 
The  bands without SOC are shifted in 
energy for clarity. Vertical bars indicate the $k_x$ values at which the corresponding 
curve has a
minimum. The results were obtained from (PBE,PBE) DFT calculations. 
\label{fig:Qcontour}
}
\end{center}
\end{figure}

Let us consider the $Q_{1}$ minimum, which can be found along the $\Gamma$--$K$ direction 
[see Figure \ref{fig:defs}(c)]. We choose  $k_x$ to be parallel to the $\Gamma$--$K$ direction,
while $k_y$ is perpendicular to it. 
Neglecting SOC for a moment, our DFT  calculations show 
 that, close to the $Q_{1}$ 
point, the energy contours are to a good approximation ellipses whose axes are 
parallel to  $k_x$ and $k_y$ [see Figure~\ref{fig:Qcontour}(a)].  
Therefore, to a first approximation the dispersion around $Q_{1}$ is 
quadratic 
with different effective masses $m_{Q,x}^{0}$ and $m_{Q,y}^{0}$ along  
$k_x$ and $k_y$: 
\begin{eqnarray}
E_{Q}(\mathbf{q})=\frac{\hbar^2 q_x^2} {2 m_{Q,x}^{0}} +
 \frac{\hbar^2 q_y^2}{2 m_{Q,y}^{0}},
\label{eq:noSO_Q-disp}
\end{eqnarray}
where the wavenumbers $q_x$ and $q_y$ are measured from the energy minimum of the dispersion  
(see Section \ref{subsec:kp-at-Q} for details). 
As one can see in Table \ref{tbl:Q-mat-params}, the ratio of the effective masses is 
$m_{Q,y}^{0}/ m_{Q,x}^{0}\approx 2 $ for MoX$_2$ {and WTe$_2$}
and $m_{Q,y}^{0}/ m_{Q,x}^{0}\approx 1.3$--$1.8$ for  WS$_2$,  WSe$_2$. 
The SOC has two major effects [see Figure \ref{fig:Qcontour}(b)]: 
\begin{itemize}
 \item[i) ] it splits the spin-degenerate levels by an energy $2\Delta_{Q}$, and
 \item[ii) ] the effective masses in the spin-split bands are different.
\end{itemize}
Similarly to Section \ref{subsec:K-point-params}, 
we introduce the notation $Q_{i}^{(1)}$ ($Q_{i}^{(2)}$)  for the higher-in-energy 
(lower-in-energy)
spin-split CB at the $Q_{i}$ point [see Figure \ref{fig:Qcontour}(b)]. 
The basic characterisation of these 
bands therefore requires two effective masses for each of the two spin-split bands and 
the spin-splitting energy $2\Delta_{Q}$. 
In addition, it is also important to know the energy difference $E_{\rm KQ}$ between 
the band extrema of the $Q_{i}^{(2)}$ and $K_{\rm cb}^{(2)}$ bands. 
\begin{table}[htb]
\caption{
\label{tbl:Q-mat-params} 
Material parameters at the $Q$ point.
$n_Q$ is the electron density above which the carriers start to populate 
the $Q$ valleys, assuming (HSE,LDA) values for $E_{\rm KQ}$.} 
\begin{indented}
\lineup
\item[]
\begin{tabular}{@{}lllllll}
\br
       & MoS$_2$  & MoSe$_2$ & WS$_2$  & WSe$_2$ & MoTe$_2$ & WTe$_2$\\
\hline
$m_{Q,x}^{(1)}/m_e$ (HSE,LDA) & $0.64$ & $0.54$ & $0.69$  & $0.73$  & $0.42$ & $0.44$\\ 
$m_{Q,x}^{(1)}/m_e$ (PBE,PBE) & $0.66$ & $0.58$ &  $0.86$ &  $0.71$ & $0.36$ & $0.44$\\ 
\hline
$m_{Q,y}^{(1)}/m_e$ (V, HSE,LDA) & $1.21$ &  $1.11$ & $0.94$ &   $0.91$ & $1.16$ & $0.922$\\ 
$m_{Q,y}^{(1)}/m_e$ (PBE,PBE)    & $1.31$ &  $1.18$ & $0.95$ &   $0.93$ & $1.18$ & $0.94$ \\
\hline
$m_{Q,x}^{(2)}/m_e$ (HSE,LDA) & $0.56$ & $0.48$ & $0.52$ & $0.42$ & $0.43$ & $0.3$\\  
$m_{Q,x}^{(2)}/m_e$ (PBE,PBE) & $0.61$ & $0.51$ & $0.54$ & $0.45$ & $0.44$ & $0.29$\\
\hline
$m_{Q,y}^{(2)}/m_e$ (HSE,LDA) & $1.13$ & $1.08$  & $0.74$ &  $0.74$ & $0.99$ & $0.81$\\
$m_{Q,y}^{(2)}/m_e$ (PBE,PBE) & $1.21$ & $1.15$  & $0.74$ &  $0.75$ & $1.22$ & $0.8$\\
\br
$2\Delta_{\rm Q}$ [meV] (HSE,LDA)  & $70$ & $21$ & $264$ &  $218$ & $22$ & $192$ \\ 
$2\Delta_{\rm Q}$ [meV] (PBE,PBE)  & $75$ & $26$ & $262$ &  $221$ & $13$ & $201$\\
\br
$E_{\rm KQ}$ [meV] (HSE,LDA)  & $207$ & $137$ & $81$ &  $35$ & $158$ & $158$\\
$E_{\rm KQ}$ [meV] (PBE,PBE)  & $246$ & $163$ & $58$ &  $32$ & $173$ & $140$\\
\hline
$E_{\rm KQ}$ [meV] (Exp)   & $\gtrsim 60^{a}$ & $150\pm30^{b}$ &  & $\approx 0^{b}$ & & \\
\br
$n_{\rm Q}$ [$10^{12}$ cm$^{-2}$] (HSE,LDA)  & $76.42$ & $54.97$ &  $17.17$ & $5.7$& $66.62$ & $37.27$\\
\br
\end{tabular}
\item[] $^{a}$ from ARPES; see \cite{priv-comm-king}.
\item[] $^{b}${\cite{shih2014b}, from scanning-tunnelling microscopy.}
\end{indented}
\end{table}
These parameters, obtained by fitting to our DFT band structures, are shown in Table \ref{tbl:Q-mat-params}.
The fitting range we used was $\approx \pm 7.5\%$  of the $\Gamma$--$K$ distance 
around the $Q$ point in the
$k_x$ direction and roughly half of that in the $k_y$ direction. 
Looking at Table \ref{tbl:Q-mat-params} one can see that the effective masses obtained 
from the two 
DFT calculations are again in good agreement, while  small differences can be seen in the 
results for $2 \Delta_{\rm Q}$. 
However, there are noticeable differences in the energy separation $E_{\rm KQ}$ between 
the bottom of the CB at the $Q$ and the  $K$ points, which we ascribe to the different 
lattice constants used in the two types of DFT calculations. 
There are no experimental results for $E_{\rm KQ}$ 
to date, except for MoS$_2$, where the data indicate that $E_{\rm KQ}\gtrsim
60$ meV \cite{king2014a}.  
Note, however, that the ARPES measurements in Reference \cite{king2014a} were 
performed on potassium-intercalated samples,
and the effects of the intercalation on the band structure of TMDCs 
have not yet been studied in detail. We also note that  {computationally} 
$E_{\rm KQ}$, in contrast to the 
band gap $E_{\rm bg}$, appears to be less sensitive 
to $GW$ corrections \cite{priv-comm-diana} 
if the latter calculations are well converged. 
{As already pointed out in Section \ref{subsec:K-point-params}, 
due to the lack of experimental evidence, it is currently difficult to tell 
how accurate these predictions for the effective masses and spin-splittings are.}

We have also calculated the carrier density $n_Q$ at which the Fermi energy, 
measured from the bottom of the $K$-point valley in the CB, reaches the bottom 
of the $Q$-point valley; see Table \ref{tbl:Q-mat-params}. 
We assumed a simple parabolic dispersion for the CB in the vicinity of $K$, 
where the effective masses of $K_{\rm cb}^{(1)}$ and $K_{\rm cb}^{(2)}$ 
are given in Table~\ref{tbl:K-dftparams-cb}. 
Our results  suggest that for MoX$_2$ it 
{would not be  easy} to achieve the doping levels needed 
to populate the $Q_{i}^{(1)}$ valleys, but for  WS$_2$ and WSe$_2$ the required doping 
levels {appear to be attainable}.

\begin{figure}[ht]
\begin{center}
\includegraphics[scale=0.55]{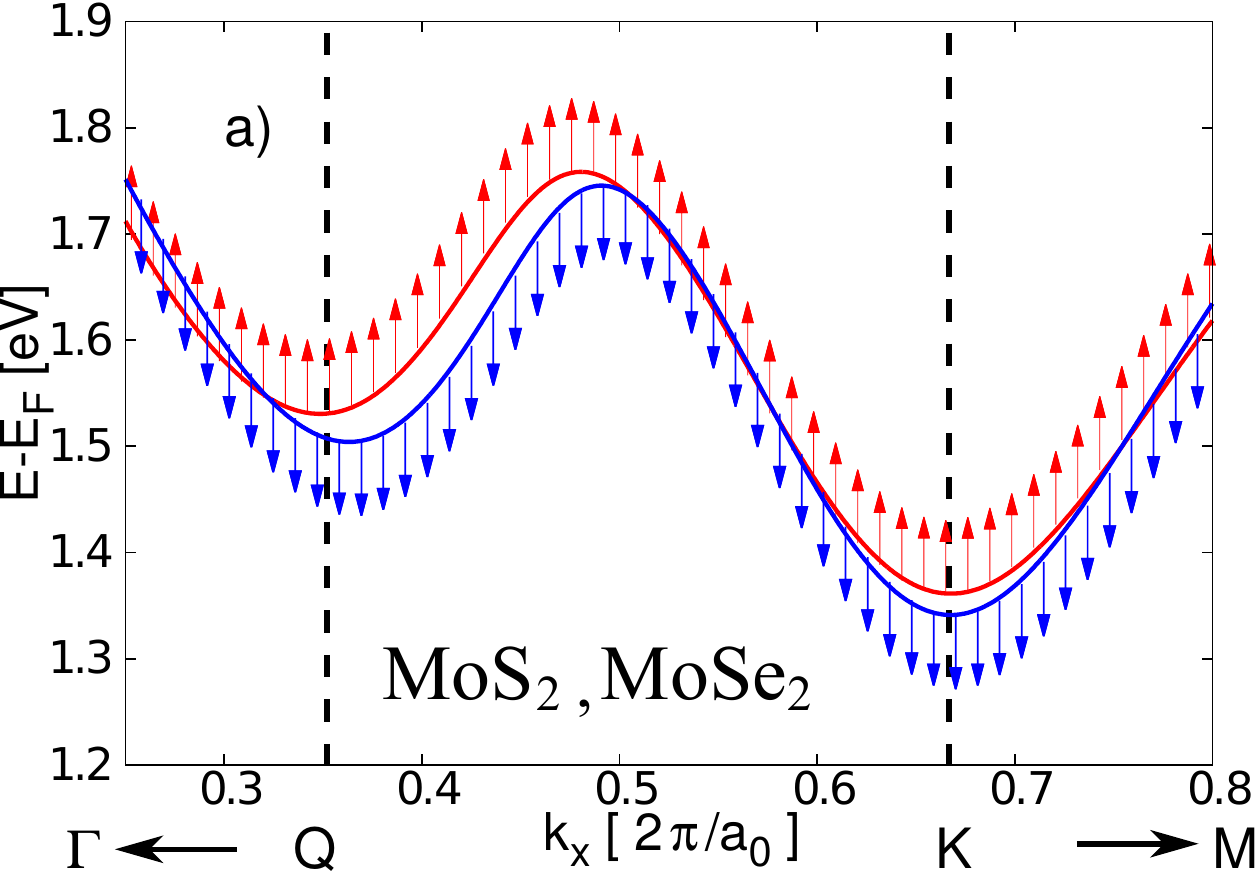}
\includegraphics[scale=0.55]{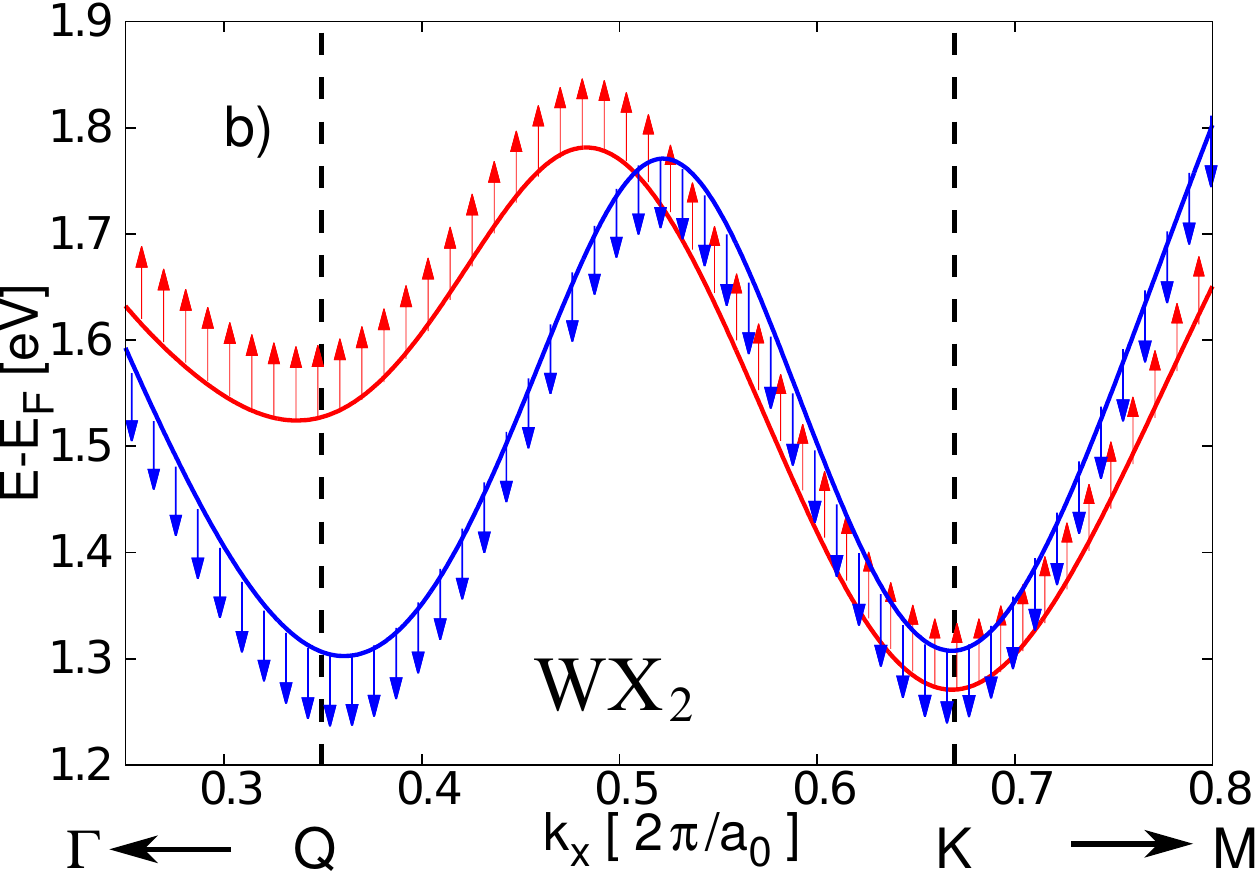}
\includegraphics[scale=0.7]{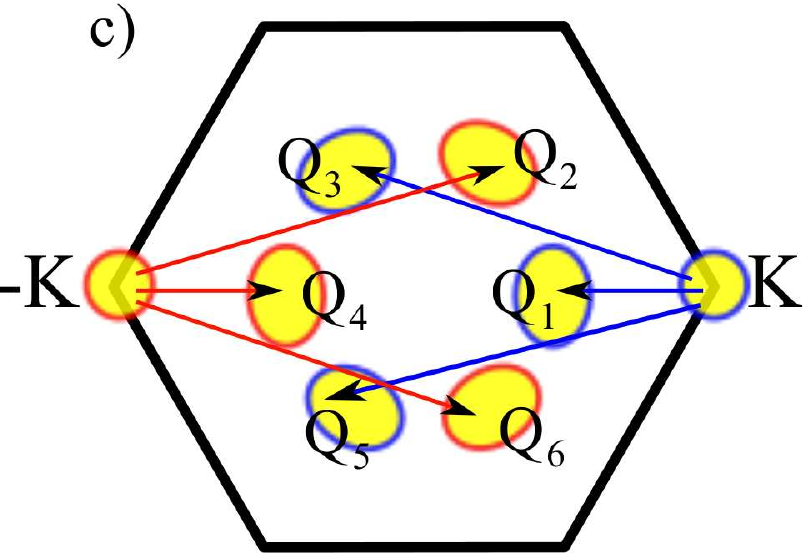}
\includegraphics[scale=0.7]{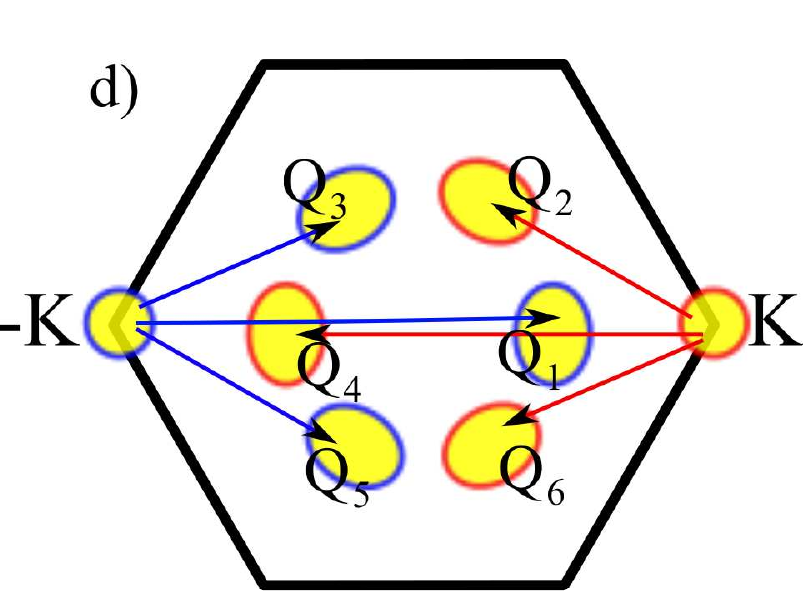}
\end{center} 
\caption{ a) and b): Dispersion of the CB  between the $K$  and $Q_1$ points obtained from DFT calculations. 
         Arrows show the direction of the spin expectation value (red: spin-up, blue: spin-down).
         a) Results for {MoS$_2$ and MoSe$_2$}; 
         b) results for WX$_2$. MoTe$_2$, strictly speaking, is different from both a) and b). 
            The actual calculations were performed for MoSe$_2$ and  
            WSe$_2$ using the (PBE,PBE) approach. c) and d): schematic illustration 
            of the lowest energy allowed 
            scattering processes between the spin-split bands at the $K$ ($-K$) and 
            $Q_{i}$ points. The situation in MoTe$_2$ corresponds to d).
 \label{fig:spins-K-Q}        
         } 
\end{figure}

As noted in Reference \cite{roldan-SOC}, the valley--spin coupling is present not 
only in the $K$ and $-K$ valleys, but also in the $Q_i$ valleys, and this may have
experimental consequences.  
The calculated spin polarisation of the CB between the $K$ and the $Q_{1}$ point is shown 
in   Figure~\ref{fig:spins-K-Q} for MoSe$_2$ and WSe$_2$. 
One  finds that despite  the  band crossing(s) between the 
$K$  and $Q_1$ points,   for {MoS$_2$ and MoSe$_2$} 
the  spin-polarisation of  the $Q_1^{(2)}$  band is 
\emph{the same} as  the spin polarisation of the $K_{cb}^{(2)}$ band 
[see Figure~\ref{fig:spins-K-Q}(a)]. 
For WX$_2$ and {MoTe$_2$}, however, due to the band crossings, 
the spin-polarisation of $K_{cb}^{(2)}$
is \emph{opposite} to the spin polarisation of   $Q_1^{(2)}$ [Figure \ref{fig:spins-K-Q}(b)]. 
The spin polarisation of the $Q_{i}^{(1)}$ and $Q_{i}^{(2)}$ bands in 
other $Q_{i}$ valleys can be deduced  by taking into account time-reversal
symmetry and whether they are along the $\Gamma$--$K$ or $\Gamma$--$(-K)$ line. 
The spin-polarisation of bands at the $Q_{i}$ points determines which scattering 
processes are allowed or suppressed between the $K$ ($-K$) and $Q_{i}$ 
valleys. This is illustrated in Figures \ref{fig:spins-K-Q}(c) and \ref{fig:spins-K-Q}(d).
For example, in the case of   {MoS$_2$ and MoSe$_2$} 
[Figure \ref{fig:spins-K-Q}(c)] 
scattering from $K_{cb}^{(2)}$ to $Q_{1}^{(2)}$, $Q_{3}^{(2)}$, 
 and $Q_{5}^{(2)}$ is allowed, while scattering to  $Q_{2}^{(1)}$, $Q_{4}^{(1)}$, 
 and $Q_{6}^{(1)}$ is, strictly speaking, also allowed but should be  suppressed with respect 
the former processes due to the relatively large spin-splitting $2\Delta_{Q}$.


\subsection{$\mathbf{k}\cdot\mathbf{p}$ Hamiltonian}
\label{subsec:kp-at-Q}

Due to the  low symmetry of the $Q_{i}$ points in the BZ and because there are 
many nearby bands in energy, there is a  large number of  band-overlap parameters that would need to
be taken into account in  a detailed multi-band $\mathbf{k}\cdot\mathbf{p}$ model. 
Therefore it is more difficult to develop such a theory and it would offer less insight. 
Nevertheless, a low-energy effective $\mathbf{k}\cdot\mathbf{p}$ Hamiltonian can 
be derived with the help of the theory of invariants \cite{BirPikus}
(for a recent discussion see, e.g., References \cite{winkler-graphene-symm} and \cite{budich}). 
The pertinent symmetry group is $C_{1h}$; for convenience,  its  
character table is shown in Table \ref{tbl:c1h} \cite{dresselhaus-book}.  
\begin{table}[hb]
\caption{Character table  and invariants for the group C$_{1h}$.}
 \label{tbl:c1h}
\begin{indented}
\lineup
\item[]
\begin{tabular}{@{}llrrr}
\br
  & C$_{1h}$ &  &  $E$ & $\sigma_{h}$ \\
\hline
$A'$  &  $k_x^2$,  $k_y^2$, $k_x k_y$  & $s_z$,  $k_x$, $k_y$ &  1 & 1 \\
\hline
$A''$ &                                & $z$, $s_x$, $s_y$         &  1  & $-1$ \\
\br
\end{tabular}
\end{indented}
\end{table}

As in the case of the $K$ and $-K$ valleys, 
the $Q_{i}$-point minima are pairwise connected by time-reversal 
symmetry and to describe this one can introduce the matrix $\tau_z$, whose eigenvalues,  
$\tau=\pm 1$ label individual members of the pairs of valleys. 
As an example, let us consider the $Q_{1}$ ($\tau=1$) and $Q_{4}$ ($\tau=-1$) minima,  
which can be found along the $\Gamma$--$K$  and 
$\Gamma$--$(-K)$ directions, respectively [see Figure \ref{fig:defs}(c)].  
This direction is parallel to  the $k_x$ component of  $\mathbf{k}$. 
{Using Table \ref{tbl:c1h}, the most general Hamiltonian, up to second-order 
in  $\mathbf{k}$ and taking SOC into account, reads:}
\begin{eqnarray}
H_{Q}^{\tau,s}=&\frac{\hbar^2 k_x^2} {2 m_{Q,x}^{\tau,s}} +
\frac{\hbar^2 k_y^2}{2 m_{Q,y}^{\tau,s}}+
\frac{\hbar^2 k_x k_y}{2 m_{Q,xy}^{\tau,s}}+\Delta_Q^{} s_z \tau_z+ 
a_1 k_x s_z + a_2 k_y s_z \nonumber\\
  &+b_1 k_x \tau_z + b_2 k_y \tau_z+ E_Q,
\label{eq:H_Q-gen}
\end{eqnarray}
${1}/{m_{(Q,x,y,xy)}^{\tau,s}}$ are effective masses,
$s=\pm 1$ are the eigenvalues of the spin Pauli matrix $s_z$. 
Furthermore, $E_Q$ is the band-edge energy 
if SOC is neglected, $\Delta_Q^{}$ is the spin-splitting at $Q$, and $a_{1,2}$ and $b_{1,2}$ are  material parameters to be 
discussed later. Since we are going to develop a theory  in which the dispersion is parabolic, 
in contrast to Section~\ref{sec:K-point-main}, we will not keep track of  
the free-electron contribution explicitly. 

To simplify the discussion, let us first neglect the spin degree of freedom. Then 
$\Delta_Q^{}=a_{1,2}=0$ and  ${m_{(Q,x,y,xy)}^{\tau,s}}={m_{(Q,x,y,xy)}^{0}}$. 
Since close to $Q_{1}$ ($Q_{4}$)  the 
energy contours are, to a good approximation, ellipses whose axes are in the 
$k_x$ and $k_y$ directions  
[see Figure~\ref{fig:Qcontour}(a)], one finds  that  $1/m_{(Q,xy)}^{0} = 0$.  
{The effect of the terms $\sim b_1, b_2$ in Equation \eref{eq:H_Q-gen} is to shift 
the minimum of the dispersion. Therefore introducing the wavenumbers 
$q_x$ and $q_y$, which are measured from 
$\mathbf{k}=(\tau k_Q,0)$, i.e., from the $Q_{1}$ ($Q_{4}$) point, one can set 
$b_1=b_2=0$ and write}
\begin{eqnarray}
 H_{Q}^{0}=\frac{\hbar^2 q_x^2} {2 m_{Q,x}^{0}} +
\frac{\hbar^2 q_y^2}{2 m_{Q,y}^{0}}+\tilde{E}_{\rm KQ},
\label{eq:noSO_H_Q}
\end{eqnarray}
where $\tilde{E}_{\rm KQ}$ measures the energy difference with respect to the 
$K$ point in the absence of SOC\@. 
The effective masses $m_{Q,x}^{0}$ and $m_{Q,y}^{0}$ are in general different.

Taking SOC into account, $H_{Q}^{\tau,s}$ [Equation \eref{eq:H_Q-gen}] 
can be re-written in the following form:
\begin{eqnarray}
H_{Q}^{\tau,s}=\frac{\hbar^2 (q_x+ s_z \cdot q_{Q,x})^2} {2 m_{Q,x}^{\tau,s}} +
\frac{\hbar^2 (q_y+s_z\cdot q_{Q,y})^2}{2 m_{Q,y}^{\tau,s}}+
\Delta_Q^{} s_z \tau_z+ E_{KQ},
\label{eq:H_Q_SOC}
\end{eqnarray}
where $E_{KQ}$ is defined in Figure~\ref{fig:defs}. 
One can see that SOC has the following effects:
\begin{itemize}
 \item[i)] it splits the bands and  opens a gap  $\Delta_Q^{}$ between the 
 spin-up and spin-down bands;
  \item[ii)] it shifts  the minima of the spin-split bands off from the  
  $\mathbf{k}=(\tau k_Q,0)$ point {by $q_{Q,x}$ and $q_{Q,y}$};
 \item[iii)] it makes the effective masses of the spin-polarised bands different, 
 so they are given by 
 ${1}/{m_{(Q,x,y)}^{\tau,s}}={1}/{m_{(Q,x,y)}^{0}}-\tau s /{\delta m_{(Q,x,y)}}$.
\end{itemize}
An illustration of i) and ii) is shown in Figure~\ref{fig:Qcontour}(b) taking  WSe$_2$ 
as an example, where these effects are most clearly seen.  
The material  parameters   $m_{Q,x,y}^{\tau,s}$, $q_{Q,x}$, $q_{Q,y}$,  $\Delta_Q^{}$, 
and $E_{KQ}$
can be obtained from, e.g.,  DFT calculations; see Section \ref{subseq:Q-point-params}.
We find that $q_{Q,y}$ is zero within the precision of our calculations and $q_{Q,x}$ 
is always very small.
%


\section{Effective  models at the $\Gamma$ point}
\label{sec:G-point-main}

\subsection{$\Gamma$ point}

Next  we  consider the band structure at the $\Gamma$ point.  
There are three main  motivations  to include the $\Gamma$ point in  our work: 
i) there is a local maximum in the VB  at the $\Gamma$ point,
which could be observed in recent ARPES measurement \cite{osgood,shen} and therefore 
it is of interest to compare the experimental and calculated effective masses;
ii) {there are several experimental reports \cite{eda2014,bolotin,kis-peaks,potemski} 
    on optical transitions over a broad energy range showing peak(s) in the 
    absorption of monolayer TMDCs at energies larger than the one corresponding to 
    the fundamental gap at the $K$ point. Theoretically, it was argued that 
    in MoS$_2$ excitons can also be formed in the  vicinity of the $\Gamma$
    point \cite{qiu}
    and that these   ``C-excitons'' are qualitatively different from the ones 
     at the $K$ point because  they arise from  an effectively one-dimensional 
    energy minimum  in the ``optical band structure'' (for the exact definition see below); and}
iii) finally, understanding of the  VB and CB  behaviour at the $\Gamma$ point is 
important in the interpretation of scanning tunnelling microscopy (STM) 
experiments \cite{crommie,shih2014a,shih2014b,lainjong,andrei,maohaixie}.


\subsection{Basic characterisation and material parameters}
\label{subsec:mat-param-at-G}

We start the discussion with the VB maximum at the 
$\Gamma$ point (VBMG)\@. 
The spin-splitting is  at $\Gamma$ point is zero, and in the VB it remains negligible  
over a  considerable region of $\mathbf{k}$ space 
[see Figure \ref{fig:defs}(a)].  
Moreover, to a good approximation the dispersion around the VBMG is isotropic 
(see Figure \ref{fig:defs}(b) and Section \ref{subsec:kp-at-Gamma}) and parabolic. 
Therefore it  can be described  by 
\begin{eqnarray}
E_{\Gamma}(\mathbf{k})=E_{\rm K\Gamma}+\frac{\hbar^2 \mathbf{k}^2}{2 m_{\Gamma}^{\rm vb}},
\label{eq:H-Gamma-VB}
\end{eqnarray}
which  is characterised by a single effective mass $m_{\Gamma}^{\rm vb}$ and 
the energy $E_{\rm K\Gamma}$,  which is the energy difference between the maximum of the 
$K_{\rm vb}^{(1)}$ band at the $K$ point and the VBMG\@. 
The values of $m_{\Gamma}^{\rm vb}$ obtained from fitting the 
results of our  DFT  calculations are given  in Table \ref{tbl:dft-at-G} and 
experimental results, where available, are also shown.
\begin{table}[htb]
\caption{\label{tbl:dft-at-G} 
Effective masses $m_{\Gamma}^{\rm vb}$ at the $\Gamma$ 
point in the VB from DFT calculations.  $m_e$ is the free electron mass. 
The energy difference $E_{\rm K\Gamma}$ between the VBMG and VBMK1 is also given. 
$n_{\Gamma}$ is the hole density where the states at the $\Gamma$ point start
to fill with holes assuming the (HSE,LDA) values for $E_{\rm K\Gamma}$.
Experimental values are shown in rows denoted by ``Exp''.
}
\begin{indented}
\lineup
\item[]
\begin{tabular}{@{}lllllll}
\br
       & MoS$_2$  & MoSe$_2$ & WS$_2$  & WSe$_2$ & MoTe$_2$ & WTe$_2$ \\
\hline
$m_{{\Gamma}}^{\rm vb}/m_e$ (HSE,LDA)  & $-2.60$  & $-3.94$ & $-2.18$ &  $-2.87$ & $-29$    & $-5.19$ \\
$m_{{\Gamma}}^{\rm vb}/m_e$ (PBE,PBE)  &  $-2.45$ & $-3.49$ & $-2.15$ & $-2.70$  & $-10.76$ & $-4.18$\\ 
\hline
$m_{{\Gamma}}^{\rm vb}/m_e$ (Exp) & $-2.4\pm0.3^{a}$ & $-3.9\pm0.3^{b}$ & &  & & \\
\br
$E_{\rm K\Gamma}$ [meV] (HSE,LDA)  & $-70$ &  $-342$ & $-252$ & $-496$ & $-540$ & $-630$\\ 
$E_{\rm K\Gamma}$ [meV] (PBE,PBE)  & $-46$ &  $-329$ &  $-269$ &  $-506$ & $-526$ & $-646$\\
\hline
$E_{\rm K\Gamma}$ [meV] (Exp)        &  $\approx -140^{a}$      &  $-380^{c}$      &  & $-880^{e}$ & & \\
                                     &                          &  $-370\pm40^{d}$ &  & $-590\pm40^{d}$ &  & \\
\br
$n_{\Gamma}$[$10^{12}$cm$^{-2}$](HSE,LDA) & $15.8$ & $130$ & $36.86$ & $81.4$ & $259$ & $124.92$\\
\br
\end{tabular}
\item[] $^{a}${\cite{osgood}, ARPES measurements, exfoliated samples  on SiO$_2$ substrate.}
\item[] $^{b}${private communication by Yi Zhang, ARPES measurements, see \cite{shen}.}
\item[] $^{c}${\cite{shen}, ARPES measurements, samples grown by MBE 
on bilayer graphene on top of SiC (0001).}
\item[] $^{d}${\cite{shih2014b}, STM measurements, samples grown by MBE  on HOPG.} 
\item[] $^{e}${\cite{dowben}, ARPES measurements, exfoliated samples  on SiO$_2$ substrate.}
\end{indented}
\end{table}
The effective masses $m_{\Gamma}^{\rm vb}$ were obtain by fitting the band structure along 
the $\Gamma$--$K$ direction 
in a range of  $\approx 21\%$ of the $\Gamma$--$K$ distance. The calculated $m_{\Gamma}^{\rm vb}$ values are in 
reasonable agreement with the available experimental results. There are, however, 
noticeable differences between the theoretical and experimental $E_{\rm KG}$  values,
which we attribute to substrate effects. Note  that the weight of 
the chalcogen $p$ orbitals is substantial at the  $\Gamma$ point 
(see Figure \ref{fig:atomic-orbs} or 
Reference \cite{capellutti}) so that one can expect a stronger  interaction between 
the substrate and the electronic states. 
Interestingly, the $m_{{\Gamma}}^{\rm vb}$ parameter does not seem to be affected as strongly 
as $E_{\rm K\Gamma}$ by the substrate. 
Comparing Tables \ref{tbl:K-dftparams-vb} and \ref{tbl:dft-at-G}, 
one would expect  the  VBMG  to be the most important  
for the transport properties of MoS$_2$ because it is probably  
quite close in energy to the maximum of the $K_{\rm vb}^{(1)}$ band 
(which we denote by VBMK1; we also introduce the label VBMK2 
for the maximum of the $K_{\rm vb}^{(2)}$ band), 
thus facilitating scattering processes in the VB that do not require spin-flips. 
Furthermore, according to our DFT calculations 
VBMG lies  in between  VBMK1  and VBMK2  
for  MoS$_2$ and WS$_2$, while it lies below VBMK2 for MoSe$_2$ and WSe$_2$.

\begin{figure}[ht]
\begin{center}
 \includegraphics[scale=0.65]{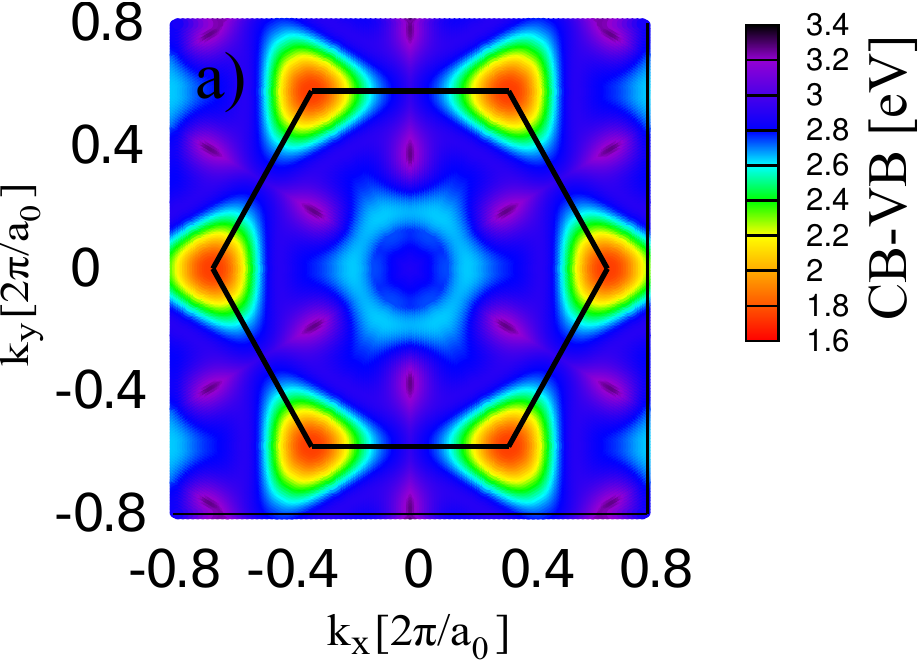}
 \includegraphics[scale=0.65]{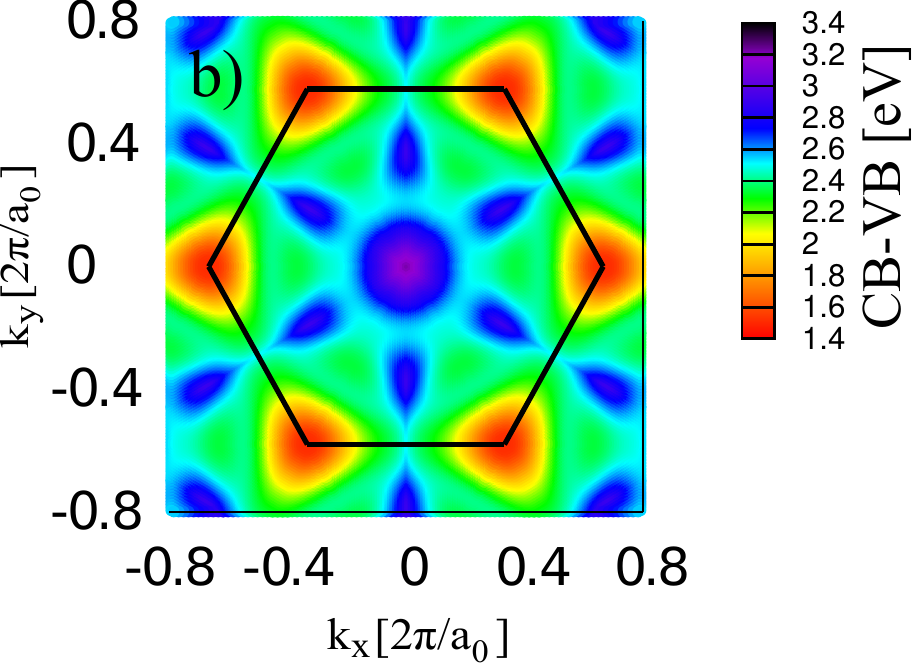}
 \includegraphics[scale=0.7]{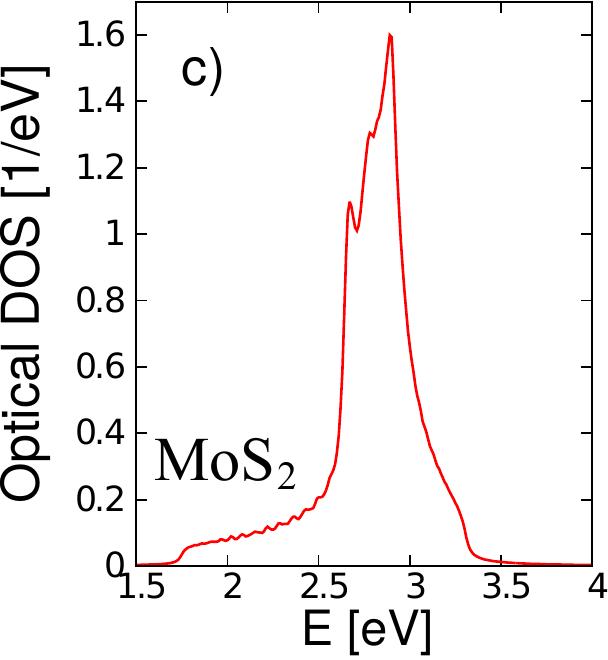}\hspace{2cm}
 \includegraphics[scale=0.7]{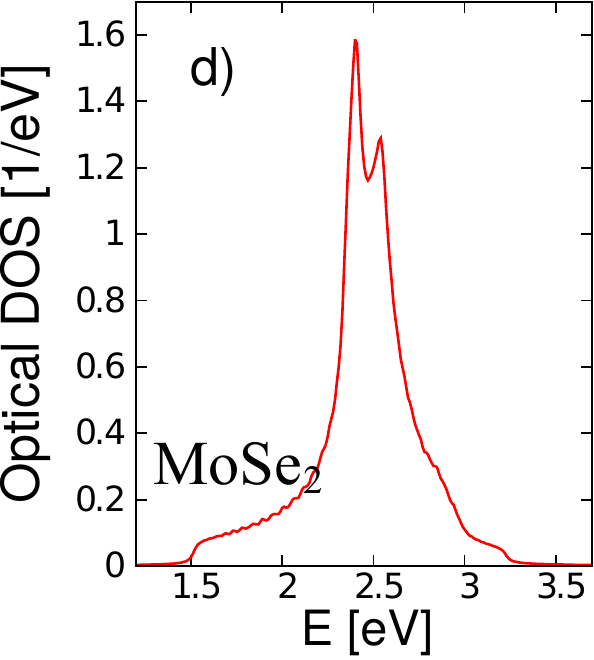}
 \end{center}
 \caption{Optical band structures  [a)  and b)]  and the corresponding
          optical densities of states  [c) and d)] obtained from (HSE,LDA) 
          DFT calculations in which SOC was neglected.   
          In a) and c) data for  MoS$_2$ are shown; in b) and d) data for MoSe$_2$ are shown. 
          In a) a lighter colour ``gear-shaped'' region around the $\Gamma$ point is clearly visible. 
 \label{fig:optDOS-1} }
\end{figure}

\begin{figure}[th]
\begin{center}
 \includegraphics[scale=0.65]{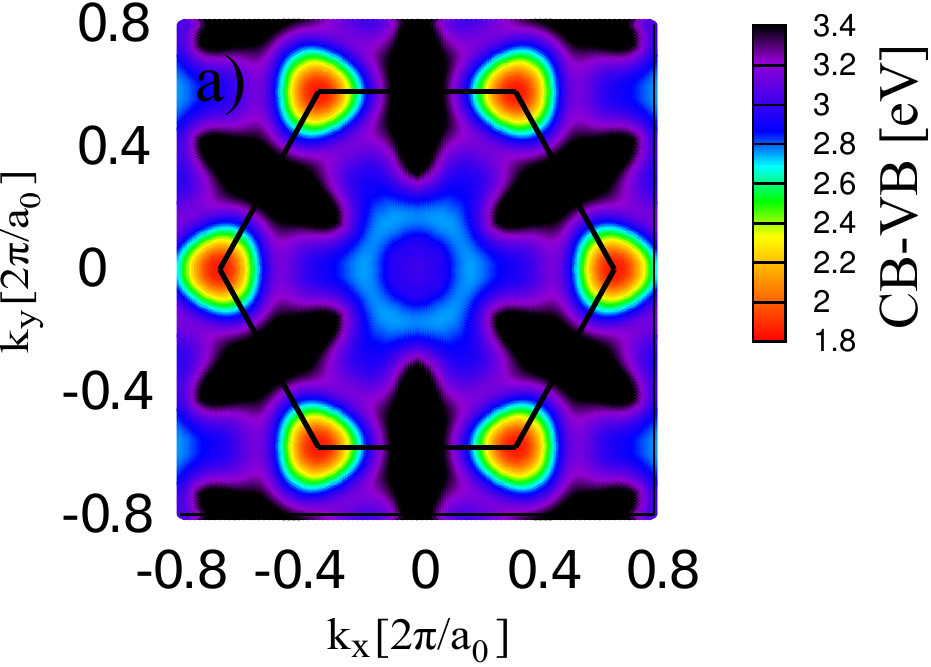}
 \includegraphics[scale=0.65]{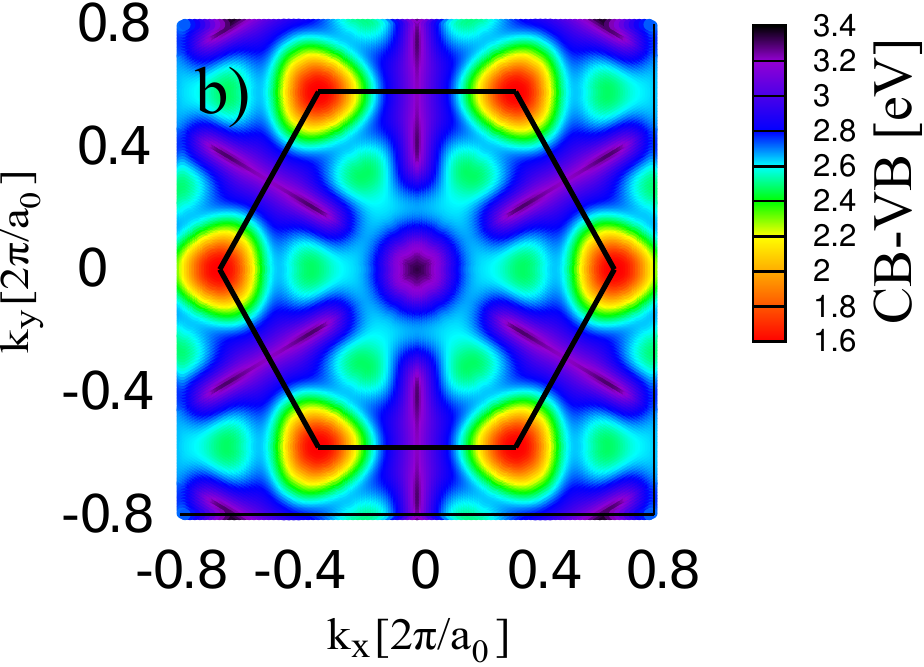}
 \includegraphics[scale=0.7]{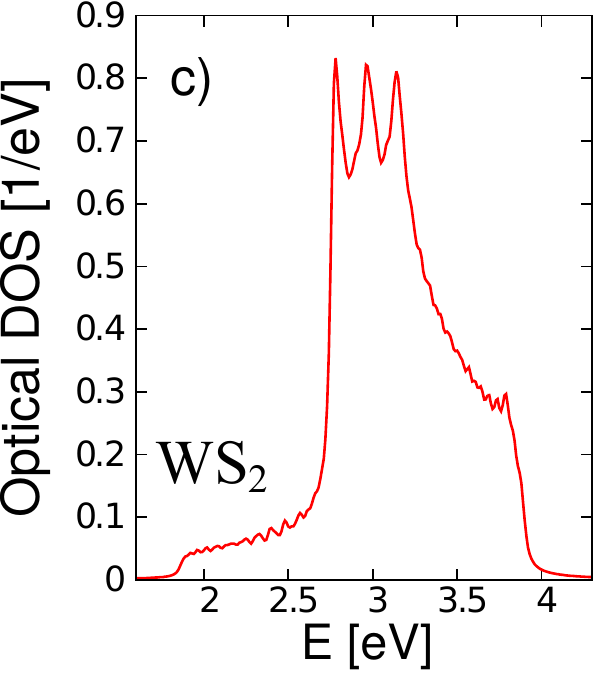}\hspace{2cm}
 \includegraphics[scale=0.7]{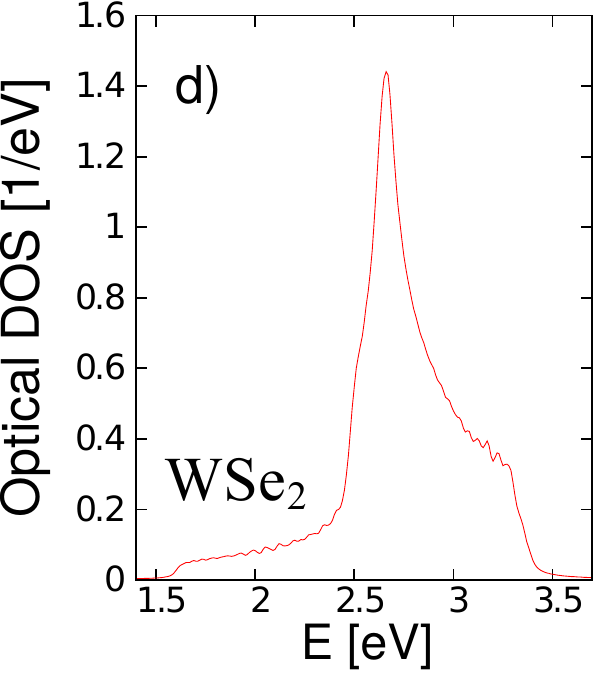}
 \end{center}
 \caption{Optical band structures [a)  and b)]  and the corresponding
          optical densities of states  [c) and d)] obtained from (HSE,LDA) 
          DFT calculations in which SOC was neglected.   
          In a) and c) data for  WS$_2$ are shown; in b) and d) data for WSe$_2$ are shown.
 \label{fig:optDOS-2} }
\end{figure}

\begin{figure}[th]
\begin{center}
 \includegraphics[scale=0.65]{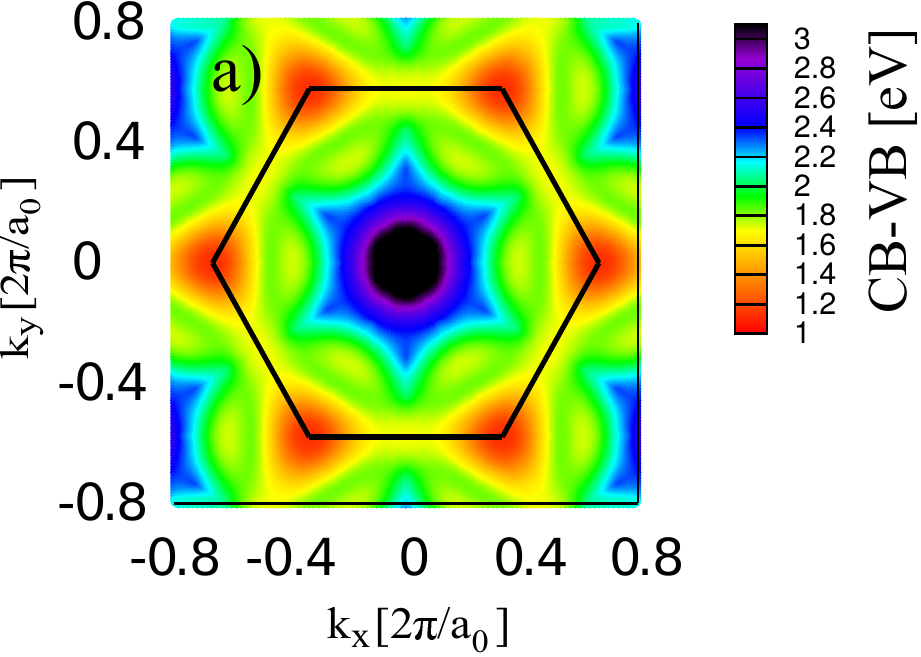}
 \includegraphics[scale=0.65]{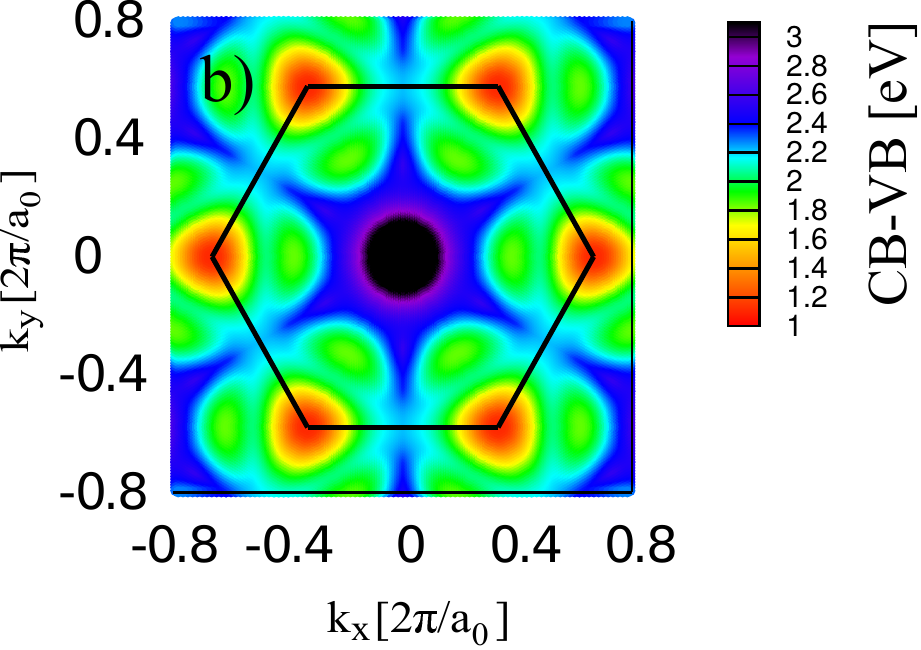}
 \includegraphics[scale=0.7]{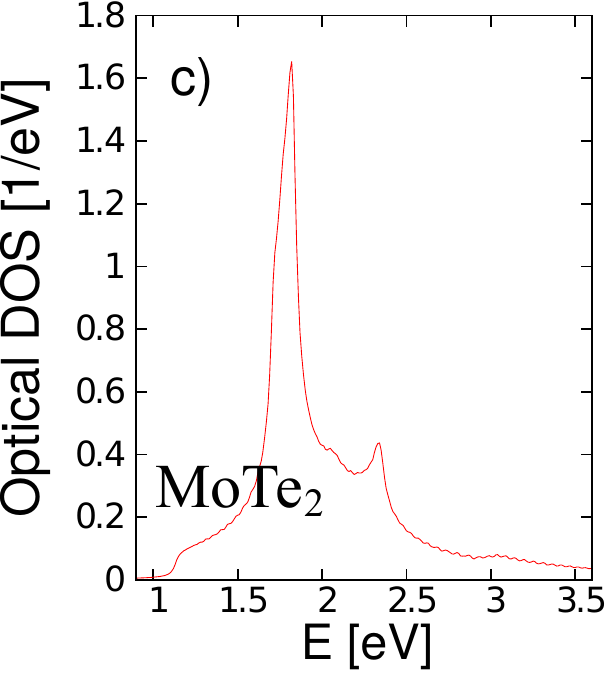}\hspace{2cm}
 \includegraphics[scale=0.7]{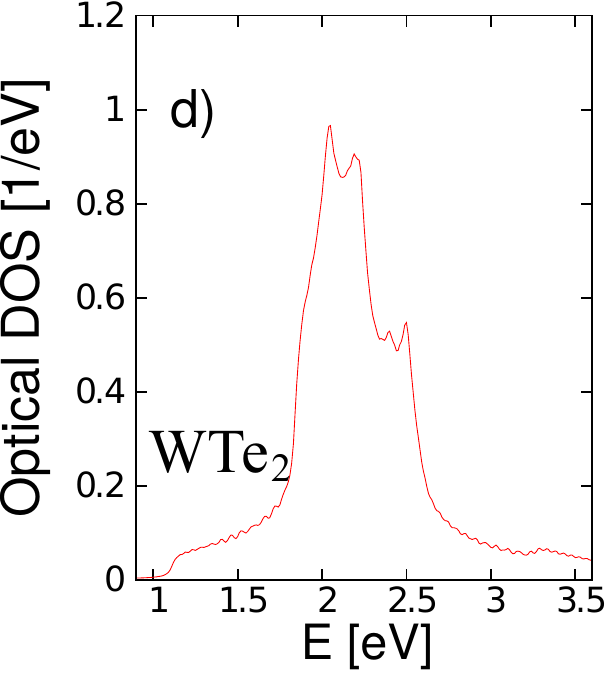}
 \end{center}
 \caption{Optical band structures [a)  and b)]  and the corresponding
          optical densities of states  [c) and d)] obtained from (HSE,LDA) 
          DFT calculations in which SOC was neglected.   
          In a) and c) data for  MoTe$_2$ are shown; in b) and d) data for WTe$_2$ are shown.
 \label{fig:optDOS-3} }
\end{figure}

Optical transitions at higher energies than the ones at the fundamental band gap 
have attracted considerable  theoretical interest \cite{qiu,carvalho,gies,polini} recently. 
To obtain an insight into the possible  transitions, we plot the optical band structure for 
monolayer TMDCs in Figures \ref{fig:optDOS-1}-\ref{fig:optDOS-3} over the whole BZ\@. 
Here, following References \cite{bolotin} and \cite{dresselhaus-exciton}, 
the optical band structure 
is defined as the difference between the dispersions of the CB and VB: 
$E_{\rm cb}(\mathbf{k})-E_{\rm vb}(\mathbf{k})$\@. 
For simplicity, SOC is neglected in these calculation. 
A clear ``gear-shaped'' minimum \cite{bolotin} is noted both  for MoS$_2$ and WS$_2$ around the 
$\Gamma$  point [Figures \ref{fig:optDOS-1}(a) and \ref{fig:optDOS-2} (a)] 
{and for each material one can also observe  saddle point(s). Both minima and 
saddle points lead to Van Hove singularities in the optical density of states 
(see below) and can have an important effect on the interband optical transitions.} 
For a more quantitative understanding of the interband transitions therefore 
one also needs to consider  the optical density of states (optical DOS),
which is defined as the density of states of the optical band structure 
(the terminology ``joint density of states'' is also used; see, e.g., 
\cite{dresselhaus-exciton}). {For 2D systems it reads}
\begin{eqnarray}
 \rho_{cv}(E)=\frac{2}{(2\pi)^2}\int 
 \delta([E_{\rm cb}(\mathbf{k})-E_{\rm vb}(\mathbf{k})]-E)\, d^2\mathbf{k}.
 \label{eq:otpDOS}
\end{eqnarray}
{The calculated  optical DOS for the monolayer 
TMDCs that we have considered are shown in Figures \ref{fig:optDOS-1}--\ref{fig:optDOS-3}. }
A peak in the optical DOS  corresponding to the minimum in the optical band structure 
is present at $\approx 2.65$ eV  for MoS$_2$ [Figure  \ref{fig:optDOS-1}(c)] 
and $\approx 2.75$ eV  for WS$_2$ [Figure \ref{fig:optDOS-2}(c)]. 
However,  other  peak(s)  and  a wide  shoulder extending into higher energies can  also be seen  in the optical DOS\@.
We attribute these features to saddle points in the optical band structure, which can be observed, e.g.,  
along the $\Gamma$--$K$ line for WS$_2$ and to the saddle points 
at the $M$ point  in the optical band structure of all four MX$_2$ materials. 
These  observations motivate us to have a closer look at the band structure at the $M$ 
point as well,  which is presented in  Section \ref{sec:M-point-main}. 

Finally, we emphasise that for a  quantitative understanding  of the optical band structure and 
 the  interband optical transitions  the effects of SOC are also important. In general,
 they lead to spin-splitting of the bands ({except along the $\Gamma$--$M$ line}), 
 or splitting of the Van Hove singularity (see Section \ref{subsec:M-point-all}).
{The energies of these splittings may be comparable to or larger  than
the linewidth of the optical transitions leading to, e.g., 
the possibility of spin-polarised optical current injection \cite{dxiao,polini}.}


\subsection{$\mathbf{k}\cdot\mathbf{p}$ Hamiltonian}
\label{subsec:kp-at-Gamma}

As in previous sections, we use group theory to obtain  effective  $\mathbf{k}\cdot\mathbf{p}$ 
Hamiltonians for the VB and CB\@. 
Similarly to the  $K$ point, it is possible to set up a multi-band $\mathbf{k}\cdot\mathbf{p}$ 
model. We  have  found, however, that the number of necessary bands, even if one neglects SOC, 
is quite large and, as will be shown later,  terms up to fourth order in $\mathbf{k}$ 
need to be taken into account
in order to capture the features of the band structure related to the C-exciton terms.  
Therefore we present here only a simplified discussion of the problem; a more complete theory is left 
for a future work. As we will show, important insight can be gained from the spinless case, i.e., 
in the discussion that follows we will neglect SOC\@.

\begin{figure}[ht]
\begin{center}
\includegraphics[scale=0.5]{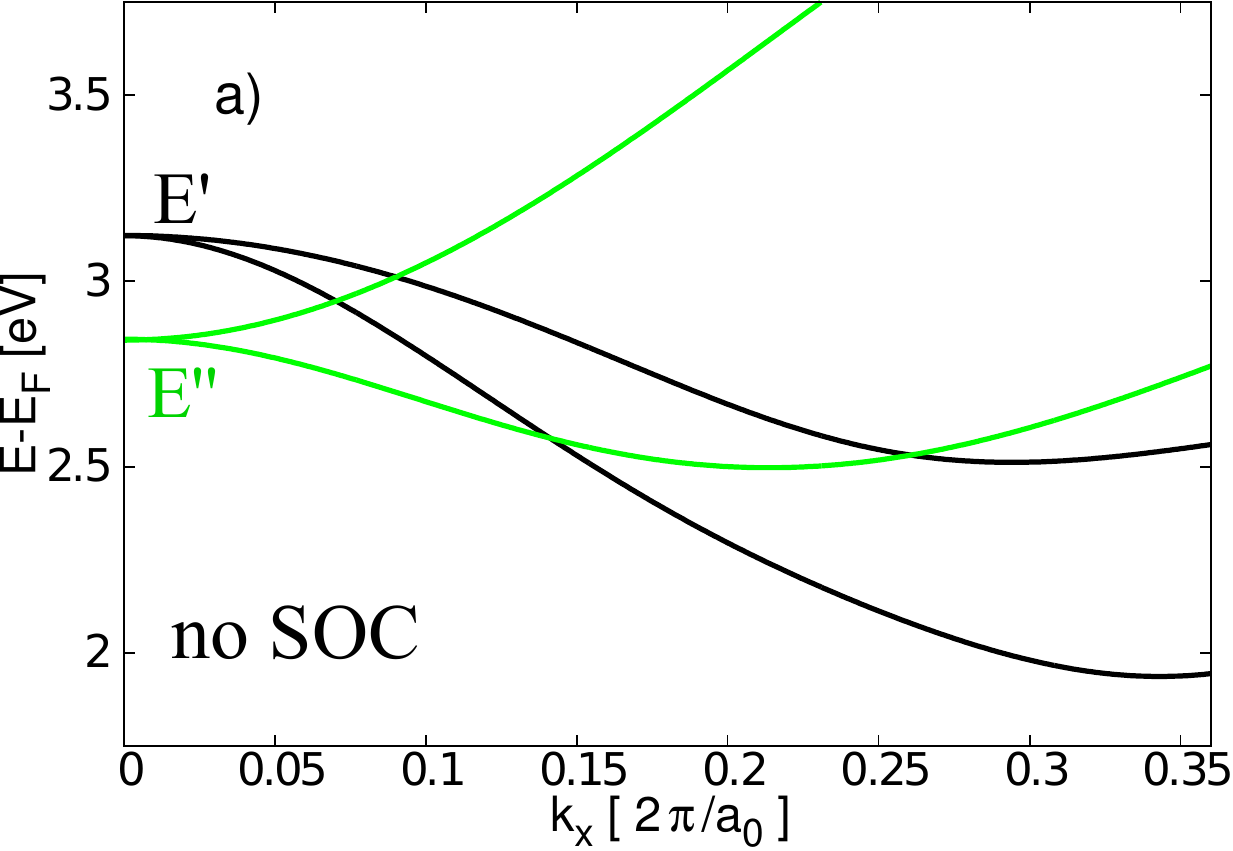}
\includegraphics[scale=0.5]{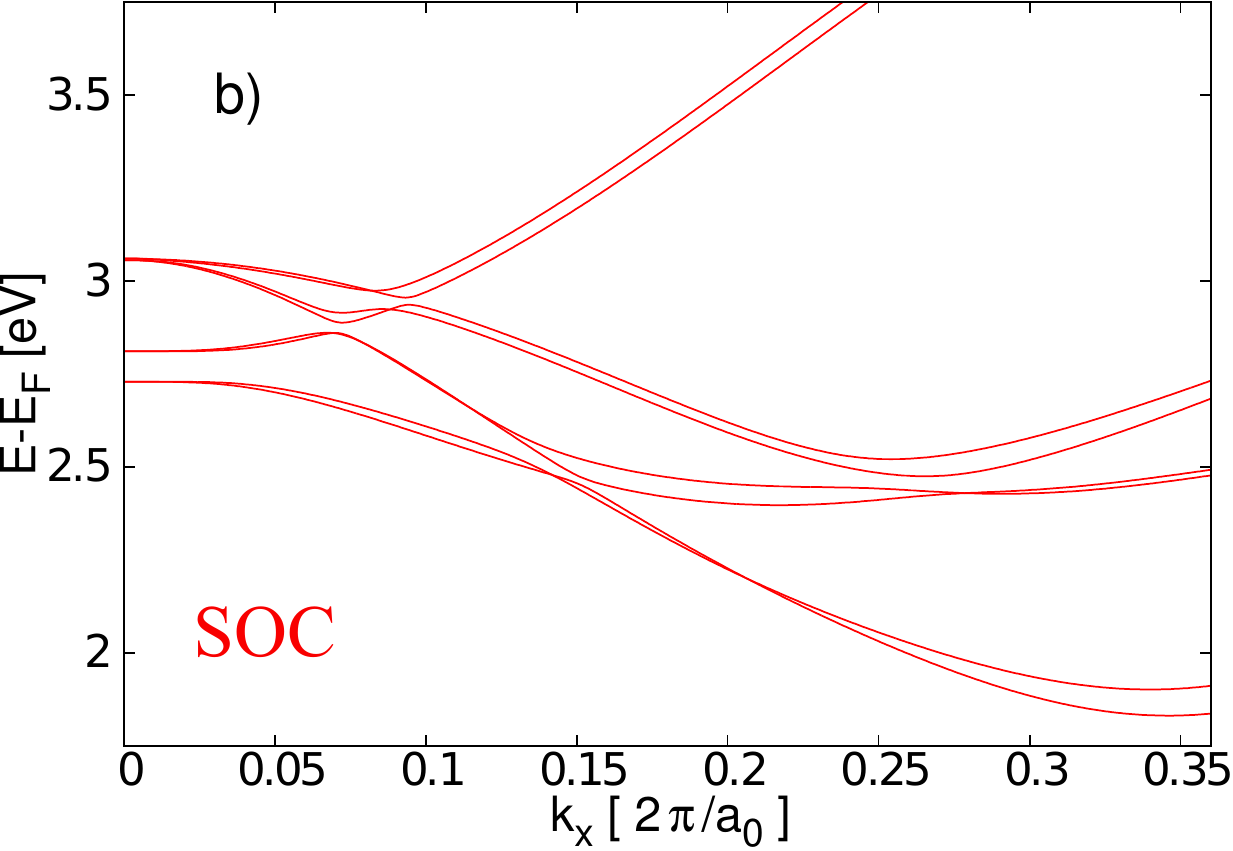}
\caption{ a) The dispersion of the CB and CB+1 bands along the $\Gamma$--$K$ direction, without 
taking SOC into account. Black lines show the symmetric $E'$  bands and
green lines the antisymmetric $E''$ bands.  
b) The same as in  a) but the SOC is  taken into account.  
The actual  DFT calculation were
performed for MoS$_2$ using the (HSE,LDA) approach; for other materials the results are qualitatively 
 similar, except that the spin-splittings  are larger.
\label{fig:bands-at-Gamma}
}
\end{center}          
\end{figure}

The bands of interest are the VB, the (doubly degenerate) CB [shown by green lines in Figure \ref{fig:bands-at-Gamma}(a)] 
and the first (doubly degenerate)  band \emph{above} the CB, which we denote by CB+1 
[black lines in Figure \ref{fig:bands-at-Gamma}(a)]. 
We  will   rely on group-theoretical arguments, which are very convenient at the  $\Gamma$ point,
where, as mentioned above, several atomic orbitals contribute with significant weight to each band. 
The pertinent symmetry group  is $D_{3 h}$ and  the character table is shown in  
Table \ref{tbl:charactd3h}. 
\begin{table}[htb]
\caption{
\label{tbl:charactd3h} Character table of the point group $D_{3h}$.}
\begin{indented}
\lineup
\item[]
\begin{tabular}{@{}llrrrrr}\br
 D$_{3h}$ &  $E$  & $\sigma_h$ & $2 C_3$  &  $2 S_3$  &  $3 C_2'$ & $3\sigma_v$  \\
 \hline
 $A_1'$  & $1$  & $1$  & $1$  & $1$  & $1$  & $1$\\ 
 $A_2'$  & $1$  & $1$  & $1$  & $1$  & $-1$ & $-1$\\
 $A_1''$ & $1$  & $-1$ & $1$  & $-1$ & $1$  & $1$\\
 $A_2''$ & $1$  & $-1$ & $1$  & $-1$ & $-1$ & $-1$\\
 $E'$    & $2$  & $2$  & $-1$ & $-1$ & $0$  & $0$\\
 $E''$   & $2$  & $-2$ & $-1$ & $1$  & $0$  & $0$\\
 \br
\end{tabular}
\end{indented}
\end{table}

Symmetry analysis of the contributing atomic orbitals implies 
(see, e.g., Table IV in Reference \cite{sajat1} and the discussion at the end of 
\ref{sec:seven-band-at-K}) 
that the  VB  at the $\Gamma$  point belongs to the $A_1^{'}$ irreducible 
representation of $D_{3h}$.  As already given  in Equation \eref{eq:H-Gamma-VB}, 
up to second order in $\mathbf{q}$,  the dispersion is parabolic and 
isotropic  [see Figure \ref{fig:defs}(b)], characterised by a single effective mass $m_{\Gamma}^{\rm vb}$. 
Values of $m_{\Gamma}^{\rm vb}$ obtained from fitting the DFT band structures are 
shown in Table~\ref{tbl:dft-at-G}.  Along the $\Gamma$--$K$ direction,  the spin-splitting of the VB is  small
up to wavevectors  corresponding to about half of the $\Gamma$--$Q$ distance. This is due to the fact that 
in the vicinity of $\Gamma$ the $d_{z^2}$ atomic orbitals of the metal  
and the $p_z$ atomic orbitals of the chalcogen  atoms contribute with large weight to the VB 
(see Figure \ref{fig:atomic-orbs} and \cite{liugb,capellutti}). 
Along $\Gamma$--$M$ all  bands remain spin-degenerate due to symmetry; see Section \ref{subsec:M-point-all}. 
The spin-splitting of the VB is therefore suppressed around the $\Gamma$ point.

Turning now to the CB [shown by green lines in Figure \ref{fig:bands-at-Gamma}(a)],  
at the $\Gamma$  point it is  doubly degenerate and  
antisymmetric with respect to the horizontal mirror plane $\sigma_h$ of the 
crystal lattice. In group theoretical terms, it corresponds to the 
$E''$ irreducible representation (irrep) of $D_{3h}$. Since the VB is symmetric with respect to 
$\sigma_{h}$, one can show using group-theoretical arguments that the optical matrix element 
between the VB, which has $A_{1}'$ symmetry,  and the CB, which has $E''$ symmetry,  is zero at the $\Gamma$ point. 

However, as shown in Figure \ref{fig:bands-at-Gamma},   
due to band crossings one of the degenerate CB+1 bands  
becomes the CB at some distance from $\Gamma$. 
The doubly degenerate CB+1 band belongs to the 2D  
$E'$ irreducible representation of $D_{3h}$. This irrep  is symmetric with respect to $\sigma_h$,   
and optical transitions between bands of $A_{1}'$ and $E'$ symmetries are allowed. 
Therefore  as a  starting  point for  studying  the  optical transitions in the vicinity of the $\Gamma$ point 
one has to describe the CB+1 bands. 

Up to second-order terms in the wavevector $\mathbf{k}$, the effective Hamiltonian describing the $E'$ bands in the 
vicinity of $\Gamma$ reads: 
\numparts
\begin{eqnarray}
 H_{\Gamma, \rm eff}^{\rm cb+1} = H_0+H^{(2)}_{d}+ H_{wr}^{(2)}, 
 \label{eq:H_Gamma_Ep}\\
 H_{d}^{(2)}=(\vareps_{cb+1}+\alpha \mathbf{k}^2) \cdot \mathbb{1}_2,  \\
 H_{wr}^{(2)}= -\beta \mathbf{k}^2
 \left(
 \begin{array}{cc}
  (1-\cos(3 \phi_{\mathbf{k}})) &  \frac{1}{2} (e^{i 6 \phi_{\mathbf{k}}}-1) e^{-2 i\phi_{\mathbf{k}} }\\
  \frac{1}{2} (e^{-i 6 \phi_{\mathbf{k}}}-1) e^{2 i\phi_{\mathbf{k}} } & (1+\cos(3 \phi_{\mathbf{k}}))
 \end{array}
 \right), 
 \label{eq:H_Gamma_Ep_warp}
\end{eqnarray}
\endnumparts 
where $\mathbb{1}_2$ is a $2\times 2$ unit matrix, and $\phi_{\mathbf{k}}$ is the argument of $k_x+i k_y$ 
(here the wavevector components $k_x$ and $k_y$ are measured from $\Gamma$). 
We also  keep explicit the free-electron term $H_0=\frac{\hbar^2\mathbf{k}^2}{2 m_e} \cdot \mathbb{1}_2$. 
The term  $ \alpha \mathbf{k}^2$  in $H_{d}^{(2)}$ describes the coupling of the CB+1 bands to other 
remote bands with the same  $E'$ symmetry, while  
 $H_{wr}^{(2)}$ captures the coupling of the CB+1 bands to 
other remote bands with $A_{1}'$ symmetry. In contrast to the VB, one can see that 
 $H_{wr}^{(2)}$ leads to a hexagonal distortion of the 
energy contours of the CB+1 bands already in second order of $\mathbf{k}$.  
Looking at Equation \eref{eq:H_Gamma_Ep_warp} one can also note that, e.g.,  along the $\Gamma$--$K$ line  the 
off-diagonal and one of the diagonal terms become zero. Therefore Equation \eref{eq:H_Gamma_Ep_warp} alone
would suggest that one of the $E'$ bands is dispersionless. 
Since the dispersion of the higher-in-energy $E'$ band is indeed very flat along  $\Gamma$--$K$
[see Figure \ref{fig:bands-at-Gamma}(a)],  we expect that $H_{d}^{(2)}$ 
largely cancels $H_0$. 

SOC, as illustrated in  Figure \ref{fig:bands-at-Gamma}(b), has two main effects:
\begin{itemize}
\item[i)] At the $\Gamma$ point it leads to a splitting of the otherwise degenerate states. 
          Therefore, instead  of four-fold degeneracies, which would follow from  taking into 
          account the spin but not the SOC  there are only two-fold degeneracies. 
          [For the $E'$ bands in MoS$_2$ the splitting is too small to be seen on  
           the scale of Figure \ref{fig:bands-at-Gamma}(b)]. 
\item[ii)] 
           Close to the $\Gamma$ point the band crossings between the $E'$ and $E''$ bands are turned 
           into avoided crossings. 
 \end{itemize}  
One can observe, however, that beyond these avoided crossings the dispersion of the 
spin-split CB follows that of the spinless CB quite closely. 
This is remarkable for the following reason: it has been argued \cite{bolotin}
that the existence of the C-exciton is related to a minimum in the optical band structure.  
Looking along the $\Gamma$--$K$ or $\Gamma$--$M$ lines, the minimum in the optical band structure 
can be found for $\mathbf{k}$ values where the spinful CB closely follows the lower-in-energy  $E'$  band.
We expect therefore that, theoretically, the starting point for describing the C-exciton physics  
in an effective-mass approximation would be to extend the model shown in Equations \eref{eq:H-Gamma-VB} and 
\eref{eq:H_Gamma_Ep} by terms that contain higher powers of  $\mathbf{k}$, especially for the $E'$ bands, 
where these corrections become important closer to the $\Gamma$ point than is the case in the VB\@. 
Neglecting the coupling between the two $E'$ bands and considering only the one lower-in-energy band
that becomes the CB,  the terms  up to fourth order in $\mathbf{k}$ that need  to be added to the dispersion are  
\begin{equation}
\fl
 H^{(3)}+H^{(4)}
 =C^{(3)} |\mathbf{k}|^{3}  (1+\cos\phi_{\mathbf{k}})
 + |\mathbf{k}|^{4} [ C^{(4)}_{1} +  C^{(4)}_{2} (1+\cos\phi_{\mathbf{k}}) + C^{(4)}_{3} (1+\cos\phi_{\mathbf{k}})^{2}],
 \label{eq:gamma-fourth-order}
\end{equation}
where the constants $C^{(3)}$ and $C^{(4)}_{1,2,3}$ can be obtained from fitting the band structure. 

Looking at Figures \ref{fig:optDOS-1} and \ref{fig:optDOS-2}, this approach appears to be most useful for 
MoS$_2$ and WS$_2$, where a  clear minimum in the optical band structure in the 
vicinity of $\Gamma$ can  be seen. 
However, the exact location of the minimum in the optical band structure would also depend on 
the SOC, which was not taken into account in  Figures \ref{fig:optDOS-1} and \ref{fig:optDOS-2}
and this introduces  additional  complexity into the problem. 
A  detailed discussion  of the optical band structure {based on $\mathbf{k}\cdot\mathbf{p}$ theory}
is therefore left for a future work.  {Numerically,  using DFT calculations combined with maximally 
localised Wannier functions, the effects of SOC on the optical transitions have
very recently been studied in Reference \cite{polini}.}


\subsection{$\Gamma$ point wave functions and STM measurements}

The  shape and extent of the
VB and CB wave functions at the $\Gamma$ point can also play an important role in 
the interpretation of STM measurements. Since there is a growing 
experimental interest \cite{crommie,shih2014a, shih2014b,lainjong,andrei,maohaixie} in STM 
studies of monolayer TMDCs, we give a  brief account of calculations 
that can be used to interpret STM measurements.

We first focus  on the STM maps that one can obtain using a tip with a curvature 
radius larger than atomic distances at scanning  distances comparable to or larger 
than the lattice constant. In this case the current is dominated by electrons 
tunnelling from the metal with the largest $k_z$ momentum component at the energy 
given by the scanning voltage. 
Therefore the in-plane momentum components of the tunnelling electron 
can be neglected: $k_{x,y} \rightarrow 0$. 
On one hand, this implies that the real space 2D maps of the tunnelling 
current should  reflect the vertical extent of the $\Gamma$-point wave functions in 
the CB and VB\@.
On the other hand, electron tunnelling into the band edges,  
which are at  the $K$ and $-K$ points,  can take 
place as a two-step process: first, the electron tunnels into a virtual state 
close to the $\Gamma$ point in the corresponding band, then it emits a 
BZ-corner phonon to scatter into the final state near the band edge.
The expected $I$--$V$ characteristic therefore should have a tunnelling gap in the current 
of  magnitude of  the phonon energy, counted from the Fermi level in a doped 2D semiconductor 
or the band edge in an undoped one.

STM images of bulk MX$_{2}$ can be simulated from first-principles\@. 
Since in  STM measurements one detects  the tail of either 
the VB or CB wave function, depending on whether electrons or holes 
are injected into the material, one  has to  
determine the decay rate of the wave function of the 
relevant states in different parts of the 
unit cell. This can be 
achieved by numerical differentiation of the logarithm of the square modulus of 
the band-decomposed wave function along the $z$ direction (i.e., 
perpendicular to the sheet). For this purpose we use a trilayer geometry to 
model the surface of the bulk material, since we do not expect the inter-layer 
interaction to affect  the tail of the wave functions severely. 
In these calculations we used the optB88 van der Waals density functional \cite{Klimes2010}.  
This functional should provide a significantly better description 
of the interlayer interaction than the LDA or PBE functionals.
\begin{figure}[ht]
\begin{center}
\includegraphics[height=0.27\paperheight]{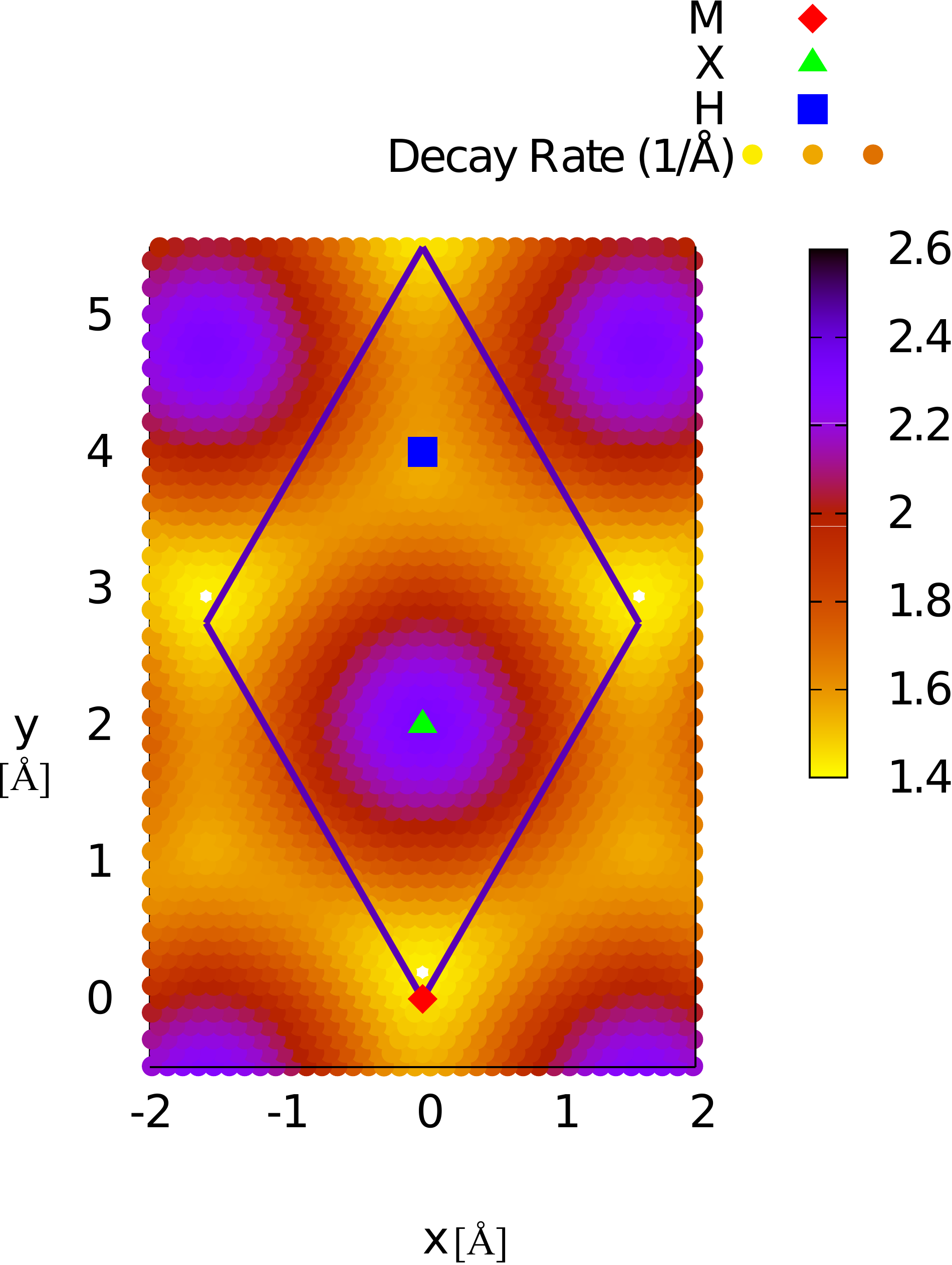}
\end{center}
\caption{\label{fig:DecayRate}Decay rate in the unit cell of MoS$_{2}$.}
\end{figure}
Figure \ref{fig:DecayRate} illustrates the decay rate of the VB of MoS$_2$. 
Three points are  highlighted: the position of the metal atom (M), 
the chalcogen atom (X), and  the centre of the hexagon formed by three 
M and three X atoms on the surface (H)\@.
Large tunnelling currents occur when the decay rate is low. For example, in the 
case of the VB, the tunnelling current is dominated by the 
contribution of the sublattice where the metal atom is located in the VB\@. Note 
that the centre of the hexagon is also quite bright; this is due to 
constructive interference between the $p_x$ and $p_y$ orbitals of the chalcogen 
atoms. Table \ref{tab:DecayRate} summarises the decay rates at the three 
notable positions in the unit cell for the four MX{$_2$} materials studied 
in this work.  It can be used to  explain which sublattice is expected to dominate the 
tunnelling current in a particular  MX$_2$ material.
{A recent experiment \cite{shih2014b} has shown a $\Gamma$-point decay
rate of $\approx 0.9$/\AA~ in monolayer MoSe$_2$ and 
$\approx 0.7$/\AA~ in monolayer WSe$_2$ 
on graphitic substrate. These values are  in good agreement with
our calculated result of $0.91$ and $1.02$ 1/\AA, respectively.}

One can also envisage an alternative STM arrangement where the tunnelling 
current is determined by coupling with a single atomic site at the end of the 
tip brought to atomic/subatomic distances from the 2D material. In this case, 
momentum transfer and momentum conservation are not problems for the tunnelling 
electron; hence, the tunnelling spectrum may reflect the structure of the 
electronic wave function at the band edges in the BZ corners. 
However, in this case, the actual current maps would be affected by the form 
of the atomic orbital of the last atom in the tip and analysis of such 
details lies outside the scope of this review. 

\begin{table}
\caption{\label{tab:DecayRate} Decay rate of monolayer MX$_{2}$ at the
$\Gamma$ point in units of 1/\AA\@. }
\begin{indented}
\lineup
\item[]
\begin{tabular}{ccccccc}
\br
~&MoS$_2$&MoSe$_2$&MoTe$_2$&WS$_2$&WSe$_2$&WTe$_2$\\
\hline
M$_{VB}$&1.44&0.91& 1.20  &1.54&1.02& 1.35  \\
X$_{VB}~$&2.31&2.19& 3.18  &2.33&2.10& 3.07  \\
H$_{VB}~$&1.53&1.20& 2.03  &1.60&1.23& 1.98  \\
M$_{CB}$&5.10&3.50& 5.44  &3.50&3.28& 5.37  \\
X$_{CB}~$&3.84&4.28& 6.01  &4.50&4.48& 5.99  \\
H$_{CB}~$&5.01&3.89& 5.50  &6.40&3.96& 5.49  \\
\br
\end{tabular}
\end{indented}
\end{table}


\section{The $M$ point: spin--orbit splitting of the Van Hove singularity}
\label{sec:M-point-main}

In Section \ref{sec:G-point-main} we have already shown  that optical transitions 
in monolayer TMDCs are expected  to occur not only at the $K$ and $-K$ points, 
but at other points in the BZ as well. 
Indeed, a strong light--matter interaction was observed in 
Reference \cite{britnell} and attributed 
to Van Hove singularities in the electronic density of states. 
Moreover, strong absorption beyond  the energy range of visible light has been 
found in  MoS$_2$ \cite{kis-peaks} and in WSe$_2$ \cite{potemski}.  
High-energy optical transitions 
(in the range of $1.5$--$9$ eV) have also been studied using ellipsometry in
Reference \cite{nguyen}. Very recently, it was argued that electron-phonon 
scattering processes involving the $M$ point may help to explain the origin of certain 
peaks in the Raman spectrum of monolayer MoTe$_2$ \cite{saito}. 
Motivated by these observations and by the fact that, according to 
Figures \ref{fig:optDOS-1} and \ref{fig:optDOS-2}, 
the optical DOS is finite at energies that correspond to transitions at the 
$M$  point of the BZ, we briefly discuss the dispersion of the VB and CB  at the $M$ point. 
{Higher energy optical transitions in TMDCs were 
studied theoretically in Refs.~\cite{carvalho,polini}, and the 
effects of a saddle point in the dispersion have also been investigated recently 
in monolayer graphene  \cite{heinz2013}.}

\subsection{Basic characterisation and $\mathbf{k}\cdot\mathbf{p}$ Hamiltonian }
\label{subsec:M-point-all}

\begin{figure}[ht]
\begin{center}
\includegraphics[scale=0.5]{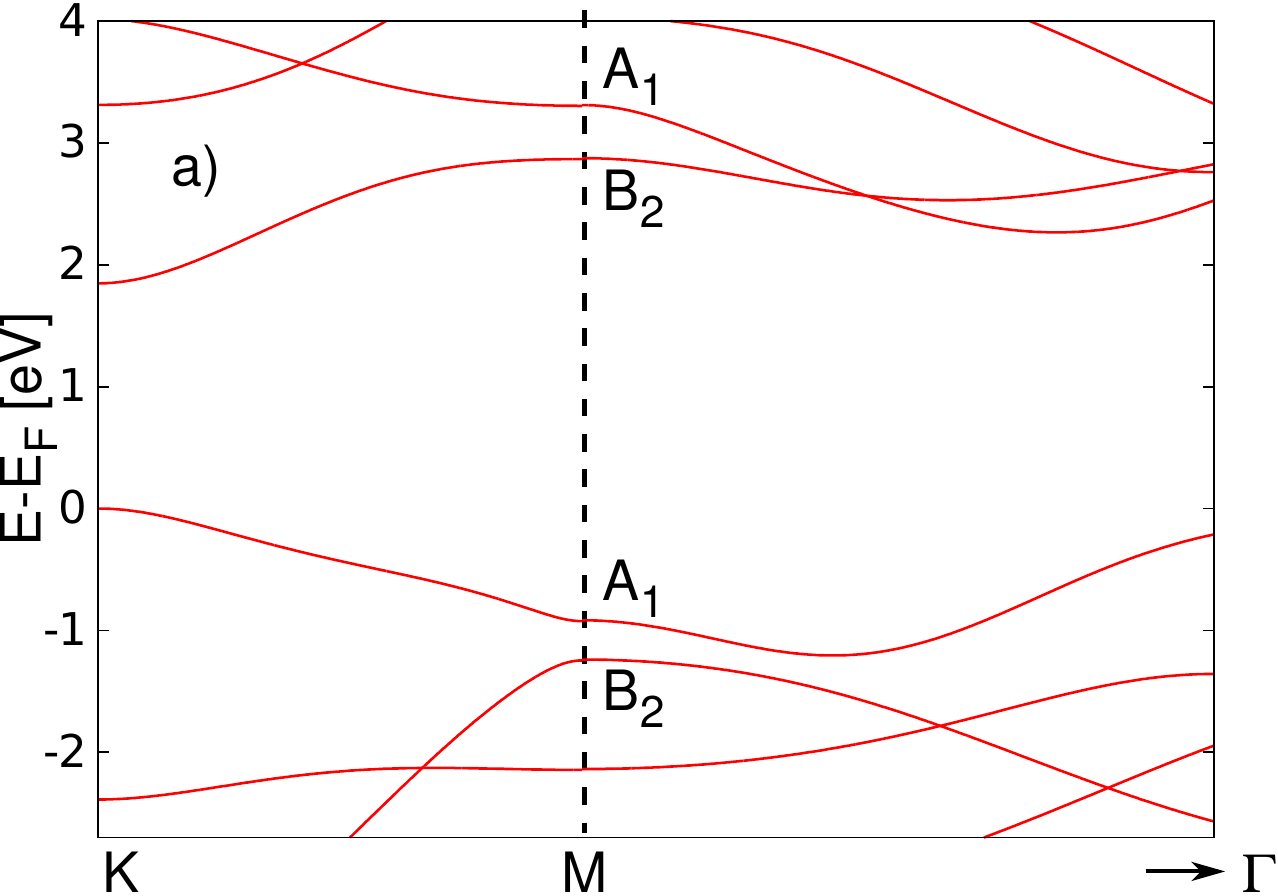}
\includegraphics[scale=0.5]{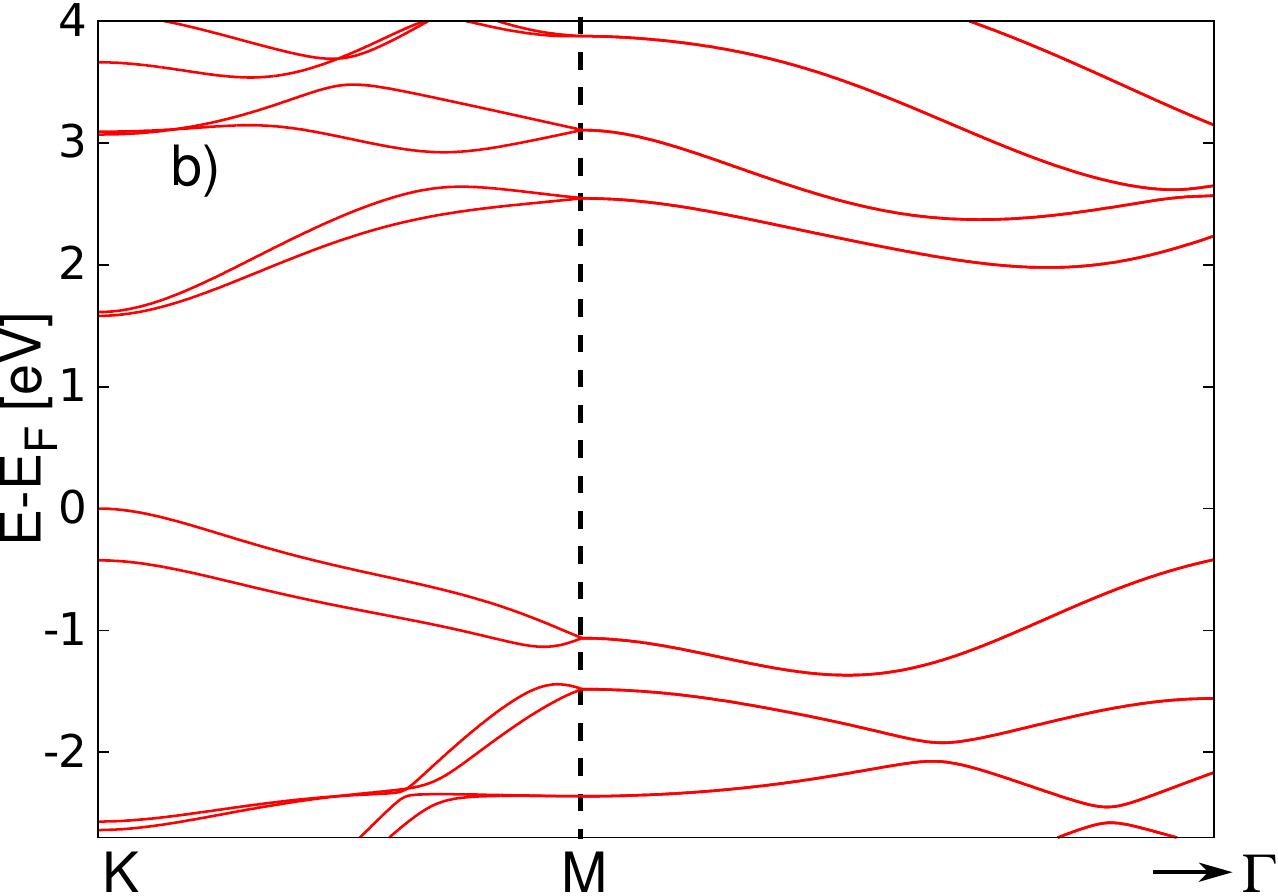}
\caption{Band structure of a monolayer TMDC at the $M$ point obtained from DFT calculations
a) without taking SOC into account; and b) with SOC\@. 
In a) the labels above the bands denote the pertinent irreps of the group $C_{2v}$. 
The actual calculations were performed for WS$_2$ using the (PBE,PBE) approach. 
\label{fig:bands-at-M}
}
\end{center}          
\end{figure}

Figure \ref{fig:bands-at-M} shows  the  band structure of a monolayer TMDC
near the $M$ point of the BZ\@. 
Looking at Figure \ref{fig:bands-at-M}(a) first, where the SOC is neglected, 
one can see that upon going from $M$ towards $K$ the energy difference
between the VB and the CB decreases, whereas along the $M$--$\Gamma$ direction it
slightly increases.  This leads to a saddle point in the optical band structure, 
as shown in Figures \ref{fig:optDOS-1} and \ref{fig:optDOS-2}. 
It is important to point out that  for all the monolayer TMDCs considered here 
the energy difference between the CB and the CB+1 
(VB and VB$-$1) bands is rather small compared to the band gap; 
the difference between the CB and CB+1 is around $0.5$ eV and the difference
between the VB and VB$-$1 is $0.15$--$0.3$ eV\@.  
Therefore, regarding optical transitions, the situation at the $M$ point is different
from  the $K$ point, where the CB and the VB are well separated in energy from
all other bands. 
It is also different from the situation encountered at the $\Gamma$ point, where 
the CB was antisymmetric, while the VB and the CB+1 bands were symmetric with respect 
to the  horizontal mirror plane.  Here all four bands are symmetric and in-plane 
polarised electromagnetic radiation can, in principle, induce transitions between them.

\begin{figure}[ht]
\begin{center}
\includegraphics[scale=0.4]{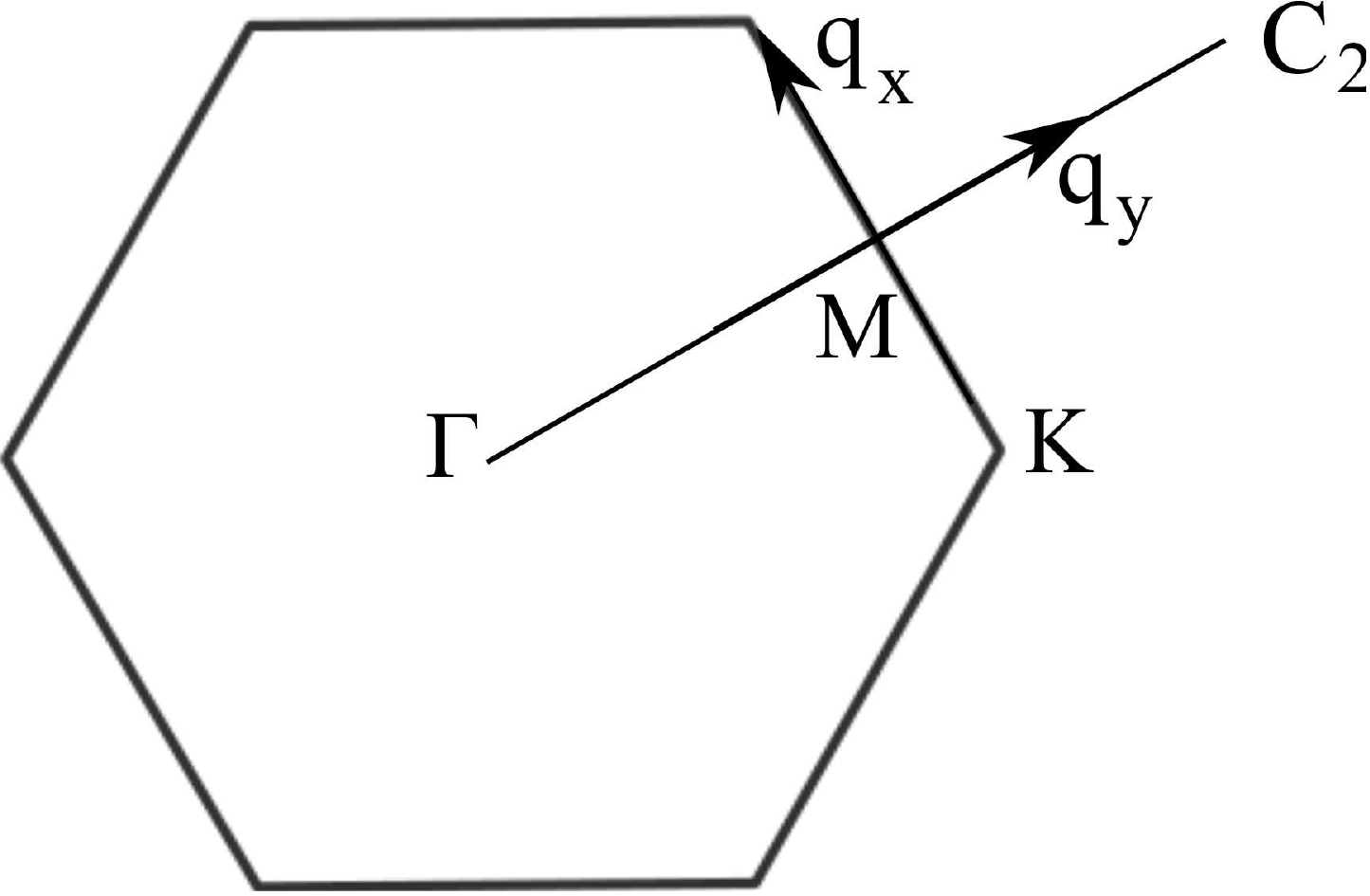}
\caption{Local coordinate system at the $M$ point of the BZ\@.  The twofold rotation 
axis $C_2$ is also shown. 
\label{fig:M-point-coord}
}
\end{center}          
\end{figure}

The discussion of the band dispersion at the $M$ point is  simplified if one 
introduces the local coordinate system  shown in Figure \ref{fig:M-point-coord}. 
Here both $q_x$ and $q_y$ are measured from the $M$ point, the former being parallel
to the $K$--$M$ direction, the latter to the $\Gamma$--$M$ direction. 
Similarly to  the $Q$ point, we content ourselves with the construction of a $\mathbf{k}\cdot\mathbf{p}$ 
Hamiltonian based on the theory of invariants.
The six $M_{i}$ points in the BZ  are pairwise connected by time-reversal 
symmetry.  To describe this one can introduce the matrix $\tau_z$, 
whose eigenvalues, $\tau=\pm 1$ label individual members of the pairs of $M$ points. 
In the simplest approximation,  the Hamiltonian of all four bands of interest is 
\begin{eqnarray}
 H_{\rm M}^{\tau,s}=\frac{\hbar^2 q_x^2}{2 m_{M,x}^{\tau,s}}+ 
 \frac{\hbar^2 q_y^2}{2 m_{M,y}^{}}+ \tau_z \Delta_{M} q_x s_z. 
 \label{eq:H_M_SOC}
\end{eqnarray}
One can see that the dispersion is  parabolic and characterised 
by different effective masses $m_{M,x}^{\tau,s}$ and $m_{M,y}$ along 
the $M$--$K$ and $M$--$\Gamma$ directions, respectively. 
To understand the implications of Equation \eref{eq:H_M_SOC}, let us first 
neglect the  SOC, i.e., we  set $\Delta_{M}=0$ and $m_{M,x}^{\tau,s}=m_{M,x}^{}$.
Looking at Figure \ref{fig:bands-at-M}(a) one can notice that for the CB  (denoted by $B_2$) the 
effective masses $m_{M,x}$  and  $m_{M,y}$ have the same sign. For the VB (denoted by $A_1$), however, 
their  sign is different. Similar conclusions hold for the CB+1 and the VB$-$1 bands.  
Therefore in the optical band structure one has a saddle point in the dispersion and consequently a
Van Hove singularity. 

SOC, as shown in Figure \ref{fig:bands-at-M}(b), has two main effects:
\begin{itemize}
 \item [i)] It leads to a linear-in-$q_x$ splitting of the bands along
 $M$--$K$, while the bands remain spin-degenerate along $M$--$\Gamma$. This means that the 
 saddle point in the optical DOS of any two bands will  also be split along $M$--$K$. 
 In addition, the effective mass $m_{M,x}^{\tau,s}$ becomes spin dependent. 
 \item[ii)] It turns band crossings into avoided crossings. 
\end{itemize}
As shown in Figure \ref{fig:bands-at-M}(b), the linear-in-$q_x$ splitting of the bands 
is a rather good approximation close to $M$. 
However, the situation is  complicated by the fact that the CB+1 (VB$-$1) 
band  is quite close in energy to the CB (VB)\@. SOC couples the  
CB and CB+1 (VB and VB$-$1) bands and leads to  avoided crossings between them. 
Therefore  a more complete description would require a model similar to the $K$ point, where 
the coupling of nearby bands is explicitly taken into account. We leave the construction 
of such a model to a future work.

The Hamiltonian of Equation \eref{eq:H_M_SOC} can be constructed using the theory of invariants. 
The symmetry group at the $M$ point (and along $\Gamma$--$M$)  is $C_{2v}$, which includes the 
following symmetry operations: a twofold rotation $C_2$ around the 
$\Gamma$--$M$ direction, a reflection
$\sigma_v$ with respect to the $q_y$--$q_z$ plane, and the reflection $\sigma_h$ with respect to 
the $q_x$--$q_y$ plane. The character table of $C_{2v}$ is shown in Table \ref{tbl:charactc2v}, where 
the relevant basis functions, in the chosen coordinate system, are also given. 
\begin{table}[htb]
\caption{
\label{tbl:charactc2v} Character table for the group $C_{2v}$. Basis functions 
for a given irrep are also shown. 
 $R_{x,y,z}$ denotes the angular momentum components. }
\begin{indented}
\lineup
\item[]
\begin{tabular}{@{}lllrrrr}\br
 D$_{2v}$ &    &  & $E$  &  $C_2$  &  $\sigma_v$ & $\sigma_h$  \\
 \hline
 $A_1$  & $q_{x}^2$, $q_{y}^2$  & $q_y$           & $1$  & $1$  & $1$  & $1$\\ 
 $A_2$  &              & $R_y$         & $1$  & $1$  & $-1$ & $-1$\\
 $B_1$  &              & $R_x$, $q_z$    & $1$  & $-1$ & $1$  & $-1$\\
 $B_2$  &  $q_x q_y$   & $R_z$, $q_x$    & $1$  & $-1$ & $-1$ & $1$\\
 
 \br
\end{tabular}
\end{indented}
\end{table}
The Hamiltonian, which can be constructed with the help of Table \ref{tbl:charactc2v} and
which is at most second order  in the wavenumbers $q_x$ and $q_y$, is given in Equation \eref{eq:H_M_SOC}. 
The symmetries  of the individual bands at the $M$ point,  which can be deduced by, e.g., 
considering which atomic orbitals contribute to a certain band, are indicated in 
Figure \ref{fig:bands-at-M}(a).

\begin{figure}[ht]
\begin{center}
\includegraphics[scale=0.7]{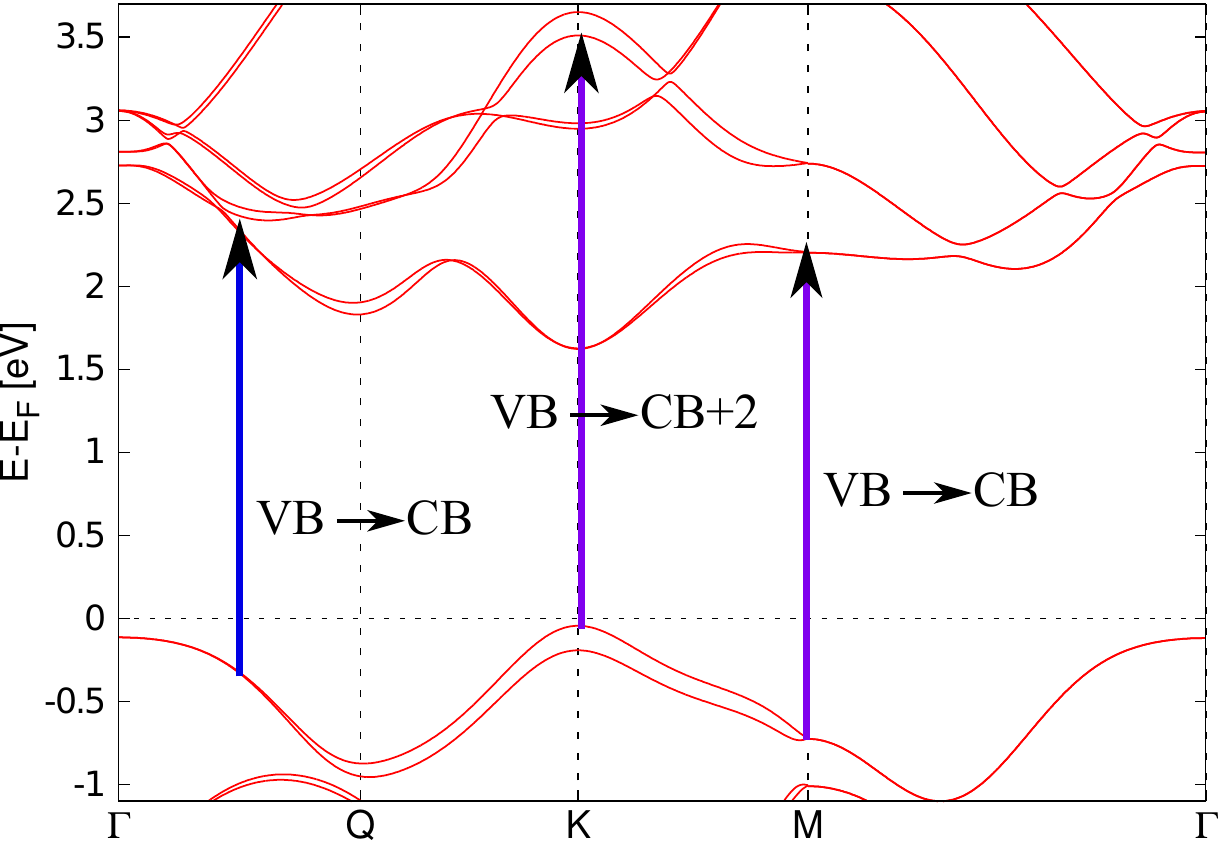}
\caption{Optical transitions discussed in Sections \ref{subsec:mat-param-at-G} and \ref{subsec:M-point-all}.
\label{fig:opt-trans}
}
\end{center}          
\end{figure}

It is important to note that there is another possible optical transition, which may 
have a similar energy to the one between the VB and the CB  at the $M$ point. 
\begin{table}[htb]
\caption{Higher-energy optical transitions in monolayer TMDCs based on DFT calculations.
SOC  is  taken into account. }
\label{tbl:high-en-transition}
\begin{indented}
\lineup
\item[]
\begin{tabular}{@{}lllllll}
\br
       & MoS$_2$  & MoSe$_2$ & WS$_2$  & WSe$_2$ & MoTe$_2$& WTe$_2$\\
\mr
$E_{\rm M, VB\rightarrow CB}$ [eV] (HSE,LDA)     & $2.93$ &   $2.52$  &  $3.60$ &  $3.03$ & $1.67$ & $2.04$\\  
$E_{\rm M,  VB \rightarrow CB}$ [eV] (PBE,PBE)  &  $2.83$ &  $2.48$ &  $3.61$ &  $3.04$ & $1.67$ & $2.07$\\
$E_{\rm M,  VB \rightarrow CB}$ [eV] ($GW$)  &  $3.87^{a}$ &   &   &   & & \\
\mr
$E_{\rm K, VB \rightarrow CB+2}$ [eV] (HSE,LDA)  & $3.56$ &     $3.02$     & $3.77$ &  $3.16$ & $2.46$ & $2.48$\\  
$E_{\rm K, VB \rightarrow CB+2 }$ [eV] (PBE,PBE) & $ 3.40$ &  $2.88$     &  $3.66$ &  $3.05$  & $2.31$ & $2.37$ \\
\br
\end{tabular}
\item[] $^{a}$ {\cite{qiu}. }
\end{indented}
\end{table}
This transition can occur between the upper spin-split VB and the lower spin-split
CB+2 (see Figure \ref{fig:opt-trans}). Our DFT results shown in Table \ref{tbl:high-en-transition} 
suggest that for  MoX$_2$ the transition at the $M$ point has lower energy, while for WX$_2$ they are nearly
degenerate. This prediction does not take into account excitonic effects, which are also expected 
to be important and may determine which transition actually has the lower energy, 
because  the exciton binding energies at $K$ and $M$ may be different. 
We expect that, e.g., 
the polarisation  of the photoluminescence  can give important information 
about these transitions. If the incident light is circularly polarised, then  
the photoluminescence related to the   transition VB$\rightarrow$CB+2, which
takes place at the $K$ and $-K$ points, should also be circularly polarised, 
as is the case for  the well known fundamental  
VB$\rightarrow$CB transition. Since the local symmetry at the $M$ point is different, 
we do not expect that the photoluminescence  due to the VB$\rightarrow$CB transition at the 
$M$ points is  circularly polarised.


\section{Conclusions}
\label{sec:conclusions}

In this  short review we have focused on the band structure of monolayer TMDCs. 
Our aim has been  to  discuss  all the details of the  band structure that we believe 
are relevant for transport and relaxation processes and optical transitions. 
The two main tools that we used were the (local) symmetries of the BZ  
(an essential ingredient of the $\mathbf{k}\cdot\mathbf{p}$ expansion)
and DFT calculations. The first of these tools  allowed us to capture  general features 
of  the band structure.  Material parameters, such 
as effective masses, spin-splittings,  and band edge 
energy differences  depend on the chemical composition of particular TMDCs and 
are important for quantitative predictions. For this reason we also performed 
extensive DFT calculations which can, in many cases,  predict  material parameters 
accurately. 
From a theoretical point of view, an important aspect of the approach used in this work 
is that it leads to explicit $\mathbf{k}\cdot\mathbf{p}$ Hamiltonians that  can 
be used to address a variety of  problems. In particular, they are expected to be accurate when
external perturbations vary on  length scales much larger than the interatomic distances. 
Therefore we believe our results will help to develop further (semi)analytical approaches to 
study, e.g.,  exciton physics \cite{rana,berkelbach,malic,glazov,mwwu-exciton,fogler,thilagam}, 
plasmons \cite{scholz,abergel}, diffusive transport \cite{ochoa-diffusive}, 
spin \cite{mwwu-spin,hatami}, noise \cite{sinitsyn}, topological properties \cite{cazalilla,law}, 
valley-currents \cite{hongyiyu,balatsky,mceuen}, proximity effect \cite{majidi,rameshti}, 
electron--electron interaction \cite{cortijo}, and quantum dots \cite{sajat2,klinovaja}. 
On the other hand, TB-based methods are  probably more appropriate for studying  the 
effects of, e.g., point or line defects \cite{ghorbani-asl,yuan} on transport. 

The picture that emerges from this study is that monolayer TMDCs in the ballistic limit 
should display a remarkable variety of optical and  electronic properties,  
many  of which are yet to be verified experimentally. Nevertheless, the recent 
reports of achieving high-transparency contacts to these materials \cite{hone,duan} are promising 
a further rapid development of this field.


\section{Acknowledgement}

Andor Korm\'anyos  acknowledges discussion with Yi Zhang, Philip King, Diana Qiu, and  Timothy Berkelbach.
We also acknowledge  support  by  
the DFG through the SFB 767 and SFB 689,  the EU through the Marie Curie ITN S3NANO, 
EC Graphene Flagship Project CNECT-ICT-604391, ERC Synergy Grant Hetero2D, 
Royal Society Wolfson Research Merit Award, and EPSRC Science and Innovation Award EP/G035954. 


\appendix

\addcontentsline{toc}{section}{Appendices}
\addtocontents{toc}{\protect\setcounter{tocdepth}{-1}}

\section{Seven-band $\mathbf{k}\cdotp\mathbf{p}$ model at the $K$ ($-K$) point}
\label{sec:seven-band-at-K}

In this section we give a brief account of the seven-band $\mathbf{k}\cdotp\mathbf{p}$ model, 
which lies behind the effective Hamiltonian of Equation \eref{eq:full-eff-Ham-at-K}.  
{The following discussion is based on our previous work published in 
References \cite{sajat2} and \cite{sajat1}; we believe it is 
helpful to present the main steps here again to make this work self-contained.}

Our seven-band model (without spin)  contains every band between the third band below 
the VB (which we denote by VB$-$3) and the second band above 
the CB (denoted by CB+2), i.e., we take the basis
 $\{ |\Psi_{\rm E_2^{'}}^{\rm vb-3},s\rangle, |\Psi_{\rm E_1^{''}}^{\rm vb-2},s\rangle, |\Psi_{\rm E_2^{''}}^{\rm vb-1},s\rangle, 
|\Psi_{\rm A'}^{\rm vb},s\rangle, |\Psi_{\rm E_1^{'}}^{\rm cb},s\rangle, |\Psi_{\rm A^{''}}^{\rm cb+1},s\rangle,|\Psi_{\rm E_1^{'}}^{\rm cb+2},s\rangle
\}$. 
The upper index $b=\{ {\rm vb-3,vb-2,vb-1,vb,cb,cb+1,cb+2} \}$ denotes the band, the lower 
index $\mu$ indicates the pertinent irreducible representation of the point group 
$C_{3h}$, which gives the symmetry of the bands at the $K$ point of the BZ 
(see Table \ref{tbl:charactc3h} for the character table of $C_{3h}$)\@.
The spinful symmetry basis functions are introduced by 
$| \Psi_{\mu}^{b}, s\rangle =  | \Psi_{\mu}^{b}\rangle \otimes | s \rangle$, 
where $s=\{ \uparrow,\downarrow \}$ denotes the spin degree of freedom. 
An important symmetry of the system is that is has a horizontal mirror plane. 
As a consequence, the basis states can be grouped into two sets: the first one 
contains states whose orbital part  {is symmetric (even) with respect to 
the mirror operation $\sigma_h$:  $\sigma_h | \Psi_{\mu}^{b}\rangle = |\Psi_{\mu}^{b}\rangle$. 
This first group contains the following states:}
$
\{ |\Psi_{\rm A'}^{\rm vb},s\rangle, |\Psi_{\rm E_1^{'}}^{\rm cb},s\rangle,
|\Psi_{\rm E_2^{'}}^{\rm vb-3},s\rangle,|\Psi_{\rm E_1^{'}}^{\rm cb+2},s\rangle
\}
$. 
The second set  contains  antisymmetric (odd) states: 
$\sigma_h | \Psi_{\mu}^{b}\rangle = -|\Psi_{\mu}^{b}\rangle$.
The corresponding states are 
$
\{|\Psi_{\rm E_1^{''}}^{\rm vb-2},s\rangle, |\Psi_{\rm E_2^{''}}^{\rm vb-1},s\rangle,
|\Psi_{\rm A^{''}}^{\rm cb+1},s\rangle
\}
$.
\begin{table}[htb]
\caption{
\label{tbl:charactc3h} Character table for the group $C_{3h}$ ($\overline{6}$).
Here $\omega=e^{\frac{2 i \pi}{3}}$.}
\begin{indented}
\lineup
\item[]
\begin{tabular}{@{}crrrrrr}\br
 $ C_{3h}$ &  $E$  & $C_3$ & $ C_3^{2}$  &  $\sigma_h$  &  $S_3$ & $\sigma_h C_3^2$  \\
 \hline
 $ A'$   & $1$  & $1$        & $1$         & $1$   & $1$         & $1$\\ 
 $A''$   & $1$  & $1$        & $1$         & $-1$  & $-1$        & $-1$\\
 $E_1'$  & $1$  & $\omega$   & $\omega^2$  & $1$   & $\omega$    & $\omega^2$\\
 $E_2'$  & $1$  & $\omega^2$ & $\omega$    & $1$   & $\omega^2$  & $\omega$\\
 $E_1''$ & $1$  & $\omega$   & $\omega^2$  & $-1$  & $-\omega$   & $-\omega^2$\\
 $E_2''$ & $1$  & $\omega^2$ & $\omega$    & $-1$  & $-\omega^2$ & $-\omega$\\
 \br
\end{tabular}
\end{indented}
\end{table}

{Two important questions can be raised at this point: 
\begin{itemize}
 \item[i)] What is the motivation  to include  seven bands in the model and not more (or less)?
  \item[ii)] How can we identify the irreducible
   representation of $C_{3h}$ according to which a given band transforms?
\end{itemize}
To answer i), we remind the reader that, as mentioned in Section \ref{subsec:kp-at-K},
a strictly two-band model, such as 
the one introduced in Reference \cite{dxiao} (see Equation~\eref{eq:gapped-Dirac})
cannot describe, e.g., the TW of the bands or the details of the spin-splitting
in the CB\@. In  $\mathbf{k}\cdot\mathbf{p}$ theory these effects can be understood 
as arising from the coupling of the VB and CB to other energy bands. As a first step,
let us neglect the SOC\@. The operator $\mathcal{H}_{\mathbf{k}\cdot\mathbf{p}}$ 
which describes the interaction of various bands in $\mathbf{k}\cdot\mathbf{p}$ theory 
(see Equation \eref{eq:kpHam} below) is symmetric with respect to $\sigma_h$: 
$
\sigma_h^{-1}\mathcal{H}_{\mathbf{k}\cdot\mathbf{p}}\sigma_h = 
\mathcal{H}_{\mathbf{k}\cdot\mathbf{p}}
$.
Therefore, non-zero matrix elements
$
\langle \Psi_{\mu}^{b}, s | \mathcal{H}_{\mathbf{k}\cdot\mathbf{p}} |\Psi_{\mu'}^{b'}, s \rangle
$
only exist between states $|\Psi_{\mu}^{b}, s \rangle$ and 
$|\Psi_{\mu'}^{b'}, s \rangle$ whose orbital parts are  either both
even   or both odd  with respect to  $\sigma_h$. 
A natural extension of a model containing only the VB and the CB is to include one 
more band, which, regarding its energy, is below the VB and one above the CB\@.   
The symmetry properties of individual bands 
can be extracted from DFT band-structure calculations.  We found that, 
at the $K$ and $-K$ points, the first 
symmetric band below the VB is the VB$-3$ and the first even band above the CB 
is the CB$+2$ band. Thus we arrive at a four-band model containing 
$
\{ |\Psi_{\rm A'}^{\rm vb},s\rangle, |\Psi_{\rm E_1^{'}}^{\rm cb},s\rangle,
|\Psi_{\rm E_2^{'}}^{\rm vb-3},s\rangle,|\Psi_{\rm E_1^{'}}^{\rm cb+2},s\rangle
\}
$. 
This four-band model can already describe the electron--hole asymmetry and the 
TW of the spectrum \cite{sajat1}. 
The next step is to take into account the SOC\@. In the atomic approximation the 
corresponding Hamiltonian is given by Equation \eref{eq:atomic-SOC}. This Hamiltonian 
can have non-zero matrix elements between even and odd states. Since our aim is,
ultimately, to obtain an effective Hamiltonian describing the coupled dynamics of 
the VB and the CB, it is  natural to enlarge  our basis of four even states by those
odd states which, regarding their energy, lie between VB$-3$ and CB$+2$: these are 
$
\{|\Psi_{\rm E_1^{''}}^{\rm vb-2},s\rangle, |\Psi_{\rm E_2^{''}}^{\rm vb-1},s\rangle,
|\Psi_{\rm A^{''}}^{\rm cb+1},s\rangle
\}
$.
In this way we set up a seven-band model, as indicated above. 
} 

{As for question ii), we next discuss how to find out the symmetries of the bands.
Some DFT codes, such 
as \texttt{WIEN2k} \cite{wien2k}, can directly provide this information. If this is 
not available, many DFT codes
can calculate the projection of the Kohn--Sham wave functions of each energy band 
onto atomic orbitals $\Phi^{\eta}_{\nu}$, where 
$\eta=\{\textnormal{M},\textnormal{X1}, \textnormal{X2} \}$
denotes whether the given orbital is centred on the metal (M) or on one of the 
chalcogen (X1, X2) atoms in the unit cell, and the lower index 
$\nu=\{s,p_x,p_y,p_z,d_{z^2},d_{xy},d_{xz},d_{yz},d_{x^2-y^2}\}$ indicates the 
type of orbital. Such a calculation is also shown in Figure \ref{fig:atomic-orbs}.
To take into account the three-fold rotational symmetry of the system, 
we use  linear combinations of these orbitals to form spherical harmonics  $Y_{l}^{m}$. 
We then consider  the transformation properties of the Bloch wave 
functions formed with these spherical harmonics:} 
\begin{eqnarray}
 \Psi_{l,m}^{\eta}(\mathbf{k},\mathbf{r}) =\frac{1}{\sqrt{N}}
 \sum_{\mathbf{R}_n} e^{i \mathbf{k}\cdot(\mathbf{R}_n+\mathbf{t}_{\eta})} \, 
 Y_{l}^{m}(\mathbf{r}-[\mathbf{R}_n+\mathbf{t}_{\eta}]).
 \label{eq:Blochwavef}
\end{eqnarray}
{Here the summation runs over all lattice vectors $\mathbf{R}_n$,  
$\mathbf{t}_{M}$ and $\mathbf{t}_{X1}=\mathbf{t}_{X2}$  give the positions of the 
metal and chalcogen atoms in the  
2D unit cell, and the wavevector $\mathbf{k}$ is measured 
from the $\Gamma$ point of the BZ\@. 
By examination one can then find out how the Bloch wave functions 
$\Psi_{l,m}^{\eta}(\mathbf{k},\mathbf{r})$
transform at, e.g., the $\mathbf{k}=K$ point of the BZ 
when the reflection $\sigma_h$ or the rotation by $2 \pi/3$ around an axis perpendicular
to the plane of the monolayer (denoted by $C_3$) is applied. 
Considering first $\sigma_h$, it is clear 
that the $d$ orbitals of the M atoms are either even ($\{d_{z^2},d_{xy},d_{x^2-y^2}\}$) 
or odd ($\{d_{xz},d_{yz}\}$). Regarding the $p$ orbitals of the $X1$ and $X2$ atoms, which
are above and below the plane of the M atoms, one can also form linear combinations of
$\Psi_{l,m}^{X1}(\mathbf{k},\mathbf{r})$ and 
$\Psi_{l,m}^{X2}(\mathbf{k},\mathbf{r})$ which are either even or odd 
(see Table \ref{tbl:bandsymK}).
}

\begin{table}[thb]
\caption{Basis functions for the irreducible representations of the $C_{3h}$ 
group of the $K$ point.
The rightmost column shows the  band to which a given  basis function contribute.
The basis functions for the $K'$ point can be obtained by complex-conjugation.
\label{tbl:bandsymK}}
\begin{indented}
\lineup
\item[]
\begin{tabular}{ccc}\br\vspace*{-0.8em}
 & \\
\vspace*{-0.8em}
irrep  &  basis functions  & band\\
 & & \\
\hline\vspace*{-0.8em} 
 &  &  \\
\vspace*{-0.8em}
 ${A'}$ &   ${\Psi_{2,-2}^{M}}$, 
 $\frac{1}{\sqrt{2}}\left(\Psi_{1,-1}^{X1} + \Psi_{1,-1}^{X2}\right)$ & VB \\
  &  &  \\  
\hline\vspace*{-0.8em} 
 &  &  \\
\vspace*{-0.8em}
 ${A''}$ &  ${\Psi_{2,1}^{M}}$, $\frac{1}{\sqrt{2}}\left(\Psi_{1,-1}^{X1}-\Psi_{1,-1}^{X2}\right)$ & CB$+1$ \\ 
 &  &\\  
  \hline\vspace*{-0.8em}
 & & \\
 ${E_1'}$ & $|{\Psi_{2,0}^{M}}\rangle$, $\frac{1}{\sqrt{2}}\left(\Psi_{1,1}^{X1}+\Psi_{1,1}^{X2}\right)$ & CB \\
 \vspace*{-1.0em}
 & & \\
\hline\vspace*{-0.8em}
  & & \\
  $E_2'$       &   $ {\Psi_{2,2}^{M}}$, $\frac{1}{\sqrt{2}}\left(\Psi_{1,0}^{X1}-|\Psi_{1,0}^{X2}\right)$ & VB$-3$ \\
  \vspace*{-1.0em}
    & & \\
     &  & CB$+2$ \\
  &  &\\  
  \hline\vspace*{-1.0em}
 & & \\   
 $E_1''$ & $|{\Psi_{1,0}^{M}}\rangle$,  $\frac{1}{\sqrt{2}}\left(\Psi_{1,1}^{X1}-\Psi_{1,1}^{X2}\right)$ &  VB$-2$\\
 & & \\        
 \hline\vspace*{-0.8em}
 & & \\   
 $E_2''$ & ${\Psi_{2,-1}^{M}}$,  $\frac{1}{\sqrt{2}}\left(\Psi_{1,0}^{X1}+\Psi_{1,0}^{X2}\right)$ &  VB$-1$\\
\br
\end{tabular}
\end{indented}
\end{table}

{The Bloch wave functions of Equation \eref{eq:Blochwavef} are also eigenstates 
of the rotation operation $C_3$ with an eigenvalue $\lambda_{l,m}^{\eta}$: 
$C_3 \Psi_{l,m}^{\eta}= \lambda_{l,m}^{\eta} \Psi_{l,m}^{\eta}$. 
At the $\mathbf{k}=K$ or $-K$ points, $\lambda_{l,m}^{\eta}$ can take on one 
of the following three values: $1$, $e^{i \frac{2\pi}{3}}$, $e^{-i \frac{2\pi}{3}}$ 
(see Table \ref{tbl:charactc3h}).
Note that $C_3$ acts on both the spherical harmonics part 
$Y_{l}^{m}(\mathbf{r})$ and on the 
plane-wave component $e^{i \mathbf{k}\cdot(\mathbf{R}_n+\mathbf{t}_{\eta})}$ in 
$\Psi_{l,m}^{\eta}(\mathbf{k},\mathbf{r})$ \cite{dresselhaus-book},
because in a rotated coordinate system the vectors $\mathbf{t}_{M}$ and
$\mathbf{t}_{X}$ are also transformed. 
For this reason the eigenvalues  $\lambda_{l,m}^{\eta}$ corresponding to 
$\Psi_{l,m}^{\eta}(\mathbf{k}=\pm K,\mathbf{r})$ depend on the 
choice of the unit cell, which determines the centre of rotation. 
In our case the unit cell is defined in Figure~\ref{fig:lattice}(b) and the 
centre of rotation is the centre of the hexagon formed by the M and X atoms.
Other possible choices are, e.g., the position of the M or the X atoms 
(see, e.g., Table 2 in Reference \cite{chemsoc-review}). Once the 
eigenvalues of $\Psi_{l,m}^{\eta}(\mathbf{k}=\pm K,\mathbf{r})$ 
under the action of $\sigma_h$ and $C_3$ are known,  
a symmetry label, e.g.,  $A'$ or $E'$,   of an irreducible representation 
can be assigned to each state. These are listed  in Table \ref{tbl:bandsymK}. 
In the single-particle picture hybridisation between different 
Bloch wave functions should preserve symmetry properties; hence, e.g., the 
CB at the $K$ point can be thought of as a linear combination of 
${\Psi_{2,0}^{M}}$ and 
$\frac{1}{\sqrt{2}}\left(\Psi_{1,1}^{X1}+\Psi_{1,1}^{X2}\right)$
(third row in Table \ref{tbl:bandsymK}). 
The corresponding Bloch wave functions at the $-K$ point can be obtained by complex 
conjugation, because the $K$ and $-K$ points are connected by time-reversal. 
}  

{The above discussion illustrates how  symmetries of each 
band at each high-symmetry point  in the BZ can be found and used to construct 
effective Hamiltonians. It is important to point out the following: although, as 
mentioned above, the 
eigenvalues with respect to $C_3$ and hence the assignment of irreducible 
representations may depend on the choice of the rotation centre,  
the  form of the Hamiltonian \eref{eq:full-eff-Ham-at-K}, up to a
unitary transformation, does  not depend on such choices.}

\subsection{$\mathbf{k}\cdot\mathbf{p}$ matrix elements}

The $\mathbf{k}\cdot\mathbf{p}$ matrix elements, {which characterise
the coupling of the bands away from the $K$ and $-K$ points,} 
are calculated using the  Hamiltonian
\begin{eqnarray}
\mathcal{H}_{\mathbf{k}\cdot\mathbf{p}}=\frac{1}{2}\frac{\hbar}{m_e} (q_{+} \hat{p}_{-} + 
                                                                        q_{-} \hat{p}_{+}) =
        \mathcal{H}_{\mathbf{k}\cdot\mathbf{p}}^{-}+ \mathcal{H}_{\mathbf{k}\cdot\mathbf{p}}^{+},                                                            
\label{eq:kpHam}
\end{eqnarray}
where $\hat{p}_{\pm}=\hat{p}_x\pm i \hat{p}_y$ are momentum operators. 
As the operator \eref{eq:kpHam} does not contain spin-operators,  the matrix elements  
are diagonal in the spin-space. 
{Furthermore, the matrix elements of $\mathcal{H}_{\mathbf{k}\cdot\mathbf{p}}$ are constrained 
by the symmetries of the states with respect to  $C_3$.  Namely,  the relation  
 $
 \langle \Psi_{A'}^{\rm vb}| \hat{p}_{+} | \Psi_{E_2^{'}}^{\rm cb+2}\rangle=
 \langle \Psi_{A'}^{\rm vb}| C_3^{\dagger} C_3 \,\hat{p}_{+}\, C_3^{\dagger} C_3 
 |\Psi_{E_2'}^{\rm cb+2}\rangle
 $ 
 should hold. Since 
 $\langle \Psi_{A'}^{\rm vb}| C_3^{\dagger}= \langle \Psi_{A'}^{\rm vb}|$, 
 $C_3 \hat{p}_{\pm} C_3^{\dagger}=e^{\mp i 2 \pi/3}\hat{p}_{\pm}$ and 
 $ C_3 |\Psi_{E_2'}^{\rm cb+2}\rangle = e^{ -i 2 \pi/3}|\Psi_{E_2'}^{\rm cb+2}\rangle $ 
 one obtains that
$
\langle \Psi_{A'}^{\rm vb}| \mathcal{H}_{\mathbf{k}\cdot\mathbf{p}}^{+} | \Psi_{E_2'}^{\rm cb+2}\rangle= 
e^{ -i 4 \pi/3} \langle \Psi_{A'}^{\rm vb}| \mathcal{H}_{\mathbf{k}\cdot\mathbf{p}}^{+} |\Psi_{E_2'}^{\rm cb+2}\rangle,
$ 
which means that this matrix element must vanish. By contrast,  
$\langle \Psi_{A'}^{\rm vb}| \mathcal{H}_{\mathbf{k}\cdot\mathbf{p}}^{-} | \Psi_{E_2'}^{\rm cb+2}\rangle=\gamma_4$
is allowed to be finite. }

The matrix elements $H_{\mathbf{k}\cdot\mathbf{p}}^{K}$ 
calculated at the $K$ point of the BZ are shown in Table \ref{tbl:kp-at-K}, 
where the  diagonal elements are the band-edge energies. 
The matrix elements at the $-K$ point can be obtained with the substitutions $\gamma_i \rightarrow \gamma_i^{*}$ and 
$q_{\pm}\rightarrow -q_{\mp}$. 

\begin{table}[htb]
 \caption{Matrix elements of  $\mathcal{H}_{\mathbf{k}\cdot\mathbf{p}}$ at the $K$ point.}
 \label{tbl:kp-at-K}
\begin{tabular}{l|ccccccc}\hline\hline\vspace*{-0.8em}
 & & & & &  & & \\
$H_{\mathbf{k}\cdot\mathbf{p}}^{K}$  & $|\Psi_{A'}^{\rm vb},s\rangle$ & $|\Psi_{E_1^{'}}^{\rm cb},s\rangle$ & $|\Psi_{E_2^{'}}^{\rm vb-3},s\rangle$ & 
       $|\Psi_{E_2^{'}}^{\rm cb+2},s\rangle$ & $|\Psi_{E_1^{''}}^{\rm vb-2},s\rangle $ & $|\Psi_{E_2^{''}}^{\rm vb-1},s\rangle$ &
       $|\Psi_{A^{''}}^{\rm cb+1},s\rangle$ \\
 \vspace*{-0.8em}
 & & & & & & & \\
 \hline \vspace*{-0.8em}
  & & & & & & &  \\
 \vspace*{-0.8em}
 $|\Psi_{A'}^{\rm vb},s\rangle$ & $\vareps_v$ & $\gamma_3 q_{-}$ & $\gamma_2 q_{+}$ & $\gamma_4 q_{+}$ & $0$ & $0$ & $0$ \\
  & & & & & & &  \\       
\vspace*{-0.8em}
 $|\Psi_{E_1^{'}}^{\rm cb},s\rangle$ & $\gamma_3^* q_{+}$ & $\vareps_c$ & $\gamma_5 q_{-}$ & $\gamma_6 q_{-}$ & $0$ & $0$ & $0$ \\
  & & & & & & &  \\         
\vspace*{-0.8em}     
$|\Psi_{E_2^{'}}^{\rm vb-3},s\rangle$  & $\gamma_2^{*} q_{-}$  & $\gamma_5^* q_{+}$ & $\vareps_{v-3}$ &  $0$ & $0$ & $0$ & $0$ \\
 & & & & & & &  \\         
\vspace*{-0.8em}     
 $|\Psi_{E_2^{'}}^{\rm cb+2},s\rangle$ & $\gamma_4^{*} q_{-}$ & $\gamma_6^{*} q_{+}$ & $0$ & $\vareps_{c+2}$ & $0$ & $0$ & $0$\\
 & & & & & & &  \\         
\vspace*{-0.8em} 
 $|\Psi_{E_1^{''}}^{\rm vb-2},s\rangle $& $0$ & $0$ & $0$ & $0$ &  $\vareps_{v-2}$ & $\gamma_8 q_{-}$ & $\gamma_7 q_{+}$ \\
 & & & & & & &  \\         
\vspace*{-0.8em} 
$|\Psi_{E_2^{''}}^{\rm vb-1},s\rangle$ & $0$ & $0$ & $0$ & $0$ & $\gamma_8^* q_{+}$ & $\vareps_{v-1}$ & $\gamma_9^* q_{-}$\\
& & & & & & &  \\         
\vspace*{-0.8em} 
 $|\Psi_{A^{''}}^{\rm cb+1},s\rangle$ & $0$ & $0$ & $0$ & $0$ & $\gamma_7^* q_{-}$ & $\gamma_9^* q_{+}$ & $\vareps_{c+1}$\\
& & & & & & &  \\ 
\hline\hline
 \end{tabular}
\end{table}

Concrete values for the parameters $\gamma_i$ can be obtained for each material  by, e.g., directly
evaluating the matrix elements $\langle \Psi_{\mu}^{b}|\hat{p}_{\pm}|\Psi_{\mu'}^{b'}\rangle$ using 
Kohn--Sham orbitals. We used this method to calculate the values denoted by $\gamma_{\rm KS}$ in 
Tables \ref{tbl:K-kp-params-1} and \ref{tbl:K-kp-params-2}.



\subsection{Spin--orbit coupling}

In the atomic approximation the SOC is given by the Hamiltonian
\begin{eqnarray}
 \mathcal{H}_{\rm so}^{\rm at}=\frac{\hbar}{4 m_e^2 c^2} \frac{1}{r} \frac{d V(r)}{d r} \,
 \mathbf{\hat{L}} \cdotp \mathbf{\hat{S}},
 \label{eq:atomic-SOC}
\end{eqnarray}
where $V(r)$ is the spherically symmetric atomic potential, $\mathbf{\hat{L}}$ is the  angular-momentum operator, and 
$\mathbf{\hat{S}}=(s_x,s_y, s_z)$ is a vector of spin Pauli matrices $s_x$, $s_y$, and $s_z$ (with eigenvalues $\pm 1$).
Note that $\mathbf{\hat{L}} \cdotp \mathbf{\hat{S}}=\hat{L}_z s_z + \hat{L}_{+} s_{-} + \hat{L}_{-} s_{+}$, 
where $\hat{L}_{\pm}= \hat{L}_{x} \pm i \hat{L}_{y}$ and 
$s_{\pm}=\frac{1}{2}(s_{x}\pm i s_{y})$. The task is then to calculate the matrix elements of Equation \eref{eq:atomic-SOC} in the 
basis introduced earlier in this section. 

The non-zero matrix elements $H_{\rm so}$ can be obtained by considering the transformation 
properties of the basis states and angular-momentum operators 
with respect to the mirror operation $\sigma_h$ and the rotation $C_3$.
Note that  in contrast to the Hamiltonian in Table \eref{tbl:kp-at-K}, the SOC 
Hamiltonian shown in Table \eref{tbl:SOC-at-K} has non-zero matrix elements between symmetric 
and antisymmetric basis states as well.
This is due to the fact that the $\hat{L}_{\pm}$ operators are themselves antisymmetric 
with respect to $\sigma_h$.
The full SOC Hamiltonian at $K$ is shown in Table \eref{tbl:SOC-at-K}.

\begin{table}[htb]
\caption{Matrix elements of $\mathcal{H}_{\rm so}^{\rm at}$ at the $K$ point.}
 \label{tbl:SOC-at-K}
\begin{tabular}{l|ccccccc}\hline\hline\vspace*{-0.8em}
 & & & & &  & & \\
$H_{\rm so}^{K}$  & $|\Psi_{A'}^{\rm vb},s\rangle$ & $|\Psi_{E_1^{'}}^{\rm cb},s\rangle$ & $|\Psi_{E_2^{'}}^{\rm vb-3},s\rangle$ & 
       $|\Psi_{E_2^{'}}^{\rm cb+2},s\rangle$ & $|\Psi_{E_1^{''}}^{\rm vb-2},s\rangle $ & $|\Psi_{E_2^{''}}^{\rm vb-1},s\rangle$ &
       $|\Psi_{A^{''}}^{\rm cb+1},s\rangle$ \\
 \vspace*{-0.8em}
 & & & & & & & \\
 \hline \vspace*{-0.8em}
  & & & & & & &  \\
 \vspace*{-0.8em}
 $|\Psi_{A'}^{\rm vb},s\rangle$ & $S_z\Delta_{v}$ & $0$ & $0$ & $0$ & $S_{-}\Delta_{v,v-2}^{}$ & 
$S_{+} \Delta_{v,v-1}^{}$ & $0$ \\
  & & & & & & &  \\       
\vspace*{-0.8em}
 $|\Psi_{E_1^{'}}^{\rm cb},s\rangle$ & $0$ & $S_z\Delta_{c}$ & $0$ & $0$ & $0$ & $S_{-}\Delta_{c,v-1}^{}$ & $S_{+}\Delta_{c,c+1}^{}$ \\
  & & & & & & &  \\         
\vspace*{-0.8em}     
$|\Psi_{E_2^{'}}^{\rm vb-3},s\rangle$  & $0$  & $0$ & $S_z\Delta_{v-3}$ &  $S_{z}\Delta_{v-3,c+2}$ & $S_{+}\Delta_{v-3,v-2}$ & $0$ & 
$S_{-}\Delta_{v-3,c+1}$ \\
 & & & & & & &  \\         
\vspace*{-0.8em}     
 $|\Psi_{E_2^{'}}^{\rm cb+2},s\rangle$ & $0$ & $0$ & $S_{z}\Delta_{v-3,c+2}^{*}$ & $S_z \Delta_{c+2}$ & $S_{+}\Delta_{c+2,v-2}^{}$ 
 & $0$ & $S_{-}\Delta_{c+2,c+1}^{}$\\
 & & & & & & &  \\         
\vspace*{-0.8em} 
 $|\Psi_{E_1^{''}}^{\rm vb-2},s\rangle $& $S_{+}\Delta_{v,v-2}^{*}$ & $0$ & $S_{-}\Delta_{v-3,v-2}^{*}$ & $S_{-}\Delta_{c+2,v-2}^{*}$ 
 &  $S_z\Delta_{v-2}$ & $0$ & $0$ \\
 & & & & & & &  \\         
\vspace*{-0.8em} 
$|\Psi_{E_2^{''}}^{\rm vb-1},s\rangle$ & $S_{-}\Delta_{v,v-1}^{*}$ & $S_{+}\Delta_{c,v-1}^{*}$ & $0$ & $0$ & $0$ & $S_z\Delta_{v-1}$ & $0$\\
& & & & & & &  \\         
\vspace*{-0.8em} 
 $|\Psi_{A^{''}}^{\rm cb+1},s\rangle$ & $0$ & $S_{-}\Delta_{c,c+1}^{*}$ & $S_{+}\Delta_{v-3,c+1}^{*}$ & $S_{+}\Delta_{c+2,c+1}^{*}$ 
 & $0$ & $0$ & $S_z\Delta_{c+1}$\\
& & & & & & &  \\ 
\hline\hline
 \end{tabular}
\end{table}

The SOC Hamiltonian at $-K$ can be obtained by making the following substitutions: 
$|\Psi_{E_2^{'}}^{\rm vb-3},s\rangle \rightarrow |\Psi_{E_1^{'}}^{\rm vb-3},s\rangle$, 
$|\Psi_{E_1^{'}}^{\rm cb},s\rangle \rightarrow |\Psi_{E_2^{'}}^{\rm cb},s\rangle$,
$|\Psi_{E_1^{''}}^{\rm vb-2},s\rangle \rightarrow |\Psi_{E_2^{''}}^{\rm vb-2},s\rangle$,
$|\Psi_{E_2^{''}}^{\rm vb-1},s\rangle \rightarrow |\Psi_{E_1^{''}}^{\rm vb-1},s\rangle$,
$\Delta_{b,b'}\rightarrow \Delta_{b,b'}^{*}$, $S_{\pm}\rightarrow -S_{\mp}$, $S_z\rightarrow -S_z $. 
The change of the wave-function symmetry notation follows from the assumption that 
orbital wave functions at $K$ and $-K$ are connected by time-reversal symmetry, i.e.,
$|\Psi_{\mu}^{b}(K)\rangle = \hat{K}_0 |\Psi_{\mu'}^{b}(-K)\rangle$, where $\hat{K}_0$ denotes 
complex conjugation.


\subsection*{Low-energy effective Hamiltonian}

The low-energy effective Hamiltonian of Equation \eref{eq:Hkp-K} can be obtained from 
$H_0+H_{\mathbf{k}\cdot\mathbf{p}}+H_{\rm so}$ by means of L\"owdin partitioning 
(see, e.g., Reference \cite{winkler-book})
by considering terms up to third order in various off-diagonal couplings.


\section{Fitting procedure at the $K$ point}
\label{sec:fitting-at-K}

The aim of this section is to explain the fitting procedure that we used to 
extract the material parameters that appear in the Hamiltonians of Equations
\eref{eq:kp-massDirac}--\eref{eq:kp-cub1} from our DFT calculations 
(see Tables \ref{tbl:K-kp-params-1} and \ref{tbl:K-kp-params-2}).   
In order  that the parameter sets obtained can be compared to other works, 
we think that it  is important to give some details of the fitting procedure.

To simplify the notation, we consider the $K$ point and suppress the $\tau$ index. 
The eigenvalues of the low-energy Hamiltonian of Equation \eref{eq:full-eff-Ham-at-K} read
\begin{eqnarray}
 \fl
 E_{\pm}^{(s)}=\frac{\tilde{\vareps}_{\rm vb}+\tilde{\vareps}_{\rm cb}}{2}+\left(\frac{\hbar^2}{2 m_e}+\frac{\alpha_{s}+\beta_{s}}{2}\right) \mathbf{q}^2 
 \pm \sqrt{\left(\frac{\tilde{\vareps}_{\rm cb}-\tilde{\vareps}_{\rm vb}}{2}+\frac{\beta_{s}-\alpha_{s}}{2} \mathbf{q}^2\right)^2+f(\mathbf{q})},    
 \label{eq:eigenvl-cub-appndx}\\
 \fl
 f(\mathbf{q})=|\gamma|^2\mathbf{q}^2+|\mathbf{q}|^3 |\gamma| |\kappa_s| 2 \cos(\theta_{\kappa_s\gamma}+3 \varphi_{\mathbf{q}})+
 \mathbf{q}^4 [|\kappa_s|^2-|\gamma||\eta_{s}|\cos(\theta_{\eta_s\gamma})]
\end{eqnarray}
where  $\tilde{\vareps}_{\rm vb}=\vareps_{\rm vb}+\tau\,s\,\Delta_{\rm vb}$  and similarly for  $\tilde{\vareps}_{\rm cb}$,
$\varphi_{\mathbf{q}}={\rm arctan}(q_y/q_x)$, 
$\theta_{\kappa_s\gamma}$ ($\theta_{\eta_s\gamma}$) are the relative phase of 
$\kappa_s$ and $\gamma$ ($\eta_s$ and $\gamma$),  
and $+$ ($-$) sign corresponds to the 
CB (VB)\@.  Since Equation \eref{eq:eigenvl-cub-appndx} depends on the parameters
$\gamma$, $\alpha_{s}$, $\beta_{s}$, $\kappa_{s}$, and $\eta_{s}$
in a non-linear way,  some care has to be taken in the fitting procedure. 

First, one can determine $|\gamma|$, $\alpha_{s}$, and $\beta_{s}$ in the following way.
For small enough $\mathbf{q}$,  the largest energy scale under the square root in Equation \eref{eq:eigenvl-cub-appndx} is the band gap 
(for a given spin $s$)
$E_{\rm bg}^{(s)}=\tilde{\vareps}_{\rm cb}-\tilde{\vareps}_{\rm vb}$. Expanding the square root   one finds 
\begin{eqnarray}
E_{+}^{(s)}\approx  \tilde{\vareps}_{\rm cb}+\left(\frac{\hbar^2}{2 m_e}+\beta_{s}+\frac{|\gamma|^2}{E_{\rm bg}^{(s)}}\right) 
\mathbf{q}^2\nonumber\\
E_{-}^{(s)}\approx  \tilde{\vareps}_{\rm vb}+\left(\frac{\hbar^2}{2 m_e}+\alpha_{s}-\frac{|\gamma|^2}{E_{\rm bg}^{(s)}}\right) 
\mathbf{q}^2.
\end{eqnarray}
In this approximation  $E_{\pm}^{(s)}$ is described by 
a simple parabolic dispersion where  the effective masses are given by
\begin{eqnarray}
\frac{\hbar^2}{2 m_{\rm cb}^{(s)}} = \left(\frac{\hbar^2}{2 m_e}+\beta_{s}+\frac{|\gamma|^2}{E_{\rm bg}^{(s)}}\right)
\label{eq:effmass-at-K-1}\\
\frac{\hbar^2}{2 m_{\rm vb}^{(s)}} =\left(\frac{\hbar^2}{2 m_e}+\alpha_{s}-\frac{|\gamma|^2}{E_{\rm bg}^{(s)}}\right)
\label{eq:effmass-at-K-2}
\end{eqnarray}
Since  $E_{\rm bg}^{(s)}$ can be directly read off from the DFT calculations and 
$2 m_{\rm cb}^{(s)}$ and $2 m_{\rm vb}^{(s)}$ can be  obtained by fitting the CB and VB in the vicinity of 
the $K$ point with a parabola, Equations \eref{eq:effmass-at-K-1}--\eref{eq:effmass-at-K-2} 
constitute four equations for 
five unknown parameters $|\gamma|$, $\alpha_{s}$, and $\beta_{s}$.  
As explained in Section \ref{subsec:K-point-params}, 
the fitting around $K$ was done  in a range that corresponds to $5\%$ of the $\Gamma$--$K$ distance.  
The dispersion over this range can be considered to be isotropic and the difference 
in the effective masses along $K$--$\Gamma$ and $K$--$M$ can be neglected. 
Over the same range in $\mathbf{q}$,  one can also fit the function   
$\frac{\tilde{\vareps}_{\rm vb}+\tilde{\vareps}_{\rm cb}}{2}+c_1^{(s)}\mathbf{q}^2+\sqrt{(E_{\rm bg}^{(s)})^2/4+c_2^{(s)} \mathbf{q}^2}$ 
to the CB and the function 
$\frac{\tilde{\vareps}_{\rm vb}+\tilde{\vareps}_{\rm cb}}{2}+c_1^{(s)}\mathbf{q}^2-\sqrt{(E_{\rm bg}^{(s)})^2/4+c_2^{(s)} \mathbf{q}^2}$
to the VB such that the  fitting parameters $c_1^{(s)}$ and $c_2^{(s)}$ simultaneously 
give the best fit to the  dispersion both in the CB and in the VB\@.  
Comparing to Equation \eref{eq:eigenvl-cub-appndx}, one can see that this corresponds to
\begin{eqnarray}
 \frac{\alpha_{s}+\beta_{s}}{2}=c_1^{(s)}-\frac{\hbar^2}{2 m_e},
 \label{eq:C1}\\
 \frac{\beta_{s}-\alpha_{s}}{2}+|\gamma|^{2}=\frac{2 \,c_2^{(s)}}{E_{\rm bg}^{(s)}},
 \label{eq:C2}
\end{eqnarray}
i.e., we have obtained four more equations for  $|\gamma|$, $\alpha_{s}$, and $\beta_{s}$.  
Using Equations \eref{eq:effmass-at-K-1}--\eref{eq:C2} one finds eight equations 
for the five unknown parameters $|\gamma|$, $\alpha_{s}$, and $\beta_{s}$, which can be solved 
as a linear least-squares problem.  The solution, however, depends on the value of  the quasiparticle 
band gap $E_{\rm bg}^{(s)}/2$ used in the least-squares problem. 
As shown in Table \ref{tbl:K-bandgap} this is significantly underestimated in  DFT calculations. 
Therefore we have performed the fitting  using both the DFT band gap and  the $GW$  gap. 
Note that in order to find $E_{\rm bg}^{(s)}$ one has to add to the $E_{\rm bg}^{}$ values shown 
in Table \ref{tbl:K-bandgap} the relevant spin-splitting energies, which can be
found in Tables \ref{tbl:K-dftparams-cb} and \ref{tbl:K-dftparams-vb}. 
The two approach lead to  two sets of parameters. 
In  both cases  the \emph{same} effective masses $m_{\rm cb}^{(s)}$ and $m_{\rm vb}^{(s)}$, 
obtained from our  DFT calculations, were used. 
Since the available experimental results suggest  that DFT can capture the effective masses quite well, 
at least in the VB (see Tables \ref{tbl:K-dftparams-vb} and \ref{tbl:dft-at-G}),  
we think that this is a reasonable approach to take into account the results of $GW$ calculations.

Finally the four remaining  parameters $\kappa_{s}$ and $\eta_{s}$ were determined in following way. 
Similarly to the previous step, a  function of the form 
\begin{eqnarray}
 \frac{\tilde{\vareps}_{\rm vb}+\tilde{\vareps}_{\rm cb}}{2}+c_1^{(s)}\mathbf{q}^2
 \pm\sqrt{(E_{\rm bg}^{(s)})^2/4+c_2^{(s)} \mathbf{q}^2+c_{3}^{(s)}|\mathbf{q}|^{3}\cos3\phi_{\mathbf q}+c_{4}^{(s)} \mathbf{q}^{4}}
\end{eqnarray}
was fitted to the VB and CB\@. Here $c_{1}^{(s)}$ and $c_{2}^{(s)}$ were kept fixed at the values  
that were obtained at the previous step
and the  parameters $c_{3}^{(s)}$ and $c_{4}^{(s)}$  were required to give the best fit simultaneously 
to both the CB and the VB\@. The fitting was performed along the $\Gamma$--$K$--$M$ directions around $K$ 
and the fitting range  corresponded to $\approx 16\%$ of the $\Gamma$--$K$ distance.  
Note, that $\cos3\phi_{\mathbf q}=-1$ ($\cos3\phi_{\mathbf q}=1$) along $\Gamma$--$K$ ($K$--$M$). 
Since $|\gamma|$, $\alpha_{s}$, and $\beta_{s}$ are  already know by this step, 
$\kappa_{s}$ and  $\eta_{s}$ is calculated as $\kappa_{s}=c_{3}^{(s)}/(2|\gamma|)$ and 
$\eta_{s}=[(\beta_{s}-\alpha_{s})^2/4-c_{4}^{(s)}]/|\gamma|$ [c.f., Equation \eref{eq:eigenvl-cub-appndx}].


\section*{References}


\begin{thebibliography}{999}
\bibitem{priv-com-effmass}
The fit results based on the ARPES measurements of Reference \cite{shen} 
were communicated to us by Yi Zhang.
\end{thebibliography}
\end{document}